\documentclass{aa}

\let\linenumbers\relax                       
\idline{A\&A, accepted (pre-print Oct 2026)}              

\usepackage{graphicx}    
\usepackage{float}  
\usepackage{cancel} 
\usepackage{txfonts}
\usepackage{orcidlink} 
\usepackage{lscape} 
\usepackage{placeins} 
\usepackage{enumitem} 
\usepackage{subcaption} 
\usepackage{xspace} 
\usepackage{longtable} 
\usepackage{lastpage} 

\usepackage{hyperref}
\hypersetup{
    breaklinks=true,
    colorlinks=true,
    citecolor=blue,
    linkcolor=magenta,
    filecolor=red,      
    urlcolor=cyan,
    pdftitle={Chromospheric Mass Loss in Giants},
    pdfpagemode=FullScreen,
    pdfauthor={Ian Baeza},%
    pdfkeywords={Chromospheres,Giants,mass-loss,H alpha, 47 Tuc},%
    }
\begin{document} 


\titlerunning{Chromospheric motion in red giants}
\authorrunning{Ian Baeza et al.}

\title{The variable mass-loss in low-mass red giant stars}
\subtitle{Chromospheric motion across stellar evolution in 47 Tuc}

\author{%
Ian Baeza  
\inst{1,2}
\fnmsep\thanks{Contact:i.baezamardones@uandresbello.edu}
\orcidlink{0000-0002-9881-6336};
L. Monaco    \inst{1} \orcidlink{0000-0002-3148-9836};
H. Frelijj   \inst{1} \orcidlink{0000-0002-3639-7877}; 
Y. Momany    \inst{3} \orcidlink{0000-0001-8946-8723};
L. Sbordone  \inst{2} \orcidlink{0000-0002-2285-8708};
S. Zaggia    \inst{3} \orcidlink{0000-0001-6081-379X};\\ 
I. Yegorova  \inst{2} \orcidlink{0000-0002-4668-4677};
I. Saviane  \inst{2} \orcidlink{0000-0002-5878-5299}; 
A. Mucciarelli \inst{4,5} \orcidlink{0000-0001-9158-8580};
K. Ohnaka   \inst{6} \href{https://scixplorer.org/search?q=author:"Ohnaka,+Keiichi"+collection:astronomy+property:refereed&sort=date+desc}{\includegraphics[width=2.75mm]{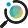}};
M. Gómez    \inst{6} \orcidlink{0000-0002-4430-9427} 
}

\institute{%
Universidad Andres Bello, Facultad de Ciencias Exactas, Departamento de Física y Astronomía - Instituto de Astrofísica, Autopista Concepción-Talcahuano 7100, Talcahuano, Chile
\and
European Southern Observatory, Alonso de Cordova 3107, Vitacura, Casilla 19001, Santiago de Chile, Chile
\and
Osservatorio Astronomico di Padova-INAF, Vicolo dell’Osservatorio 5, I-35122 Padova, Italy
\and
Dipartimento di Fisica e Astronomia, Università degli Studi di Bologna, via Gobetti 93/2, 40129 Bologna, Italy
\and
INAF – Osservatorio di astrofisica e scienza dello spazio di Bologna, Via Gobetti 93/3, 40129 Bologna, Italy
\and
Universidad Andres Bello, Instituto de Astrofísica, Facultad de Ciencias Exactas, Departamento de Física y Astronomía, Av. Fernández Concha 700, Las Condes, Santiago, Chile
}

\date{Received July 31, 2026; Accepted September 27, 2026}


\abstract{%
%
Mass-loss ({${ML}\ $})is a critical process that remains poorly understood despite its broad implications across astrophysics. Our current knowledge of the underlying physics is still insufficient to reproduce {${ML}\ $}consistently across stars, particularly in the cool low-mass giants regime. A major limitation is likely the intrinsically time-variable nature of chromospheres, the region where most of the physical phenomena powering {${ML}\ $}in this specific regime are expected to occur.
%
%
In this work we focus on the variable nature of {${ML},$} aiming to characterize the relevant timescales of chromospheric variability in red giants, and the correlations with intrinsic stellar properties.
%
%
We analyzed the first large temporal spectroscopic monitoring of chromospheric variability using the {${H\alpha}\ $}line in red giant stars belonging to the globular cluster NGC-104 (47 Tuc). Our dataset consisted of multi-epoch FLAMES/GIRAFFE multi-object spectroscopy spanning one year and covering 115 cluster stars.
%
%
We found that after $\sim$4 days, for $\log(L/L_\odot)\gtrsim2.4$, variations in {${H\alpha}\ $} on average increase and widen their range of values with increasing luminosity, together with a systematic blueshift of the line core --which is indicative of outflows-- and deeper cores --which are consistent with increased chromospheric optical depth. The characteristic timescale required to observe variations in the {${H\alpha}\ $} profile is only about one day, showing a flat trend for all stars with luminosity. While shorter timescale variations remain consistent with the statistical uncertainties of our measurements, we found a possible increase toward dimmer stars. All of these trends appear similarly correlated with luminosity and effective temperature. No further correlations were found with the different cluster populations, nor were significant differences observed between asymptotic giant branch and red giant branch stars. The outflow trends closely match previous observations across a wide metallicity range.
%
%
We discuss whether the observations reflect changes in {${H\alpha}\ $}sensitivity to the winds, a genuine atmospheric transition driven by changes in heating efficiency, or continued chromospheric expansion with geometrically dominated {${H\alpha}\ $}formation.
}

%
%
%
%
\maketitle


\section{Introduction}
Mass loss ({${ML}$}) is a fundamental process experienced by all stars throughout their evolution. Its impact extends well beyond stellar physics, influencing galaxy evolution, supernova rates, the mass function of compact objects, and the chemical enrichment of the interstellar and intergalactic medium. Since the formation of the first stars, {${ML}\ $}has played a central role in shaping the Universe.

Evidence of stellar {${ML}\ $}has long been established through multiple independent approaches, including horizontal branch (HB) morphology \citep{1973-Rood-0.2ML-in-RGB,2008-Caloi-0.2ML+-0.001-RGB,2016-Salaris-response-Heyl-47Tuc-constrains-ML-RGB,2021-Tailo-new-ML-law-RGB-SSP,2020-Tailo-ML-on-46-GC-first-generation,2025-Tailo-differential-mass-loss-in-populations}; initial--final mass relations \citep[i.e., the mass difference between white dwarfs and their progenitors;][]{1965-Auer-white-dwarfs-on-hyades-ML,2018-Cummings-initial-final-mass-relation}; intracluster medium studies \citep[see][and references therein]{2024-Pancino-diferential-redenning-intracluster-mass}; white dwarf kinematics in clusters \citep[e.g.,][]{2015-Heyl-white-dwarfs-indicate-ML-on-AGB}; maser, radio, infrared, and ultraviolet diagnostics \citep[e.g.,][]{1988-De-Jager-ML-across-HR,1999-Lamers-book-stellar-winds,2011-McDonald-spitzer-spectra-dust,2012-Momany-ending-discussion-origlia,2022-Vink-ARAA-mass-loss-blue-giants}; asteroseismology \citep{2022-Howell-ML-on-RGB-bump,2024-Howell-astereosismic-different-ML-subpopulation,2025-Howell-asteroseismic-masses-integrated-ML}; and spectral line diagnostics \citep[e.g.,][]{1992-Dupree-He-line-missing-mass-intracluster,1994-Dupree-split-of-halpha-magnesium-line,2004-Cacciari-Flames-template-star,2009-Meszaros-III}. Despite this extensive observational evidence, the physical mechanisms driving {${ML}\ $}remain poorly understood. Current models struggle to reproduce {${ML}\ $}consistently across the {Hertzsprung–Russell diagram (HRD)}, particularly during transitions between different driving regimes and across varying stellar parameters. This difficulty largely reflects our incomplete understanding of the underlying physics -- including fluid dynamics, convection, and magnetic field generation -- where simplified approximations often fail. As a result, stellar evolution calculations commonly rely on (semi-)empirical prescriptions \citep[e.g.,][]{1975-Reimers-law-I,1977-Reimers-law-II,2005-Schroder-new-version-reimers-law}, which are known to inadequately reproduce {${ML}\ $}trends across different populations and metallicities \citep[e.g.,][]{2009-Meszaros-III,2025-Li-trend-with-metallicity}.

Several physical origins have been suggested to sustain {${ML}\ $}in single stars, with their relative dominance usually being connected to structural parameters such as stellar mass, metallicity, surface gravity, and effective temperature. For simplicity, the physical drivers can be grouped into three broad classes:

\begin{itemize}[topsep=3pt, partopsep=1pt]
\item {Mechanical lifting:} Momentum from internal stellar processes, such as pulsations, accelerates material outward \citep{2008-Van-Loon-pulsation-relation-wind}. This can occur on global scales and/or locally (e.g., via convective overshooting; \citealt{2023-Freytag-simulation-dust-wind-AGB}).\vspace{1mm}
\item {Radiation pressure:} Photons transfer momentum to material in the outer layers. This mechanism depends strongly on chemical composition and has been successfully applied to both atomic line driving \citep{1975-Castor-Radiation-driven-winds-O-stars,2022-Vink-ARAA-mass-loss-blue-giants} and molecules and dust grains \citep{2016-Hofner-PEDRO-model-ALSIO-able-to-drive-pulsation}.\vspace{1mm}

\item {Chromospheric \& Coronal winds:} Magneto-acoustic processes transfer energy to the plasma in chromospheres and coronas through heating and mechanical work, creating the outward acceleration of the material \citep{2011-Cranmer-mass-loss-theory-cool_stars,2007-suzuki-MHD-RED-giant-ML-variable,2005-Schroder-new-version-reimers-law}.
\end{itemize}

These processes are not mutually exclusive and most likely operate simultaneously, with their relative contributions on {${ML}\ $}rates evolving as the star evolves. 

For low-mass dwarf stars, {${ML}\ $}is dominated by chromospheric/coronal winds%
\footnote{Chromospheres and coronas disappear alongside the convective envelope. A clear threshold has been observed at spectral types earlier than late A/early F \citep{1991-Simon-onset-of-chormospheric-activity,2004-Gudel-review-stellar-corona}. On the cool side, coronas vanish beyond the ``dividing line'' identified by \cite{1979-Linsky-Haisch-Dividing-line}, which appears near K1 in giants and F0 for supergiants. \citep[see Fig. 1 of][]{2010-Bennet-wind-on-all-stars-image}. This is thought to occur because extended chromospheres bury magnetic regions that are expected to be responsible for heating plasma to coronal temperatures, while low surface gravity allows particles to escape before reaching such temperatures \citep{2003-Ayres-buried-corona-in-chromosphere,2007-suzuki-MHD-RED-giant-ML-variable,2011-Cranmer-mass-loss-theory-cool_stars}.}%
, whose theoretical description broadly follows the theoretical framework introduced and further developed by \cite{1958-Parker-wind,1964-Parker-wind-series-paper-1}, although important caveats remain \citep[see][for current considerations]{2025-Song-how-winds-reach-supersonic-speed}. In massive hot stars, {${ML}\ $}is instead driven by radiation pressure on spectral lines, a mechanism strongly dependent on metallicity that has at least two regimes that behave differently \citep[optically thin and thick winds, see][]{2022-Vink-ARAA-mass-loss-blue-giants,2025-Vink-50-years-of-CAK}. While the theoretical framework is comparatively mature and successful, model predictions still differ by about an order of magnitude \citep{2026-Romagnolo-models-ML-order-of-mag-diff}.

In evolved cool stars with extended atmospheres, the {${ML}\ $}picture becomes more complex. For red/yellow supergiants, it remains unclear under which conditions individual processes dominate \citep[see][for a recent review]{2025-Van-Loon-review-red-supergiants}. In low- and intermediate-mass stars, evidence points to distinct {${ML}\ $}regimes that depend on mass and evolutionary stage. At the bright end of the {red giant branch (RGB)} and the {asymptotic giant branch (AGB)}, {${ML}\ $}can be attributed to pulsation-driven or pulsation-assisted dusty winds: Pulsations lift material into the outer atmosphere, dust condenses, and radiation pressure accelerates the dust, which subsequently drags the gas \citep[the {``dusty''} or {``dusty-pulsation''} mechanism;][]{2018-Hofner-review-AGB,Habing_Olofsson_2004_book_AGB}.

The relative roles of pulsation and radiation pressure driving the wind have been explored by \cite{2018-McDonald-AGB-lowL-ML-dominated-pulsation-mass-of-the-star} \citep[see also][]{2016-Mcdonald-zijilstra-60-pulsation-limit,2021-Yu-60-days-period-massloss-miras-does-not-dpend-L,2025-Schirmer-radiation-pressure-not-enough}, who found that {${ML}\ $}correlates with pulsation once regular variability with periods $\gtrsim 60$ days develops, approximately coinciding with the onset of first-overtone pulsation. For giants with $L \lesssim 4000\,L_\odot$, the lack of a clear correlation between luminosity and {${ML}\ $}suggests that stellar mass (hence still pulsation) may be the dominant factor. At higher luminosities ($L \gtrsim 4000\,L_\odot$) or for pulsation periods $\gtrsim 300$ days, {${ML}\ $}correlates with luminosity, which is consistent with fundamental-mode pulsation. At even longer periods ($\sim 700$ days) or $L \sim 10^4\,L_\odot$, {${ML}\ $}reaches a plateau \citep{1993-Vassiliadis-ML-on-AGB-model}, likely reflecting the limit of optically thick winds where most of the available opaque material is ejected.

\begin{figure*}[t]
\centering
     \begin{subfigure}[t]{0.68\textwidth} 
        \caption{}
        \label{fig:CMD}
        \includegraphics[width=1.0\linewidth]{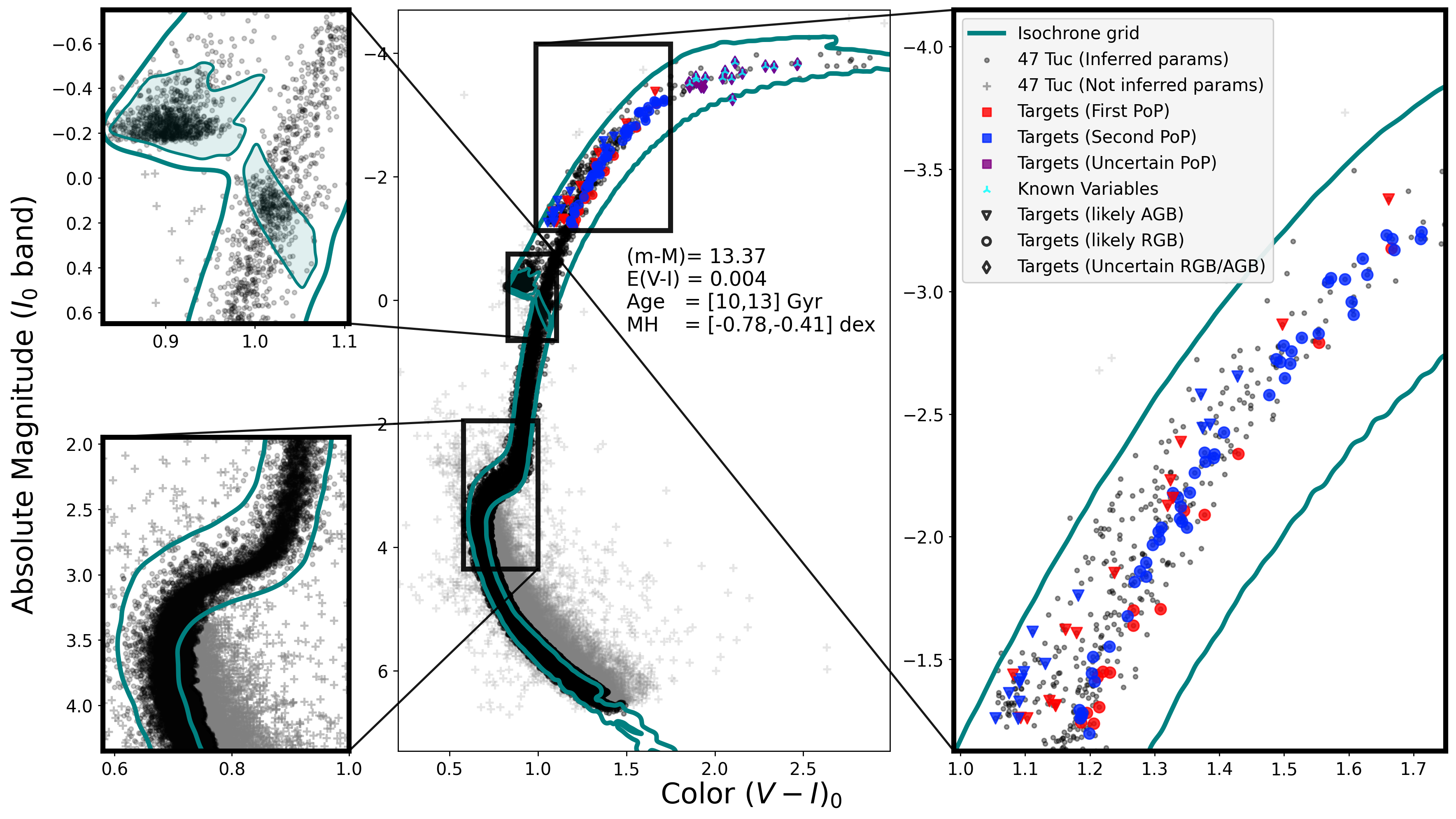}
    \end{subfigure}%
     \begin{subfigure}[t]{0.28\textwidth} 
        \caption{}
        \label{fig:kielDG}
        \includegraphics[width=1.0\linewidth]{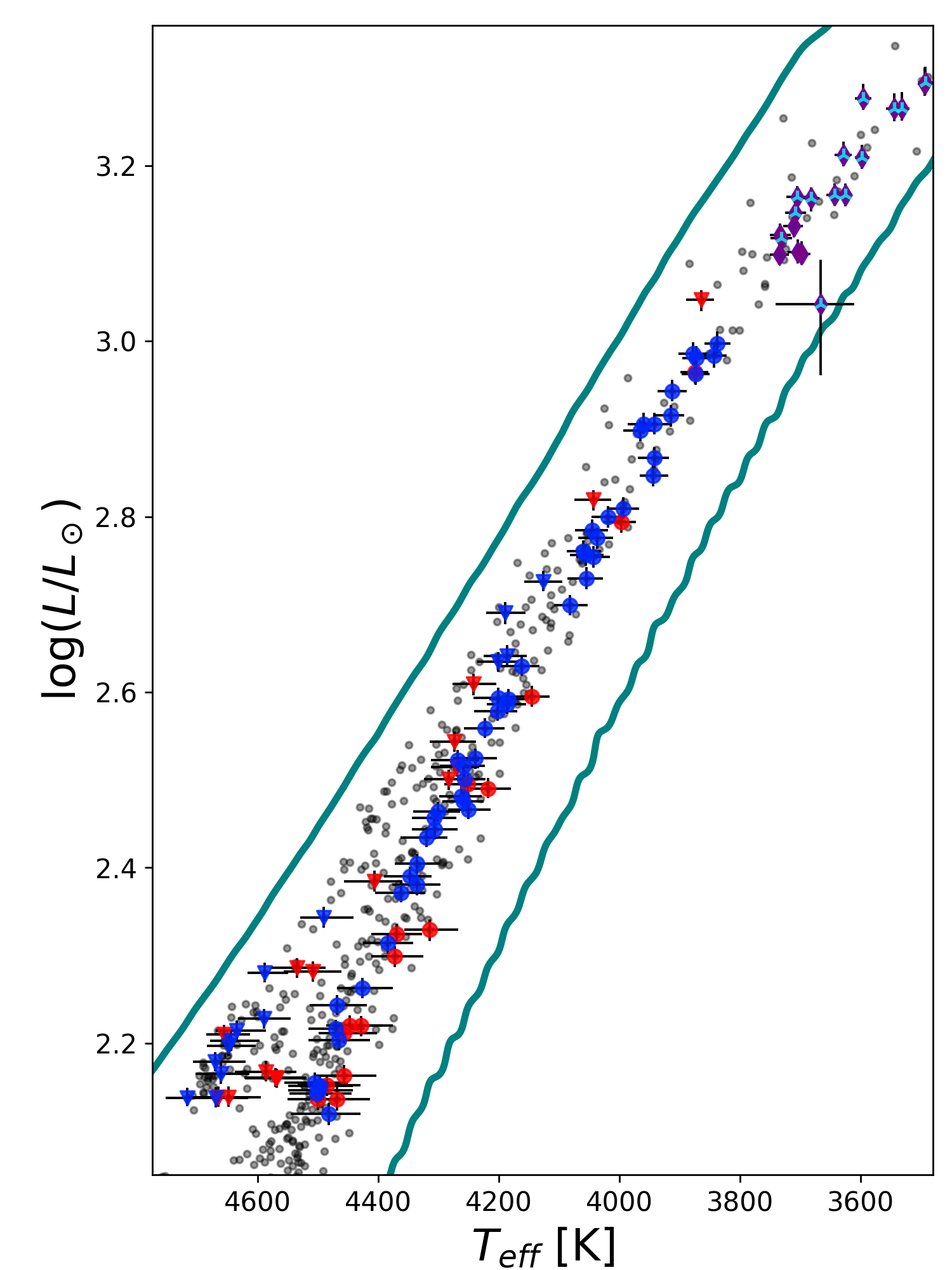}
    \end{subfigure}%
    \caption{
    \label{fig:CMD-kiel}
    {\bf (a):} CMD from the photometry of \citet{2019-Stetson-Photometry} for members with a probability $>0.95$ \citep{2021-vasiliev-likely-members-GC}. The dark green curve approximately encloses the adopted isochrone grid (\S.~\ref{sec:isochrone-fitting-section}, App.~\ref{appendix-ngc104}), shifted by the distance modulus and reddening quoted in the panel. The two smaller contours enclose the horizontal branch (upper) and RGB bump (lower) predicted by the isochrones with the used shift. Red, blue, and purple mark first-population, second-population and uncertain-population stars; triangles, circles, and diamonds mark AGB, RGB, and uncertain RGB/AGB classification; and cyan marks known variables. Left insets show zoom-ins on the HB/RGB-bump and turn-off regions. The right inset shows a zoom-in on the region of stars where we could separate populations.
    {\bf (b):} HRD for our stars with values and uncertainties inferred through the multiple-isochrone fitting (see, e.g., Fig. \ref{iso-fitting}). Symbols are as in (a). The dark green shows the space of parameters covered by the isochrone grid. 
    }
\end{figure*}

At lower luminosities ---RGB and {early AGB (E-AGB)}--- {${ML}\ $}is instead expected to be dominated by chromospheric winds driven by the dissipation of magneto-acoustic waves. The behavior of {${ML}\ $}in this regime has been explored much less, with theoretical predictions suggesting that it should become significantly stronger after the RGB bump%
\footnote{RGB bump: Excess in the luminosity function of RGB stars originating from the encounter of the hydrogen burning shell of these stars with the chemical discontinuity left behind by the deepest penetration of their convective envelope \citep[first dredge-up;][]{2002-Salaris-RGB-review}.}
\citep{2025-Macdonald-mullan-plasmoid-mass-loss,2018-Valle-simulation-ML-after-bump}, while observational evidence indicates enhanced activity roughly above $\log(L_\star/L_\odot)\gtrsim2.5$ \citep{2009-Meszaros-II}. Among the proposed mechanisms, magnetosonic and Alfvén waves are considered likely contributors. Owing to their large damping lengths, these waves could efficiently transfer energy and momentum to the outer atmosphere \citep{1983-Holzer-damping-length-alfen-waves-in-giants-not-able-to-create-slow-winds,2010-Bennet-wind-on-all-stars-image,2011-Cranmer-mass-loss-theory-cool_stars,2019-Yasuda-types-of-wind---dusty-slow-more-ML-rate-than-windy-fast,2025-Suzuki-alfven-waves-driving-ML}. This process can be referred to as the {``gaseous''} {${ML}\ $}mechanism. Therefore, chromospheric diagnostics --such as {${H\alpha}\ $}-- provide powerful insights into {${ML},$} as they trace mass motions at large heights above the photosphere and are directly linked to the physical processes driving the wind \citep{1994-Dupree-split-of-halpha-magnesium-line,1981-Vernazza-original-model-chromosphere-lines,2004-Cacciari-Flames-template-star,2008-Meszaros-I,2009-Meszaros-II,2009-Meszaros-III,2008-Hall-review-stellar-activity,2017-Linsky-ARAA-chromospheric-lines}. In this study, we focus on this group ({evolved low-mass stars}, $M_{\rm init} \sim 0.85 \pm 0.2 \, M_\odot$). Hereafter, the terms ``giants'' and ``stars'' refer exclusively to this population.

Understanding the efficiency of this gaseous mechanism is crucial because although the AGB phase exhibits more intense winds, most of the {${ML}\ $}in these stars occurs along the RGB due to its much longer evolutionary timescale \citep{2018-Valle-simulation-ML-after-bump}%
.\footnote{
This behavior contrasts with more massive stars, for which the bulk of {${ML}\ $}occurs during the AGB \citep{2018-Hofner-review-AGB}.
} The prevailing view is that these stars lose $\sim0.2\pm0.1\,M_\odot$ along the RGB \citep{1973-Rood-0.2ML-in-RGB,2021-Tailo-new-ML-law-RGB-SSP}, with an additional $\sim0.1\pm0.1\,M_\odot$ on the AGB \citep{1993-Vassiliadis-ML-on-AGB-model,2011-McDonald-ML-0.2RGB-0.1-AGB,2018-Valle-simulation-ML-after-bump}. Such estimates are derived primarily from studies of {globular clusters (GCs)}, which provide ideal laboratories for {${ML}\ $}investigations. Their stellar populations share nearly identical values of age, distance, and overall metallicity, differing mainly in initial mass, thereby enabling direct evolutionary comparisons. However, GCs are not strictly single stellar populations. They exhibit star-to-star abundance variations in light elements, producing well-established (anti)correlations and {multiple stellar populations} \citep[{MPs}; the origin of these chemical patterns remains debated;][]{2018-Bastian-and-lardo-anual-review-MP}. Recent studies suggest that these subpopulations must also be considered in {${ML}\ $}investigations, as they seem to be linked to the integrated amount of mass lost \citep{2024-Howell-astereosismic-different-ML-subpopulation,2020-Tailo-ML-on-46-GC-first-generation,2021-Tailo-new-ML-law-RGB-SSP,2025-Tailo-differential-mass-loss-in-populations} and may even prevent some stars from reaching the AGB phase \citep[AGB manqué;][]{2016-MacLean-AGB-stars-only-from-1sGen-on-M4,2021-Lagiola-star-avoid-AGB-multiple-population}.

Despite this coherent global picture, a long-standing tension remains. Attempts to reconcile instantaneous {${ML}\ $}diagnostics with the time-averaged rates required by stellar evolution models/observations have consistently failed \citep[e.g.,][]{2009-Meszaros-III,2025-Li-trend-with-metallicity}. A likely contributor to this discrepancy is the intrinsically variable nature of {${ML}.$} Although variability has been reported for decades \citep[e.g.,][]{1983-Cacciari-Freeman-variability-few-days,1988-Smith-Dupree-wing-emision-h-alpha-variable,1998-Mullan-variations-on-winds-masive-and-low-mass-giant,2009-Meszaros-III}, it remains poorly constrained. In contrast to the steady winds assumed in classical prescriptions \citep{1999-Lamers-book-stellar-winds}, both observations and recent (magneto)hydrodynamic (MHD) models indicate that stars experience episodic and maybe stochastic mass ejections \citep{2025-Suzuki-alfven-waves-driving-ML,2023-Freytag-simulation-dust-wind-AGB}. The mismatch between instantaneous tracers and evolutionary averages likely reflects this temporal complexity, underscoring the need for observational studies capable of directly studying/constraining variability.

In this paper, we investigate the gaseous {${ML}\ $}mechanism from an observational perspective through a multi-epoch analysis of the chromospheric {${H\alpha}\ $}line. Our primary aim is to characterize the variable nature of {${ML}\ $}in the relatively controlled environment of the GC 47 Tuc.


\section{Targets, observations, and data reduction}
\label{sec:obs_data} %

\subsection{NGC 104 (47 Tucanae)}
\label{subsec:ngc104}

For this work, the GC \object{NGC-104}/47 Tuc was selected due to its suitability for our objectives. As one of the nearest GCs, it allows us to obtain high-quality, high {signal-to-noise ratio (S/N)} spectra for stars down to luminosities of $\sim100${}$L_{\odot}$\xspace within a reduced amount of observation time. This makes it an ideal target for a multi-epoch study such as ours.

Despite its proximity, several of its fundamental properties remain highly debated. Its distance is reported to be in the range of 4.29 kpc to 5.29 kpc. The iron abundance [Fe/H]\xspace has been measured to be between $-0.78$ and $-0.66\ \text{dex}$, while the overall metallicity [M/H]\xspace, has been estimated between $-0.85$ and $-0.41\ \text{dex}$. However, the most debated parameter for 47 Tuc is arguably its age, with reported values ranging from $9.6\ Gyr $ to $13.06\ Gyr$, although older estimates even suggested ages as high as $17-20\ Gyr$ (see Appendix \ref{appendix-ngc104} and Table \ref{table-ng104-properties} for a brief review of references for these values). These literature values are primarily relevant for our study as a guide for the range of parameters to consider when fitting isochrones for stellar parameter estimation, as discussed further in Section \ref{sec:isochrone-fitting-section}.

\subsection{FLAMES/GIRAFFE observations}
\label{subsec:flames_obs}

Observations were conducted using the Fibre Large Array Multi Element Spectrograph (FLAMES) coupled with the GIRAFFE spectrograph \citep{2002-Paquini-paper-FLAMES} on the Very Large Telescope (VLT) at the ESO La Silla Paranal Observatory. We utilized the HR14B grating in MEDUSA mode, covering the wavelength range $638.3-662.6\ nm$ with a spectral resolution $R  \sim 37000$. FLAMES was chosen for its multi-object capability, which is ideal for observing large samples of stars simultaneously. The HR14B setup was selected to target the {${H\alpha}\ $}line, providing higher S/N in shorter exposure times for these cool giants, offering a more suitable chromospheric diagnostic than the commonly used {${Ca}{II}\ {H,K}$} lines \citep{1985-Cram-mullan-halpha-in-giants,1990-Strassmeier-Halpha-better-in-cool-stars}.

Our FLAMES/GIRAFFE data were downloaded from the ESO Science Archive Facility under program IDs 0102.D-0046, 0103.D-0557, and 0104.D-0040. This dataset comprises 12 nights of {${H\alpha}\ $}observations spread over approximately one year, targeting 115 stars identified as members of 47 Tuc, including several known variable stars and one star considered not a member of the cluster. Additionally for each observation 8-9 fibers were dedicated to measure the sky (Plate 1 and 2 respectively). Typically, each observing block consisted of three to five exposures of approximately 15.5 minutes each ($S/N_{\ continuum} \gtrsim 40$). This strategy was adopted to optimize the study of the {${H\alpha}\ $}line profile and to minimize the impact of cosmic rays in the critical wavelength regions by combining the multiple exposures.

This setup was used on all nights except Jun 25, 2019, when only a single exposure with very low S/N was obtained. Data from this night were therefore excluded from the analysis. In addition, stars  MED\_RGB\_402 and MED\_RGB\_370 (N=4 and N=36 in table \ref{tab:targets}) observed on Oct 20, 2018 were found to be affected by fiber issues and this epoch for these stars were consequently discarded. Furthermore the target MED\_RGB\_2430 (N=27 in table \ref{tab:targets}), was identified as a binary star due to its RV variations and because its spectral profile differs from similar stars. Therefore, as this star requires a more detailed study, we have discarded it from our current analysis.

On Sep 09, 2019 and Oct 07, 2019, several observations were obtained during the same night, including plate changes. We identified a likely systematic effect between observations taken with different plates. Specifically, after switching plates, the depth of spectral lines changed proportionally to their intrinsic depth for a subset of stars. This effect is clearly visible when computing the ratio between spectra obtained with different plates. No clear explanation for this discrepancy was found, as returning to the original plate during the same night caused the differences to disappear within the level expected from photon noise. It is worth mentioning that these observations were not obtained in simultaneous calibration mode, which is known to potentially produce such effects.

A similar trend was also found between observations obtained on Oct 07, 2019 and Dec 03, 2018, with the most "extreme" case showing residual intensity variations of approximately $\sim0.04$ in the deepest point of the cores of the strongest photospheric lines. In this case the observations were taken with the same plate (1), but the effect was present in most of the stars.

The discrepancy in the line cores can be approximately minimized either by applying an additive correction of $\sim \pm 40$ counts
\footnote{While the values in the extracted one-dimensional spectra are not direct counts from individual CCD pixels --- although throughout our analysis they are treated as such --- due to the nature of multi-fiber spectrograph such as FLAMES/GIRAFFE/MOS, each wavelength bin represents a weighted combination of multiple detector pixels and is subsequently resampled during wavelength calibration rather than corresponding to a single counting experiment. However, because photon shot noise appears to propagate approximately linearly through the extraction process, the resulting variance remains well described by Poisson statistics. Empirically, repeated observations from the same plate show a dispersion consistent in shape with the expected Poisson level, indicating that this approximation provides an adequate description of the noise scaling properties for the purposes of our analysis.} 
(assuming a scenario of differential scattered-light contribution between fibers) or by convolving the spectra displaying the deeper lines with a Gaussian kernel of $\sigma\sim2\ km/s$ (consistent with a scenario in which the effective line-spread function changes because of different fiber illumination conditions or differential fiber transmission). We investigated several potential instrumental effects (e.g., fiber position on the CCD and sky), environmental conditions (differential sky, moon illumination, seeing, airmass, integrated water vapor, wind direction), and data-reduction aspects (e.g., the bias frames used during the nights), but found no clear source or systematic trend capable of explaining these discrepancies. We therefore proceeded with the analysis by explicitly considering these apparent systematic uncertainties and by cross-checking whether specific epoch pairs exhibit this behavior when doing sensitive analyses or drawing conclusions.

\begin{figure*}[ht]
\centering
     \begin{subfigure}[t]{0.6\textwidth} 
        \caption{}
        \label{fig:CN-Band}
        \includegraphics[width=1.05\linewidth]{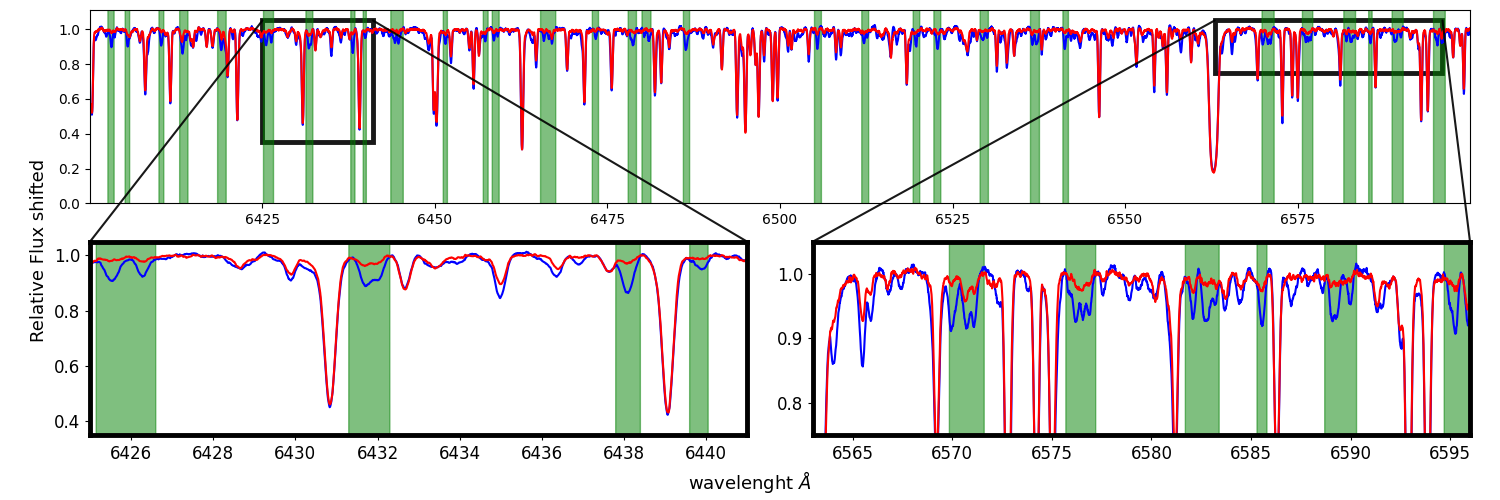}
    \end{subfigure}\hspace{9mm}%
    %
    \begin{subfigure}[t]{0.16\textwidth} 
        \caption{}
        \label{fig:integrated-CN}
        \hspace{0mm}\includegraphics[height=1.35\linewidth,trim=0 0 0 0, clip]{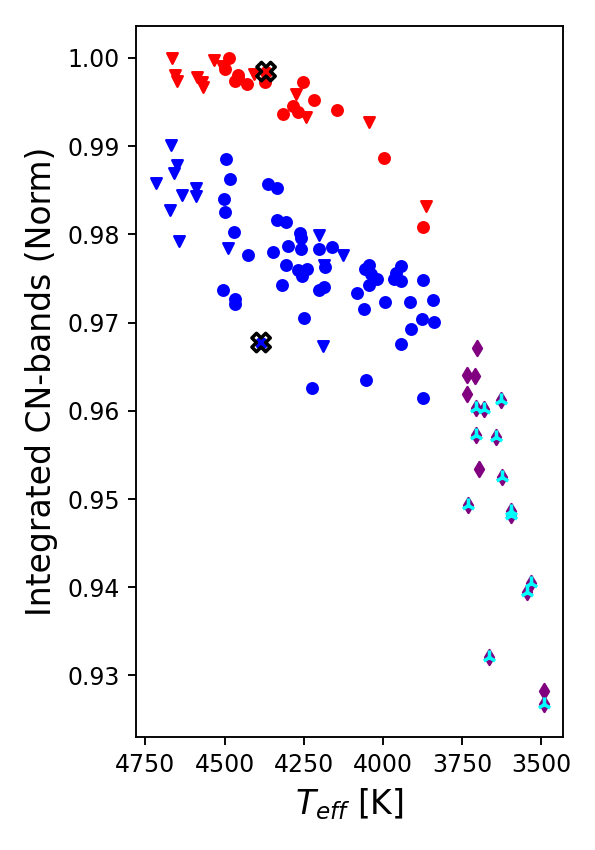}
    \end{subfigure}%
    \begin{subfigure}[t]{0.16\textwidth} 
        \caption{}
        \label{fig:integrated-CN-lum}
        \hspace{0mm}\includegraphics[height=1.35\linewidth,trim=25 0 0 0, clip]{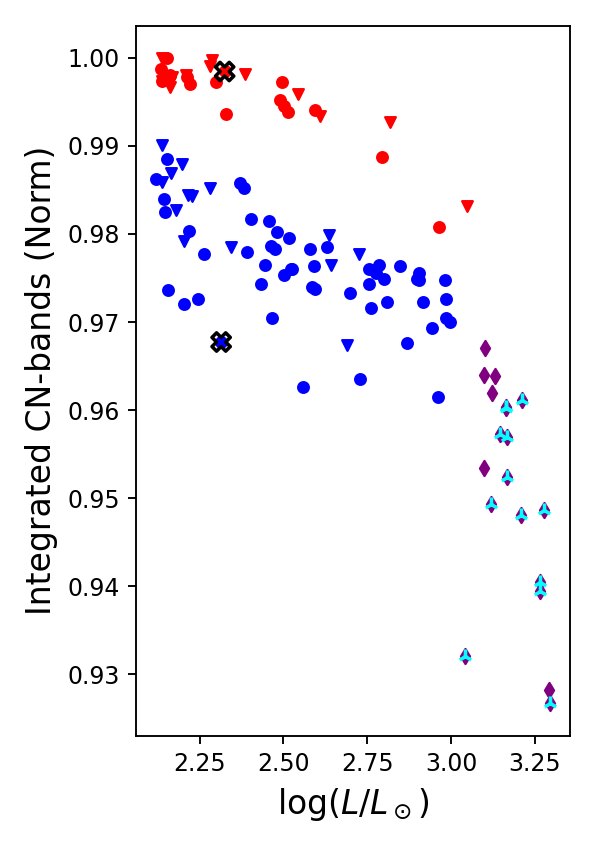}
    \end{subfigure}%
    \caption{
    \label{fig:CN-bands-MPs}
    {\bf (a):} Median spectra across all epochs for two stars with similar $T_\mathrm{eff}$ and $\log(g)$ but different populations (stars N=35 and N=36 in Table \ref{tab:targets}). Green shaded regions indicate the CN bands used for the population separation. {\bf (b,c):} Integrated flux in these bands, normalized to the maximum value (higher values indicate weaker CN absorption), versus effective temperature and luminosity for the full sample. Colors and symbols are as in Fig. \ref{fig:CMD-kiel}; crossed symbols mark the stars shown in panel (a).}
\end{figure*}

\subsection{Data reduction pipeline} 
\label{subsubsec:data_reduction}

The data were downloaded from the ESO archive as Advanced Data Products (ADP) of type ANCILLARY.MOSSPECTRA,\footnote{ANCILLARY.MOSSPECTRA files are 2D arrays where each column corresponds to a fiber in GIRAFFE's MEDUSA mode, including both science and sky spectra. For details, see \url{https://www.eso.org/sci/facilities/paranal/instruments/flames/doc.html}.} implying that they were processed using the standard ESO Common Pipeline Library (CPL) recipes. To manage the large volume of data (hundreds of spectra across multiple epochs), we developed custom software using \texttt{Python}. This software, described in Appendix \ref{app:data_reduction_software}, is optimized for FLAMES data but can be readily adapted to data from other instruments.

Our \texttt{Python} scripts perform several key tasks, including sky subtraction, separation of individual stellar spectra from the MOS data, and radial-velocity (RV) estimation. The RVs were determined using the cross-correlation function (CCF) method \citep{1979-Tonry-Davis-CCF} against a single synthetic template spectrum. This template was generated using the AUTOKUR wrapper for the ATLAS9 model atmosphere code \citep{2003-Castelli-ATLAS9,1993-Kurucz-SYNTHE} together with the SYNTHE spectrum synthesis code \citep{1993-Kurucz-SYNTHE}. For the synthetic template, we adopted stellar parameters of $T_{\text{eff}}$\xspace$= 4100\ K$, $\log g$\xspace$= 1.00$, $V_{\text{turb}} = 1.50${}$km\,s^{-1}$\xspace(microturbulence), and [M/H]\xspace$= -0.60$.

Continuum fitting is performed interactively using the method described in Appendix \ref{app:continuum_fitting}, and is further refined through a sigma-clipping algorithm \citep{2018-Sanchez-sigma-clipping}. This procedure incorporates an initial ``first guess'' for the continuum, obtained by slightly shifting the spectrum and measuring a high percentile of the flux values at each wavelength, effectively smoothing over sharp features. The RV estimation and continuum-fitting modules of our software are designed to be broadly applicable to other spectroscopic datasets with minimal modification, see Appendix \ref{app:data_reduction_software} for further details.

\subsection{Small sky absorption lines}

Additionally to the standard sky subtraction, we have masked regions of small sky absorptions. To identify the affected regions, we performed a procedure to determine, for each observation, which wavelength ranges were contaminated. The procedure consisted of using the already normalized rest-frame spectra to compute a median spectrum for each star. Then, each epoch was divided by its corresponding median spectrum, and the resulting ratio was shifted back to the observed velocity frame (corrected for the intrinsic stellar velocity and the barycentric velocity of the specific epoch). We subsequently coadded the spectra of all stars observed during a given epoch, enhancing any residual signal correlated with the observed velocity and that does not depend on any star.

This procedure successfully identified regions where telluric spectral templates indicate the presence of ``strong'' water-vapor lines \citep{2020-Baker-sky-lines,2024-Malherbe-solar-spectrum-sky-lines,2025-Malherbe-data}, allowing us to use the observed sky lines to establish a threshold for which features in the telluric templates were sufficiently strong to affect our observations. The affected regions were masked in all epochs. Furthermore, we made use of the outlier star in our sample that possesses an intrinsic velocity of $\sim+69.5\ [km/s]$, which, given the mean cluster velocity of $\sim-17.3\ [km/s]$, provided a suitable window to directly assess whether the core and blue wing of {${H\alpha}\ $}in our target stars showed any appreciable sky contamination. Fortunately for our purposes, no significant contamination was detected in any epoch.


\section{Inferred properties and definitions}
\label{sec:methods} %

\subsection{Stellar parameters via multiple isochrone fitting}
\label{sec:isochrone-fitting-section} 

To derive stellar parameters, we utilized photometric data from two publicly available catalogs for NGC 104. First, the deep wide-field photometry from \citet{2019-Stetson-Photometry}. Second, astrometric and photometric data from the catalog of \citet{2021-vasiliev-likely-members-GC}, focusing on likely cluster members identified using proper motions and parallaxes from Gaia Data Release 3 (DR3). Furthermore, we used the catalog of \cite{2024-Prudil-variables-in-GC} to identify known variables in the cluster. It is worth mentioning, however, that star N=70 MED\_RGB\_3399 (see Table \ref{tab:targets}) is identified as V22 in that catalog, but after cross-checking with \cite{2005-Lebzelter-47tuc-variables-ML} we found that the reported RA and DEC do not match. Consequently, we do not consider this star to be a confirmed variable.

The catalogs were cross-matched with our spectroscopic targets. The resulting {color-magnitude diagram (CMD)} was then fitted with theoretical isochrones to estimate effective temperatures ($T_{\text{eff}}$), surface gravities ($\log g$), luminosities ($\log\ L/${}$L_{\odot}$) and Masses for our sample stars. Photometric data used and some of the stellar parameters inferred are displayed in Fig. \ref{fig:CMD-kiel}. The theoretical isochrones were sourced from the PAdova and tRieste Stellar Evolution Code (PARSEC) version 2.0 \citep{2022-Nguyen-PARSEC-isochrones}, downloaded from the CMD 3.8 web interface\footnote{\url{http://stev.oapd.inaf.it/cgi-bin/cmd_3.8}}.

To estimate uncertainties in the derived stellar parameters, we performed a multiple isochrone fitting by varying the input parameters (age, metallicity, with fixed distance modulus and reddening) within the ranges reported in the literature (summarized in Table \ref{table-ng104-properties}), prioritizing matching the key evolutionary stages (i.e., the turn-off, RGB bump and horizontal branch). The process to derive these values and their errors are based in a Bayesian approach which assign probability to the isochrone points in the nearby area of each data point, details on these determinations can be found on the Appendix \ref{appendix-ngc104}.

\subsection{Multiple stellar populations, RGB, and AGB}
\label{sec:MPS}
To explore possible connections with the MPs in 47 Tuc, we inferred the population membership of our targets using CN molecular bands across the spectrum (shown in Fig.~\ref{fig:CN-Band}). MPs in GCs exhibit well-established anticorrelations between N–C and Na–O abundances \citep{2018-Bastian-and-lardo-anual-review-MP,2025-Villanova-47Tuc-CMD-mult-pop}, where N-rich/C-poor and Na-rich/O-poor stars correspond to the so-called second population. The N/C anticorrelation exceed that expected from internal mixing during giant branch evolution \citep[see Fig.~10 of][]{2000-Gratton-N-C-changes-due-mixing}, while Na/O anticorrelation seem to be a property exclusive of GCs.

The chemical pattern of the second population enhances the CN bands because CN formation depends on the availability of both carbon and nitrogen. Although carbon varies between populations, it remains sufficiently abundant to sustain efficient CN formation, consistent with its higher cosmic abundance relative to nitrogen \citep{2025-Bergemann-the-chemical-composition-of-the-sun,2019-Maiolino-Cosmic-chemical-evolution-of-galaxies}. At the same time, oxygen depletion effectively reduces the fraction of carbon locked in CO. Even if oxygen remains more abundant than carbon (C/O < 1), the combination of enhanced nitrogen and reduced oxygen will favor CN formation. Due to the relatively high dissociation energy of CN compared to many other non-oxygen-bearing diatomic molecules \citep{2016-Barklem-Partition-function-molecules-binding-energy}, CN absorption features are favored in this scenario and well suited to separate populations when comparing similar stars. In 47 Tuc, nitrogen and oxygen abundances vary by approximately 1 dex between populations \citep{2013-Carreta-abundances-47-tuc,2025-Villanova-47Tuc-CMD-mult-pop}.

This method allowed us to clearly separate populations in our sample for stars with $T_{\rm eff} \gtrsim 3800\,\mathrm{K}$ and $\log(L/L_\odot) \lesssim 3.1$. At lower effective temperatures (i.e., in cooler and more luminous stages), other molecular bands increasingly dominate the opacity, making the separation between populations unfeasible \citep{2009-Lederer-rosseland-opacities-low-temp,2016-Boberg-limitations-CN-CH-saturation}.

Once separated by population, we performed an additional separation of the giants into RGB and AGB stars within the same region where the populations can be distinguished. This classification was carried out through visual inspection of the individual populations, identifying stars that appeared relatively bluer at a given magnitude, consistent with previous findings showing that different populations display slight but systematic color differences in the CMD \citep[e.g., see Fig. 5 in][]{2025-Villanova-47Tuc-CMD-mult-pop}. Performing this separation independently for each population minimizes the overlap between stars, particularly between bluer RGB stars from the second population and redder AGB stars from the first population. Nevertheless, it is worth mentioning that a clear separation is only evident for temperatures hotter than $T_\text{eff}\gtrsim4400\ K$. Although the populations still appear separated at cooler temperatures, we recommend caution when interpreting the RGB/AGB classification for the cooler stars.

\begin{figure}[ht]
    \centering
    \includegraphics[width=0.92\linewidth]{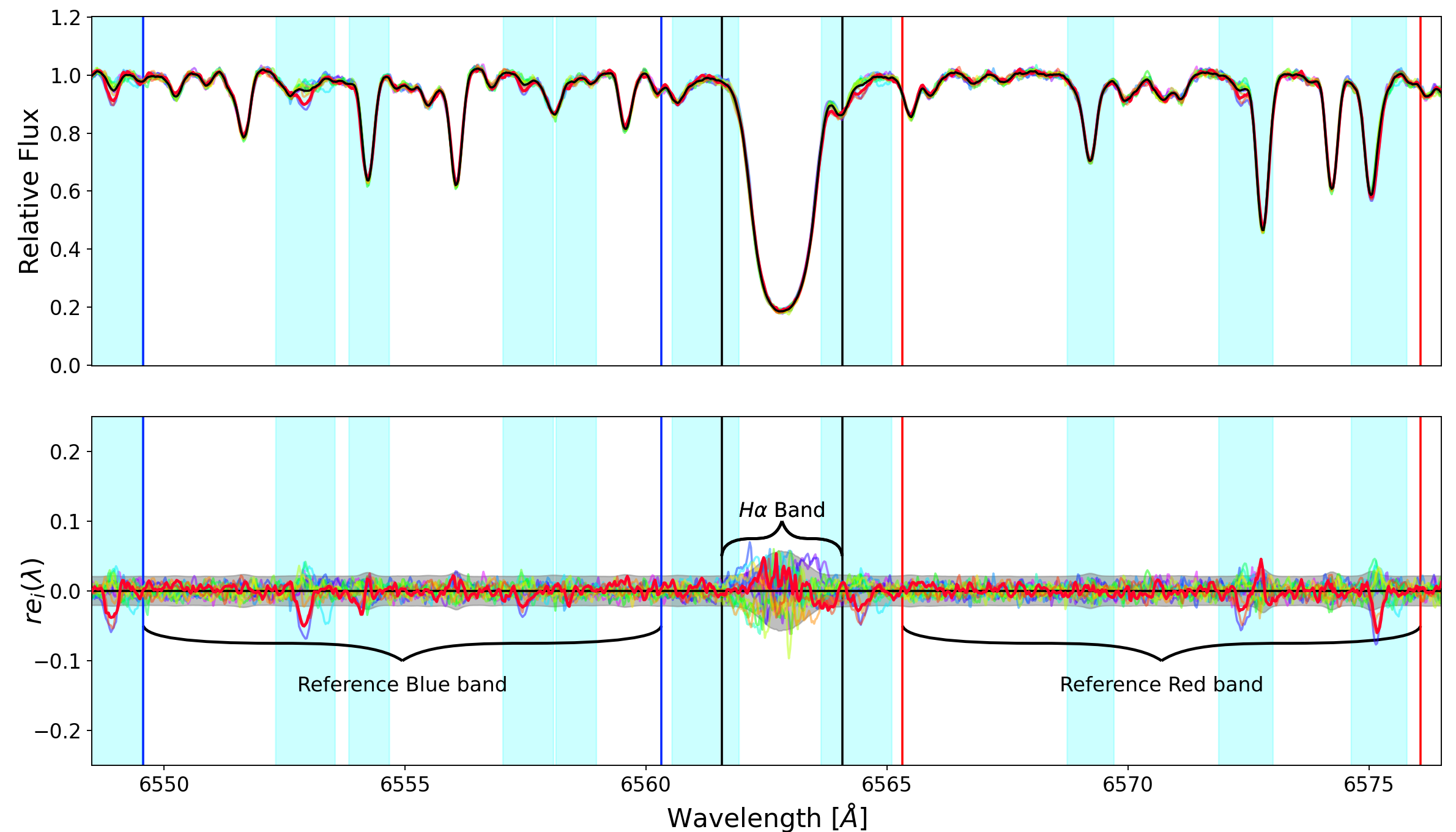}
    \includegraphics[width=0.92\linewidth]{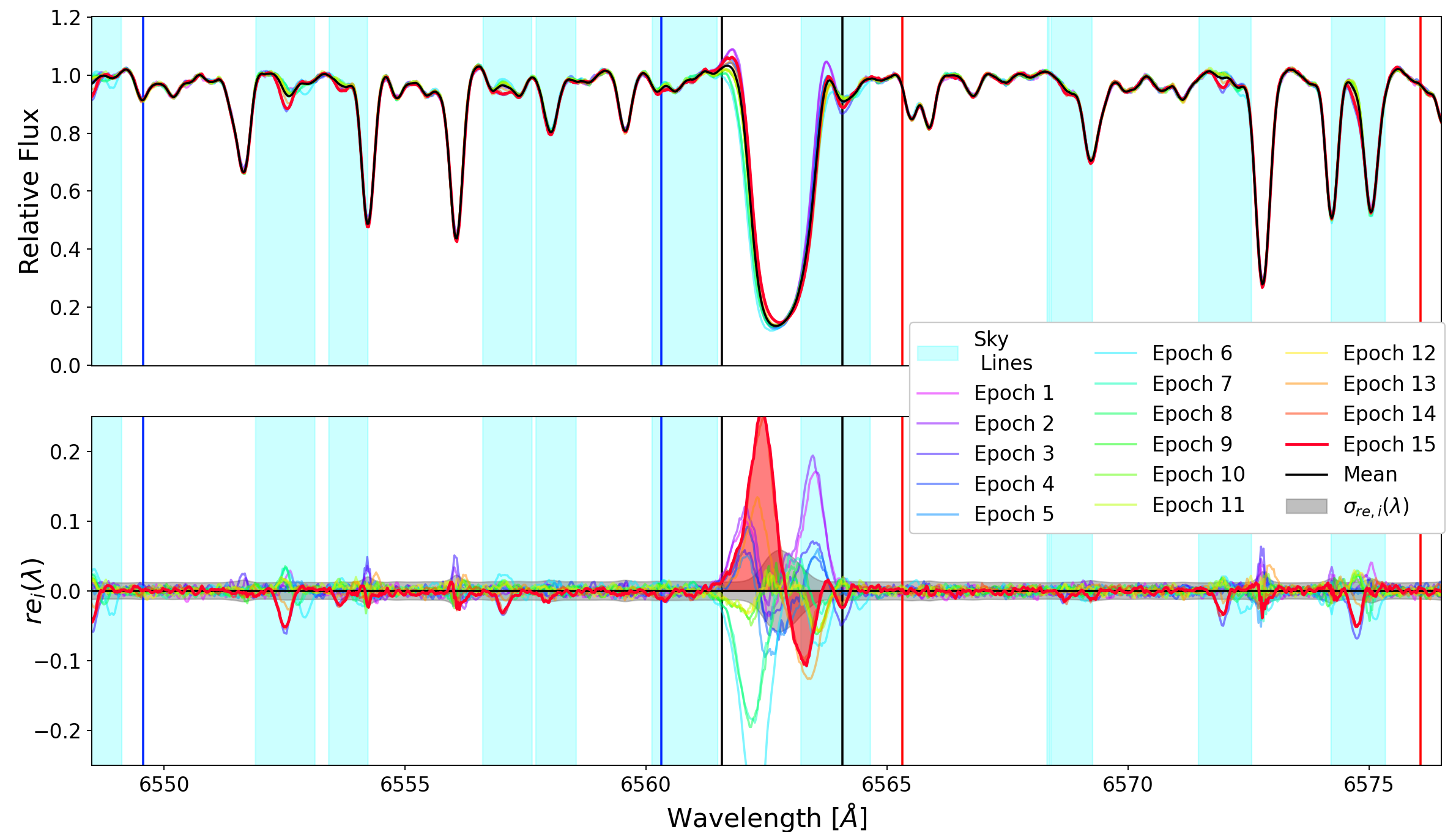}
    \caption{Variability metric illustrated using two stars across 15 epochs (all separated by more than one day). We have highlighted epoch 15 for illustration. Below each spectrum, the residual (Eq.~\ref{eq:residual}) is shown together with the expected residual error (shaded; Eq.~\ref{eq:total-err}). The upper panels show one of the least variable stars, with line core fluctuations consistent with the expected uncertainties, while the lower panels show one of the most variable stars. The reference bands ($\Lambda_\text{Blue}$ and $\Lambda_\text{Red}$) and the {${H\alpha}\ $}band ($\Lambda_{H\alpha}$) are indicated. Light-blue shaded regions mark wavelengths affected by sky absorption in at least one epoch; the masked interval is smaller in individual epochs (see Fig.~\ref{fig:variability-on-time}).}
    \label{fig:variability-spectra}
\end{figure}

\begin{figure*}[ht]
    \centering
    \resizebox{1.005\textwidth}{!}{%
    \includegraphics[height=0.47\linewidth]{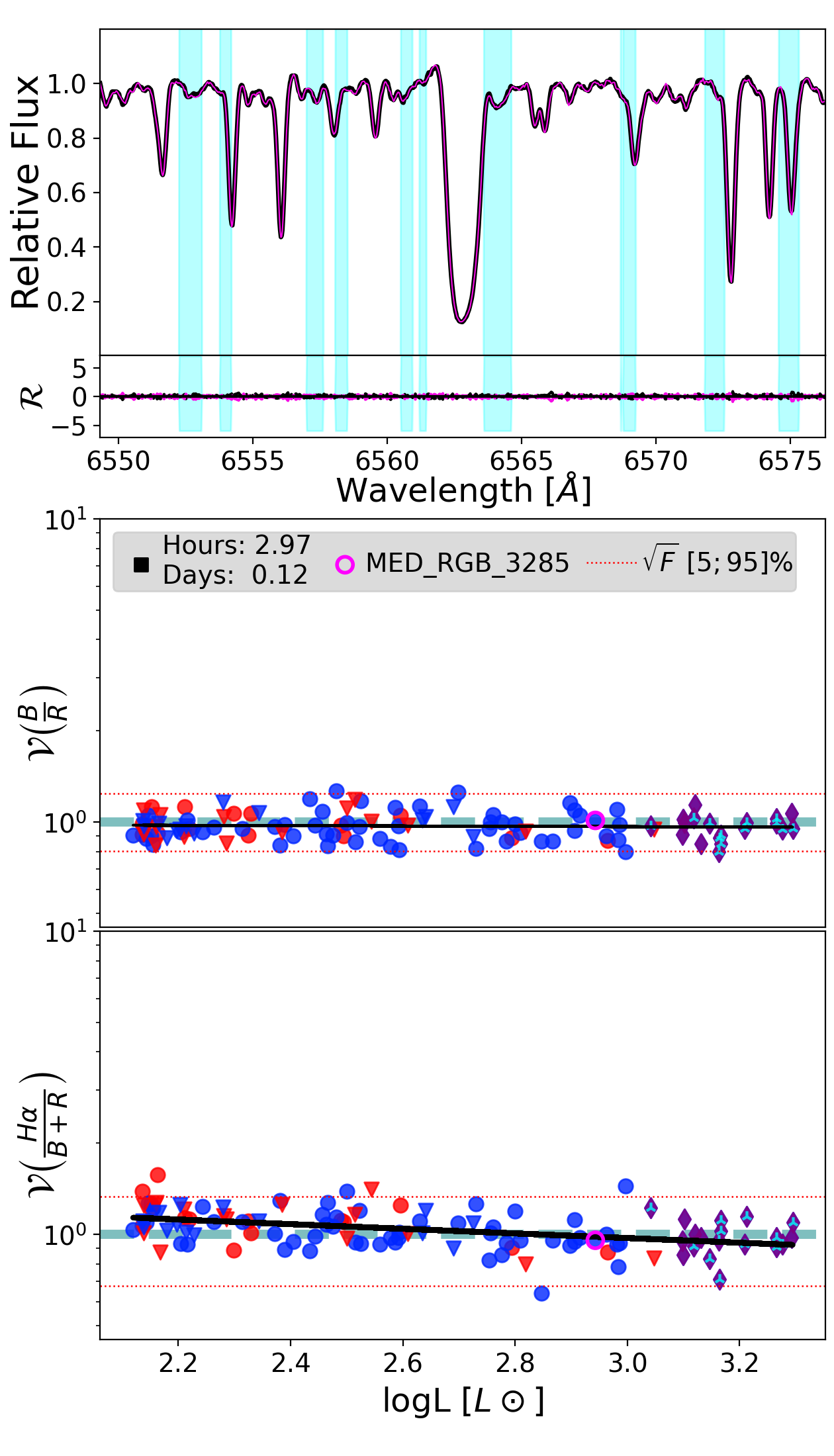}%
    \includegraphics[height=0.47\linewidth,trim=50 0 0 0, clip]{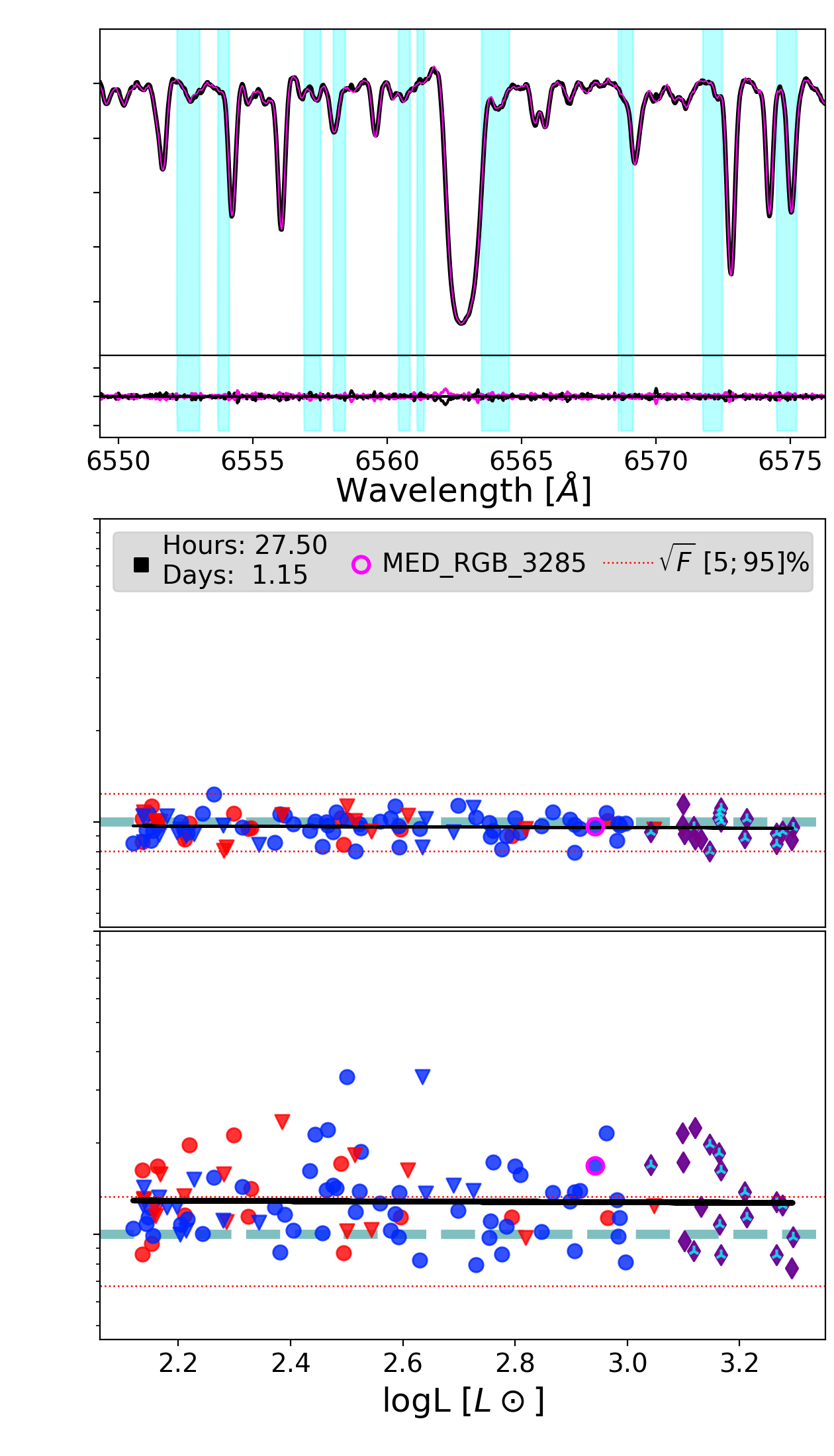}%
    \includegraphics[height=0.47\linewidth,trim=50 0 0 0, clip]{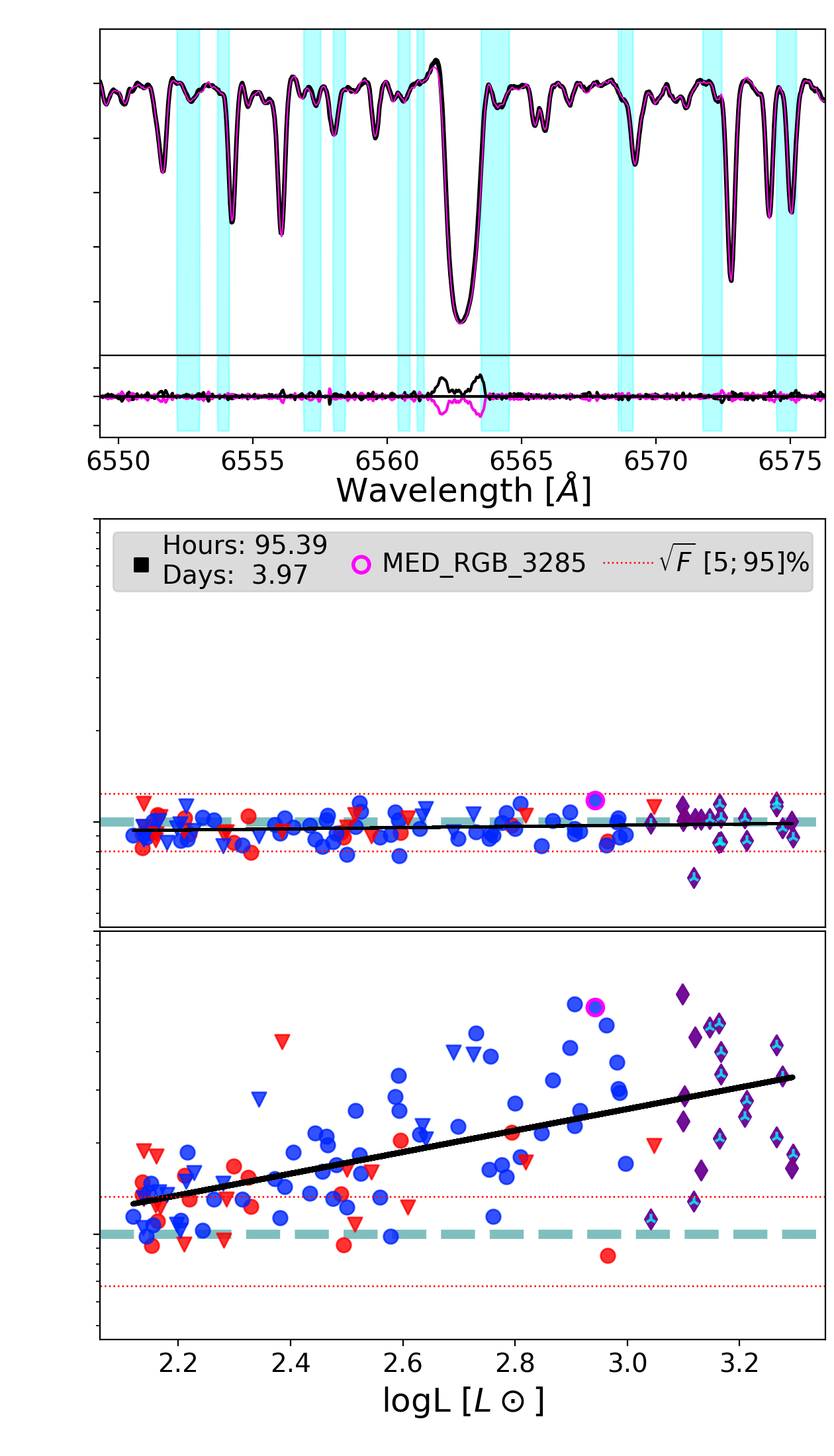}%
    \includegraphics[height=0.47\linewidth,trim=50 0 0 0, clip]{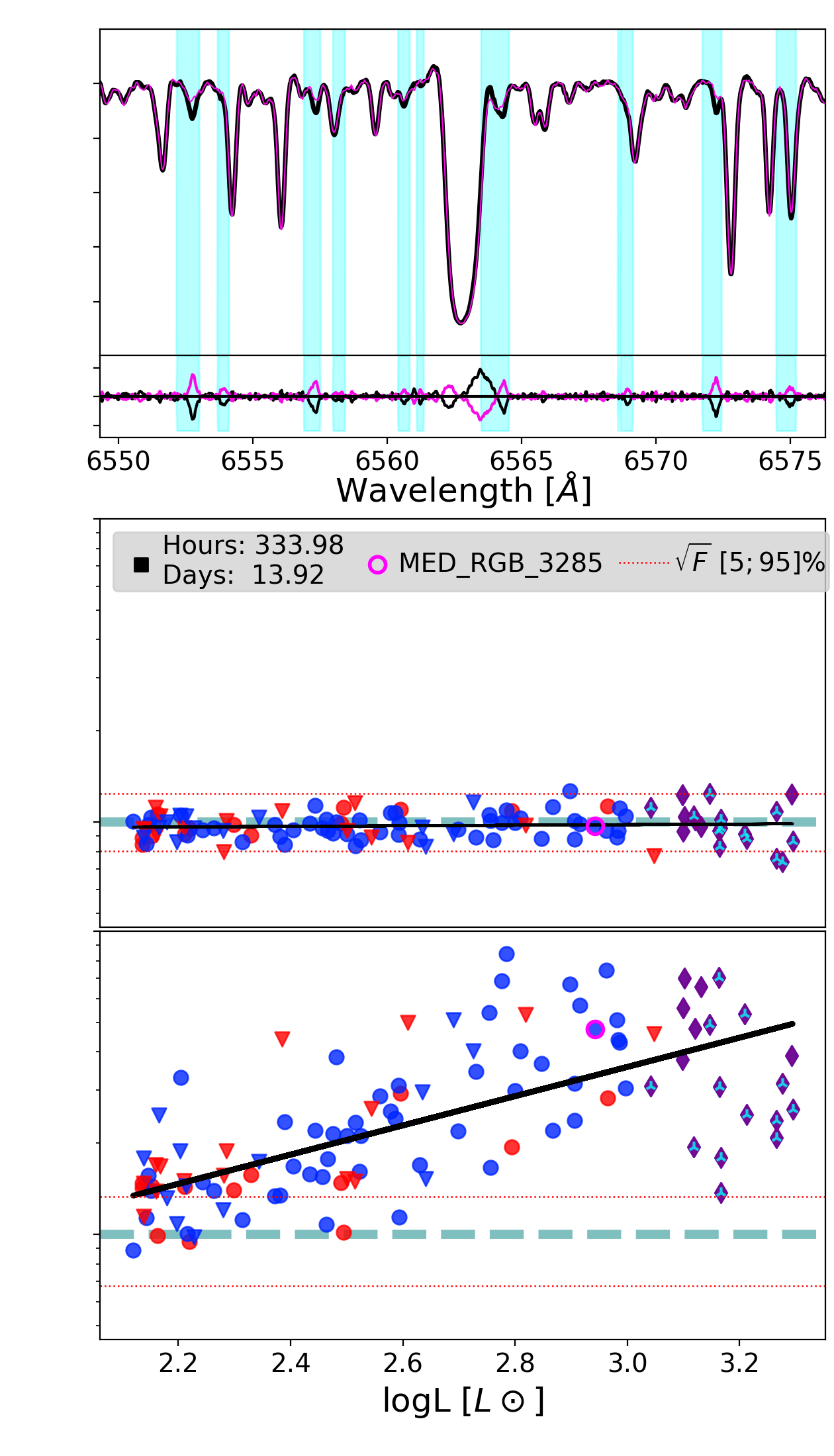}}
    \caption{Variability of {${H\alpha}\ $} across different timescales (each column). {\bf Top row:} Pairs of spectra for the same star (N = 89 in Table \ref{tab:targets}). The time separation is indicated in the label of the second row. The reduced residual, $\mathcal{R}$ (Eq.~\ref{eq:reduced-residual}), is shown below, and light-blue regions mark wavelengths masked by sky absorption for each pair. {\bf Middle and bottom rows:} Variability of the full sample as a function of luminosity for different time intervals (columns) showing the variability of the blue relative to the red band (middle) and of the {${H\alpha}\ $} band relative to the blue$\ \cup\ $ red bands $(B+R)$ (bottom). The dashed cyan line marks the 1:1 ratio (equal variability between bands). Red lines indicate the 5--95\% confidence interval of the F distribution, using 55 and 58 degrees of freedom for the blue/red ratio and 14 and 113 for {${H\alpha}/(B+R)$}. Colors and symbols are as in Fig.~\ref{fig:CMD-kiel}. 
}
    \label{fig:variability-on-time}
\end{figure*}

\begin{figure*}[t]
\centering
    \includegraphics[width=0.94\linewidth,trim=0 10 0 0, clip]{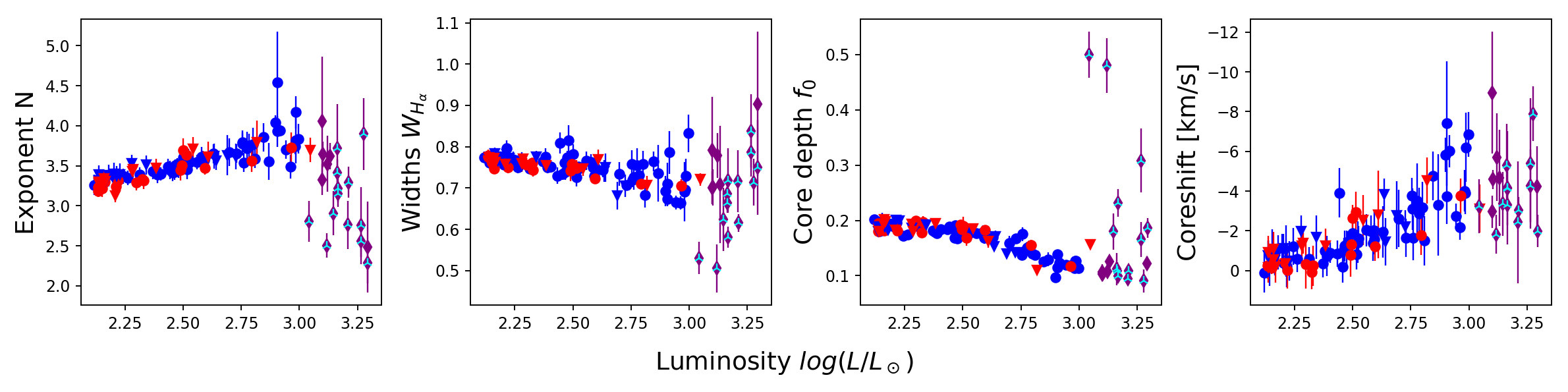}
    \includegraphics[width=0.94\linewidth,trim=0 0 0 0, clip]{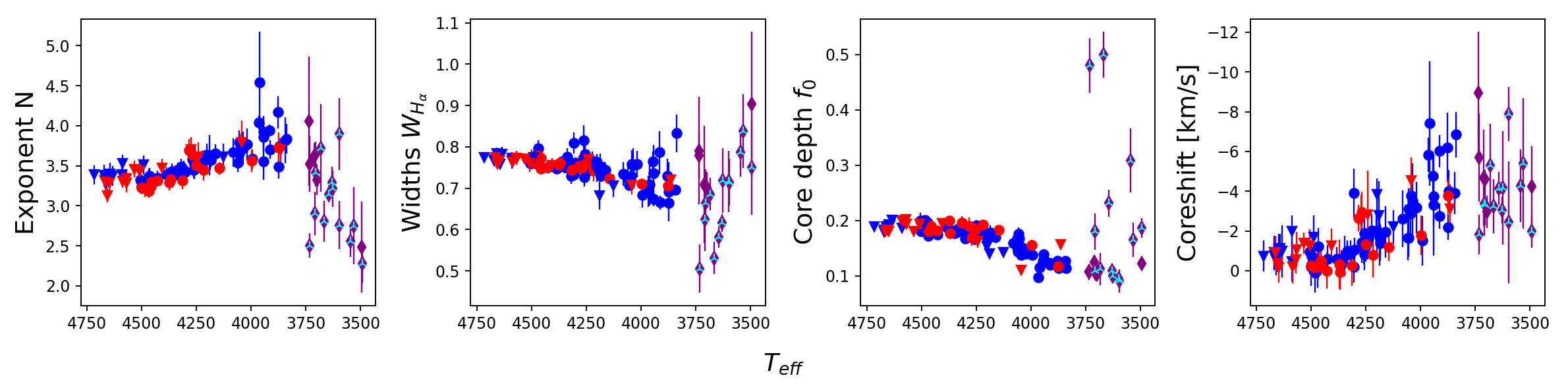}
    \caption{
    \label{fig:halpha-profiles}
    Trends of the blue wing of {${H\alpha}\ $} across the full sample. Shown are the fitted parameters from Eq.~\ref{eq:profile} and the measured coreshift velocity. The parameter $N$ describes the profile steepness, $W_{H_\alpha}$ is its width, $f_0$ is the line core minimum flux, and the coreshift velocity is the displacement of the core from the Ritz wavelength. Points and error bars give the mean and standard deviation over all epochs for each star, illustrating the overall trends and variability of each parameter. The top row shows the parameters as a function of luminosity and the bottom row as a function of $T_\text{eff}$. Colors and symbols are as in Fig.~\ref{fig:CMD-kiel}.}
\end{figure*}

\subsection{Measurement of relative variability}
\label{sec:variability-on-time}

To quantify intrinsic chromospheric variability, we developed an empirical metric that measures spectral fluctuations relative to the level expected from observational noise. At large time separations the variations between epochs are visually evident, but on short timescales it becomes difficult to determine whether the observed differences are intrinsic to the stars or simply produced by the measurement process itself.

The most commonly used {${H\alpha}\ $}activity indices are based on narrow bandpasses centered on the line core \citep[e.g., see fig. 1 of][]{2023-He-halpha-index}. Although useful for activity studies, these indices are not optimal for our purposes because they discard information contained in the adjacent wings and can become highly sensitive to small statistical fluctuations when working in the low-variability regime.

For these reasons we developed a metric based on broad spectral regions around {${H\alpha}.$} The complete formalism is presented in Appendix \ref{app:variability-formalism}. Here we summarize the main idea. For a given set of epochs, each spectrum is first compared against the mean spectrum of time series, producing a residual spectrum $re_i(\lambda)$. These residuals are then normalized by their expected uncertainties $\sigma_{re,i}(\lambda)$, including both photon-noise fluctuations and uncertainties associated with continuum normalization, resulting in a reduced residual $\mathcal{R}_i(\lambda)$.

The root mean square (RMS) of $\mathcal{R}_i(\lambda)$ is subsequently computed within selected spectral bands (i.e., $RMS(\mathcal{R}_i(\forall\lambda\in\Lambda_x))$), providing a measure of the typical fluctuation amplitude once the expected flux dependence has been removed. Finally, the ratio between the RMS measured in the bands defines our variability metric $\mathcal{V}$. In this formulation, values of $\mathcal{V}\approx1$ indicate fluctuations consistent with the expected noise scaling, while $\mathcal{V}>1$ indicates excess variability.

Because the metric compares simultaneous spectral regions within the same observation, global scaling effects and most systematic trends largely cancel out. Confidence intervals and uncertainties are estimated assuming an F-distribution \citep{1995-johnson-distribution-book1,1996-Walck-handbook-distributions}, which provides a reasonable approximation given the ratio-of-variances formulation adopted here, and allows us to propagate the expected fluctuation by comparing bands of different widths.

A schematic view of the adopted bands and of the quantities involved in the calculation is shown in Fig. \ref{fig:variability-spectra}. Examples of the method applied to individual pairs of spectra are presented in Fig. \ref{fig:variability-on-time}, while the resulting variability trends across the full temporal coverage are shown in Figs. \ref{fig:Trends} and \ref{fig:rvs}.

\subsection{Blue line profile}

To characterize the morphology of the {${H\alpha}\ $}line and quantify both temporal and stellar-to-stellar variations, we adopt the parametric profile introduced by \cite{2016-Bergemann-Gaia-ESO-survey-h-alpha}. In that work, it was shown that the blue wing of the line can be accurately reproduced by the functional form

\begin{equation}
    \label{eq:profile}
    f_B(\lambda) =1-
    (1 - f_0)\,
    \exp\!\left(-
        \left[
        \frac{|\lambda - \lambda_0|}
             {W_{H_\alpha}}
        \right]^N
    \right).
\end{equation}

This parametrization provides a flexible description of the line morphology. Variations in $f_0$ trace changes in the core depth. Changes in $W_{H_\alpha}$ reflect modifications in the overall width of the absorption profile. Variations in $N$ provide a diagnostic of the steepness of the line, tracing the transition between wings and core (i.e., if the wing was in emission the profile will be steeper). $\lambda_0$ is the central wavelength and is kept fixed.%
\footnote{In this work we adopt the Ritz air wavelength $\lambda_0 = 6562.819$\,\AA\ from \cite{2025-NIST-ASD}, updated according to the CODATA 2022 values \citep{2025-Mohr-CODATA-2022}.}

In \cite{2016-Bergemann-Gaia-ESO-survey-h-alpha}, the parameter $N=3$ was fixed, and under those assumptions $W_{H_\alpha}$ was found to correlate with stellar mass. Here, instead, all parameters are allowed to vary freely in order to capture temporal changes in the line profile that may reflect chromospheric variability.

Although this profile does not enable us to disentangle different physical effects, by fitting this parametric model to each spectrum (see Fig. \ref{fig:halpha-profiles}) we obtain a compact set of parameters that describes how the (blue wing of the) {${H\alpha}\ $}line evolves. This approach provides a global view of the variability pattern, complementing the integrated variance analysis by summarizing profile changes in terms of core depth, width, and steepness.

Furthermore, we measured the coreshift displacement relative to the theoretical center of {${H\alpha}.$} The estimation of this displacement is tied to the measurement of the position of the deepest point of the line. We estimate the position of the line core using three methods: the direct minimum of the line, the minimum of a smoothed profile obtained using a cubic spline, and the mean wavelength of the shallowest $5\%$ portion of the line. By averaging these methods, we obtain a robust estimate of both the depth of the line and the observed central wavelength at maximum depth ($f_0$). This value is later used in equation \ref{eq:profile}. The fitting parameters were obtained through the nonlinear least-squares method implemented in the widely used function {\small\sf scipy.optimize.curve\_fit} \citep{2020-scipy}. Probable errors for the coreshift are assumed to be standard for all our stars and tied to be half of the expected precision in RV (see Appendix \ref{app:cross_correlation}).   

The red wing appears to be unsuitable for drawing conclusions about the {${H\alpha}\ $}behavior because this region is blended with at least three transitions of Co I line with hyperfine structure \citep{1996-Pickering-Co-line}, which may affect stars differently depending on their abundances -- although this is expected to be a minor effect within our GC sample, where significant Co abundance variations are not expected. However, the shape of the photospheric line will change across different luminosity and $T_\text{eff}$ regimes. At the edge of the red wing, a CN-band feature is also present, differentially affecting the multiple populations. On top of that, a small sky line appears near the red wing region, which may affect any comparison or variability estimate across epochs.

\subsection{Rescaling normalization factor}

As the continuum cannot be directly identified in the observed spectra, we fitted a pseudo-continuum. We then used synthetic spectra to estimate a correction factor that accounts for the depression of the continuum by molecular bands, thereby improving the estimate of the relative core depth of {${H\alpha}.$}

For each target, we generated synthetic spectra using the AUTOKUR wrapper for the ATLAS9 model atmosphere code \citep{2003-Castelli-ATLAS9,1993-Kurucz-SYNTHE} with the SYNTHE spectrum synthesis code \citep{1993-Kurucz-SYNTHE}. By including TiO molecules in the synthesis of all spectra, the output is normalized relative to the expected theoretical continuum. In this way, by measuring the relative offset between the observed pseudo-continuum --of the synthetic spectra-- and the unity level, we derive a constant correction factor that is used to shift the pseudo-continuum toward a more realistic continuum estimate, this value is obtained through the estimation of the percentile 97th of the normalized synthetic spectrum. The motivation for this procedure is that we are interested in obtaining a better approximation of the true depth of the {${H\alpha}\ $}profiles in order to compare their morphology and variability both between epochs and among different stars.

All normalized spectra were rescaled using a single constant factor for each star, and this correction should therefore be regarded only as an improved approximation to the true continuum level. The warmer/faintest stars in our sample are barely affected by this correction, while the coolest/brightest stars require the largest rescaling factors due the enhanced TiO bands, with values reaching approximately $\sim0.7$, corresponding to a downward shift of the fitted pseudo-continuum.


\section{Results}
\label{sec:results} %



\subsection{Temporal threshold}
We have used the variability definition to identify the time threshold at which significant variation appears in our stars (see Fig. \ref{fig:variability-on-time}). For this analysis, we have used measurements from plate 1 only, in order to avoid the apparent systematic differences between plates. Furthermore, the epochs used in this analysis were visually inspected and show no evident systematics in the core of the photospheric lines. For reference, we have included $\mathcal{V}(\Lambda_\text{Blue}/\Lambda_\text{Red})$ (see \ref{eq:variability-def}), which confirms that the reference bands show no significant variability at any point. The observed distribution falls within the expected F-distribution confidence interval, although with fewer outliers than predicted for the considered confidence interval. This is consistent with our conservative noise estimate. Since we expect the same trend to be present in $\mathcal{V}(\Lambda_{H\alpha}/(\Lambda_\text{Blue}+\Lambda_\text{RED}))$, the F-distribution confidence interval provides a conservative threshold propagating the expected variations that will arise by comparing bands with different widths, and detections above it represent robust evidence of the rejection of the null hypothesis that the variances are equal, demonstrating that excess variance is present in the band, which we associate with intrinsic variability.

On the shortest timescale that we are able to test reliably (a couple of hours), most stars fall within the confidence interval. Nevertheless, in this timescale the fainter stars of our sample fall above the 1:1 ratio, in contrast with the more luminous ones. Further studies with higher precision will be needed to confirm the nature of these variations.

After approximately one day, stars consistently display significant variations, with what appears to be a generally flat trend across all luminosities, establishing a floor of variability experienced by the giants. It is worth noting, however, that a considerable group of the fainter stars ($\log(L_\star/L_\odot)<2.2$) seem to reach the same variability threshold already seen after $\sim$3 hours.

After $2-4$ days, we found a similar floor of variability for stars with $\log(L_\star/L_\odot)\lesssim2.4$, while brighter stars show increasingly larger variations with luminosity (see third column in Fig.~\ref{fig:variability-on-time}). The physical origin of this timescale threshold has not been reported in the literature and, to the best of our knowledge, no model has predicted such an increase. This inflection point at $\log(L_\star/L_\odot)\sim2.4$ suggests that atmospheres of these giants undergo a significant change above this luminosity threshold.

On timescales of weeks, the typical variations appear to consolidate and resemble those observed after $2$--$4$ days, although with a slightly larger scatter. A knee-like structure remains present around $\log(L_\star/L_\odot)\sim2.4$, defining two distinct regimes experienced by the giants in this cluster.

\subsection{\texorpdfstring{{${H\alpha}\ $}}{Ha }blue wing profile}

Having established the temporal limit where our sample start to display significant variations ($\sim1$ day), we can examine in detail what is specifically changing in the {${H\alpha}\ $}profile. For this we have used the profile described in Eq.~\ref{eq:profile}, along with the relative displacement of the core, these values are shown in Fig.~\ref{fig:halpha-profiles}, it is worth noting that the trends here display a similar correlation with luminosity and $T_\text{eff}$ considering errors (see App \ref{app:tightness}). 

The first panel shows that the profiles become steeper up to $\log(L_\star/L_\odot)\sim2.9$, the point where stars become cool enough for the spectra to begin to be dominated by molecular bands ($T_\text{eff}\sim3800$ K). This growing term is consistent with what we can see by visual inspection, which displays increasingly prominent emission wings as the stars are brighter and cooler. 

In the second panel, the fitted line width appears to clearly decrease with luminosity. This corresponds to the narrower profiles observed visually. 

The core of the line in the third panel appears to deepen for most stars, if examined carefully (see \ref{fig:tightness}), one may notice that the deepening of the core begins at approximately the same point as the inflection point observed in the variability, within $2.3<\log(L_\star/L_\odot)<2.5$. The core becomes progressively deeper to reach its greatest depth at around $3.0<\log(L_\star/L_\odot)<3.2$. The data points showing shallower cores are strongly dominated by molecular bands and correspond to well-known variables in the cluster (see Table~\ref{tab:targets}).

Looking at the fourth panel, we observed the relative shift of the core with respect to its Ritz wavelength. Here a clear discrepancy is apparent between stars above and below $\log(L_\star/L_\odot)\sim2.4$. Moreover, compared with the other parameters, the coreshift appears to be the most variable quantity measured from the line (as seen through the error bars that here represent the standard deviation of the measures across the epochs). Interestingly, in this case the most variable stars do not display an evidently different regime, unlike the other parameters where a clear distinction is perceived for the variables.

\begin{figure}[t]
    \centering
     \includegraphics[width=0.95\linewidth,trim=0 5 0 0, clip]{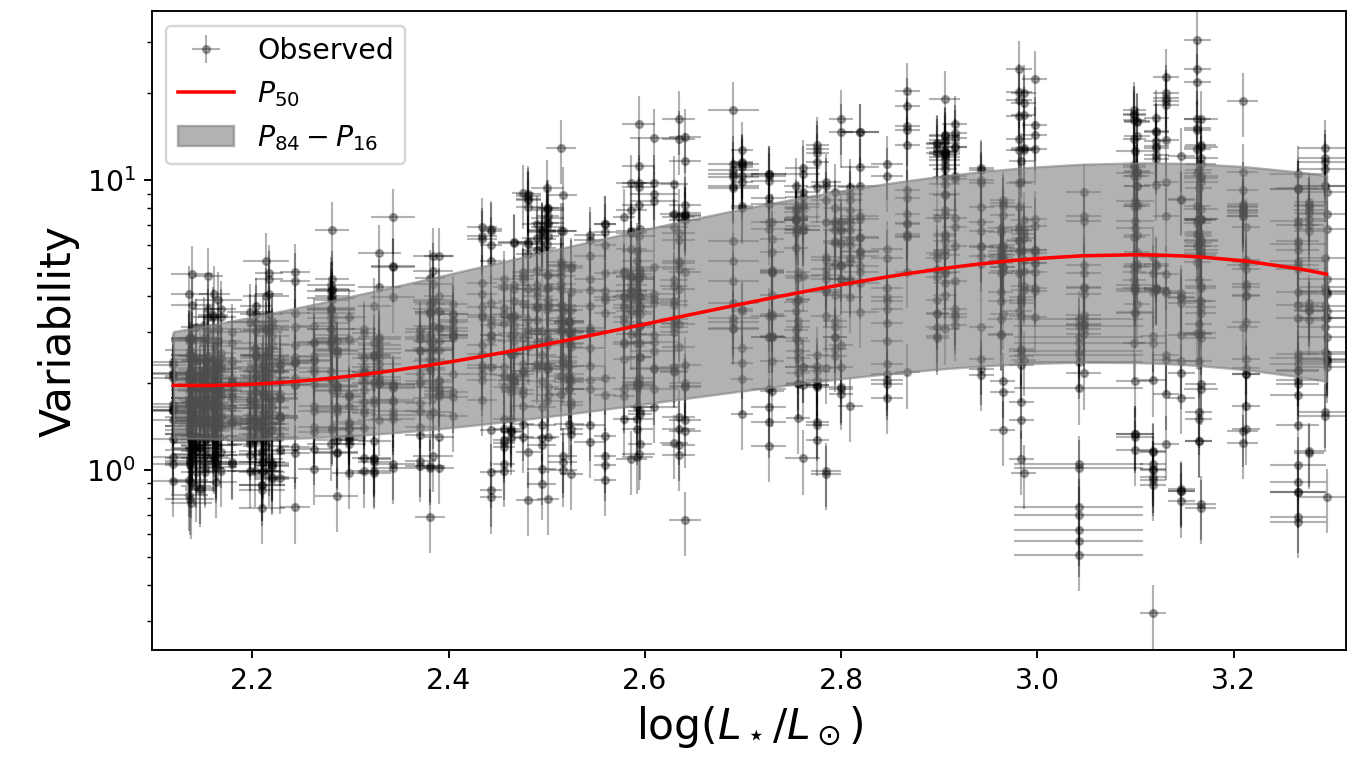}
      \caption{
      Variability of the {${H\alpha}\ $}line across all epochs measured relative to the median spectrum of each star. The 16th, 50th, and 84th percentiles within luminosity bins (boxcar) are shown. The displayed trends have been smoothed for illustrative purposes.}
    \label{fig:Trends}
\end{figure}

\subsection{Structural trends}

By combining measurements from all epochs, we can investigate whether the observed quantities exhibit general trends with the intrinsic properties of the stars. As the trends with luminosity shows a similar correlation with other stellar parameters (see App \ref{app:tightness}), in this section we show only the trends with respect to the luminosity.

The overall variability, as shown in Fig.~\ref{fig:Trends}, increases in both mean and amplitude. The fainter stars show more stable variations confined to a narrower range, while the brighter ones can be observed to be either relatively quiescent or extremely different from the median profile. This means that if we take a spectrum of a fainter star, though we are unlikely to capture the median behavior of that star, we will most likely obtain a profile representative of the typical variations seen in the {${H\alpha}\ $}profile. On the other hand, for the brighter stars this is no longer true, as the variations are so large that only after several exposures are we likely to recover a representative sample of the kind of variations that these stars experience.

\begin{figure}[ht]
    \centering
     \includegraphics[width=0.515\linewidth,trim=0 27 0 0, clip]{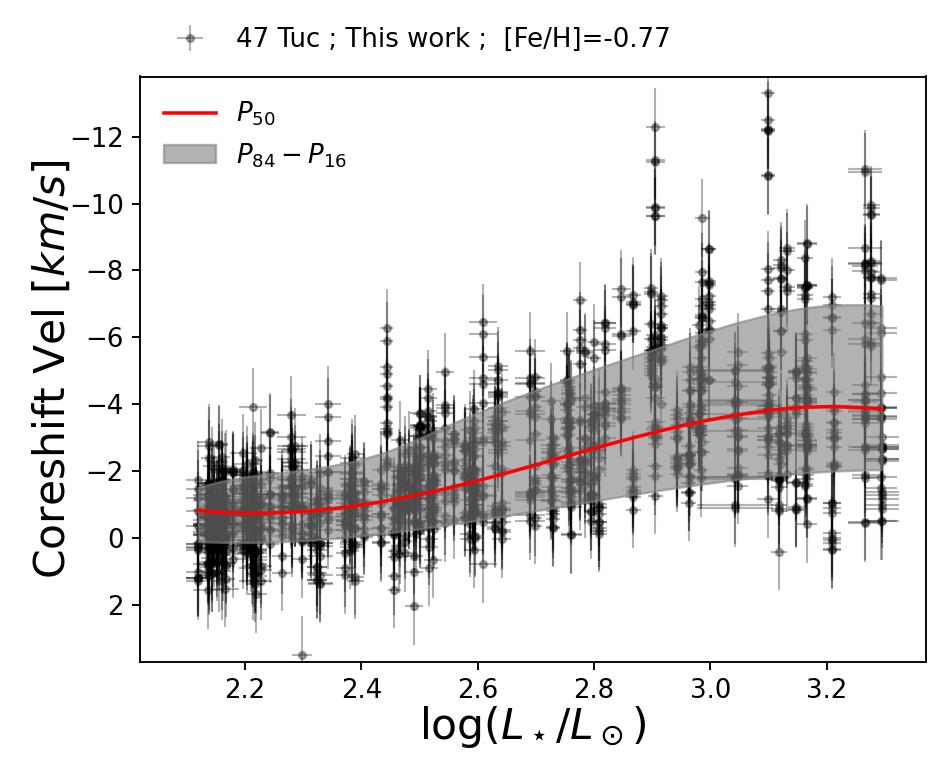}
     \includegraphics[width=0.475\linewidth,trim=30 27 0 0, clip]{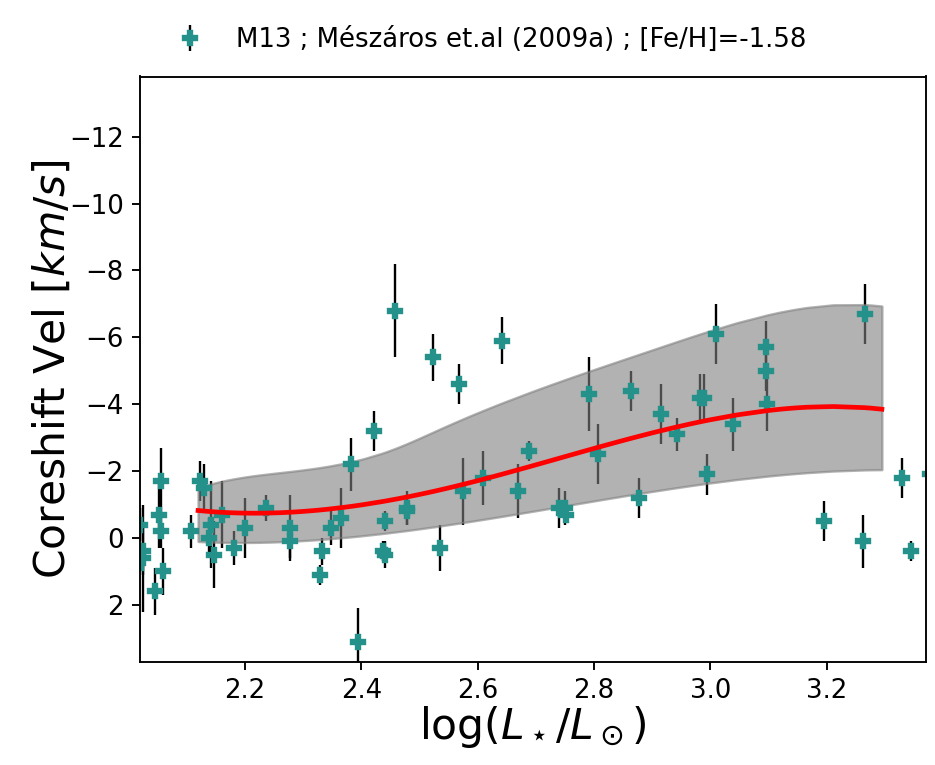}
     \includegraphics[width=0.515\linewidth,trim=0 0 0 0, clip]{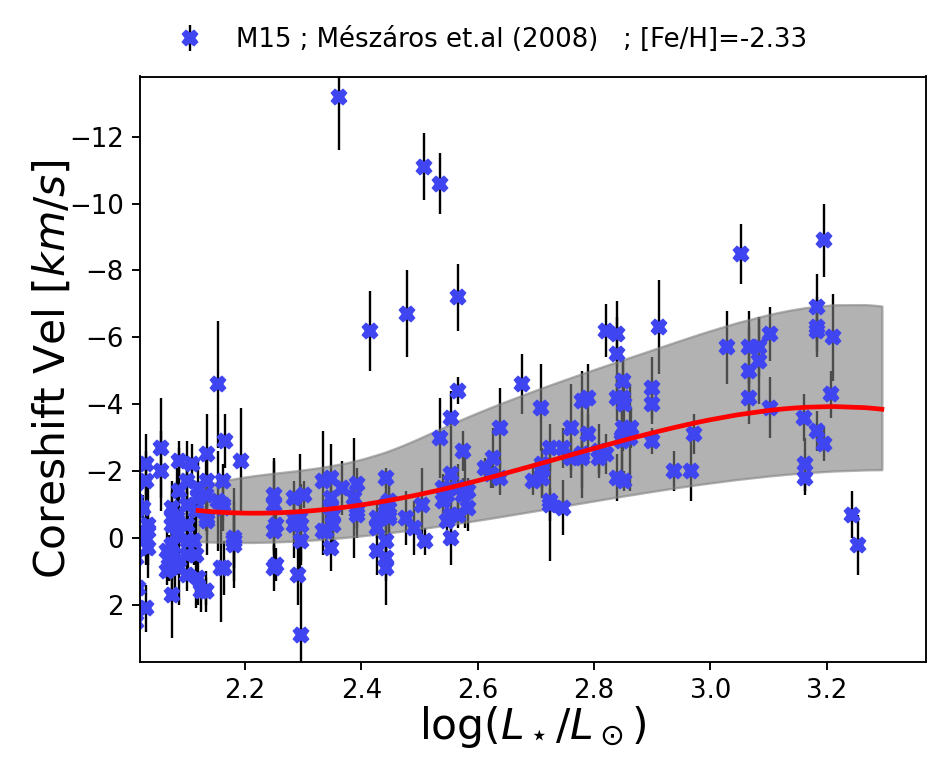}
     \includegraphics[width=0.475\linewidth,trim=30 0 0 0, clip]{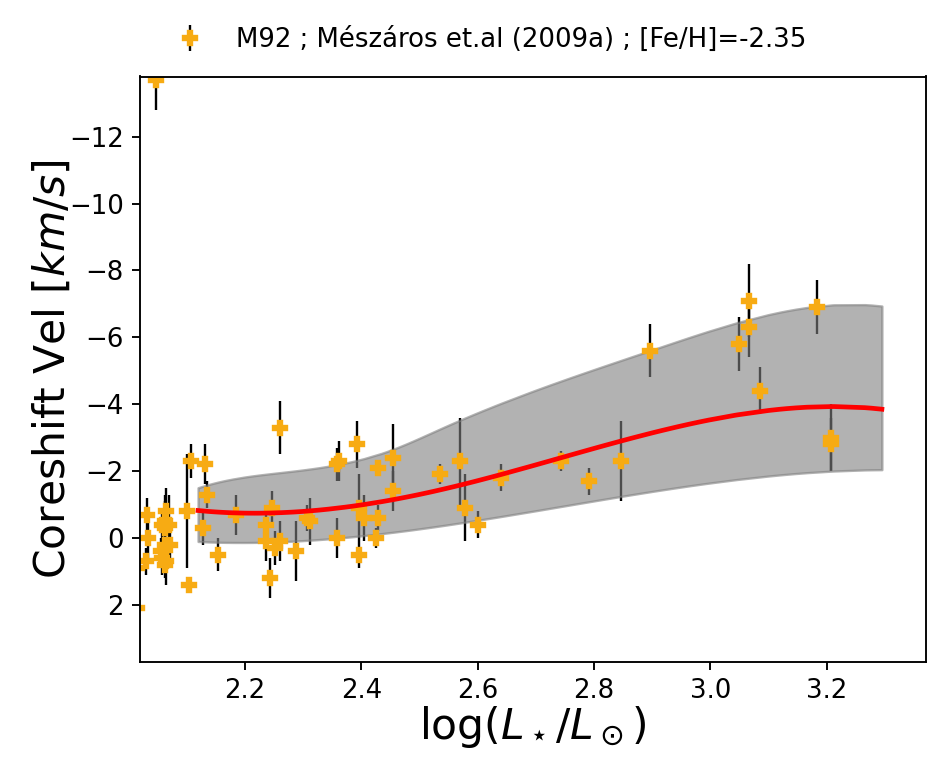}
      \caption{
      Comparison of the {${H\alpha}\ $}coreshift trends derived in this work with those reported for the GCs M15 \citep{2008-Meszaros-I}, M92, and M13 \citep{2009-Meszaros-II}. Negative core shifts indicate outflow motions relative to the photosphere. Each panel corresponds to a different cluster. For 47 Tuc, core shifts were measured at every epoch, and the 16th, 50th, and 84th percentiles are shown in all panels using the same boxcar binning and smoothing as in Fig.~\ref{fig:Trends}. Metallicities are adopted from \citet{2026-Carreta-GC-homogeneous-metallicities,2009-Carreta-metallicity-scale-GC}. For clarity, only the luminosity range overlapping the Mészáros samples is displayed.}
    \label{fig:Trends-mezsaros}
\end{figure}

The variability on longer timescales resembles the pattern already shown on the timescale of a couple of weeks. The most populated trend shows that all stars in our sample display a floor of variability at long timescales, and that for luminosities above $2.4\lesssim\log(L_\star/L_\odot)\lesssim2.6$, the expected variations increase until reaching a maximum at $\log(L_\star/L_\odot)\sim3.0$.

Examining the general behavior of the core shifts displayed by the stars (see the first panel in Fig.~\ref{fig:Trends-mezsaros}), we find measurements consistent with the idea that around $\log(L_\star/L_\odot)\gtrsim2.4$ the stars undergo a change that increases the coreshift, presumably associated with larger outflow velocities in their atmospheres. In this figure, we also present a direct comparison with the previous works of \cite{2008-Meszaros-I,2009-Meszaros-II}. Within the luminosity range where a comparison is possible, their measurements are broadly consistent with the trend found in our data. Since these measurements correspond to different clusters spanning a wide range of metallicities (-0.77;-2.35 [Fe/H]), this may suggest that the mechanism responsible for the observed core shifts has only at most a weak dependence on metallicity. Further measures in other clusters can be used to properly rule-out the dependence.

\subsection{RV variation connection}

\begin{figure}[ht]
    \centering
     \includegraphics[width=0.5\linewidth,trim=0 50 0 0, clip]{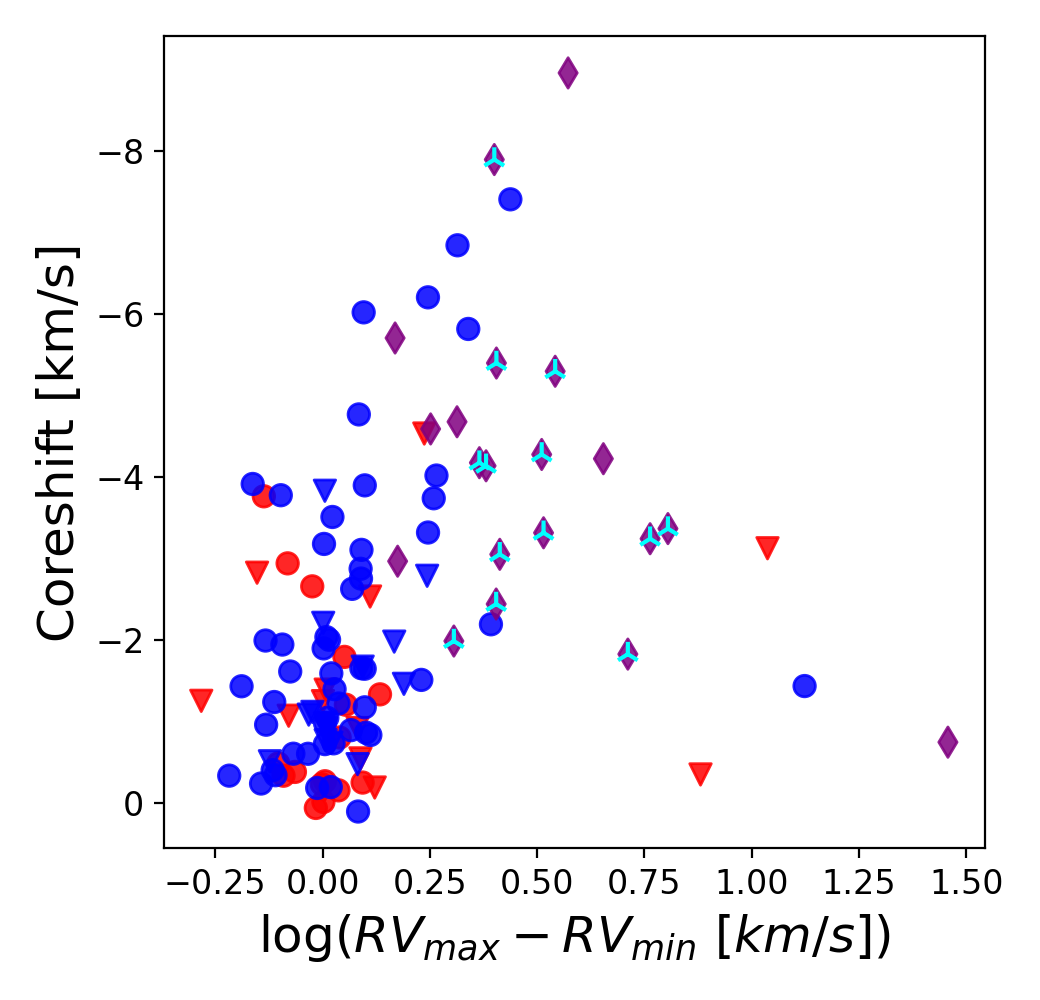}
     \includegraphics[width=0.428\linewidth,trim=54 50 0 0, clip]{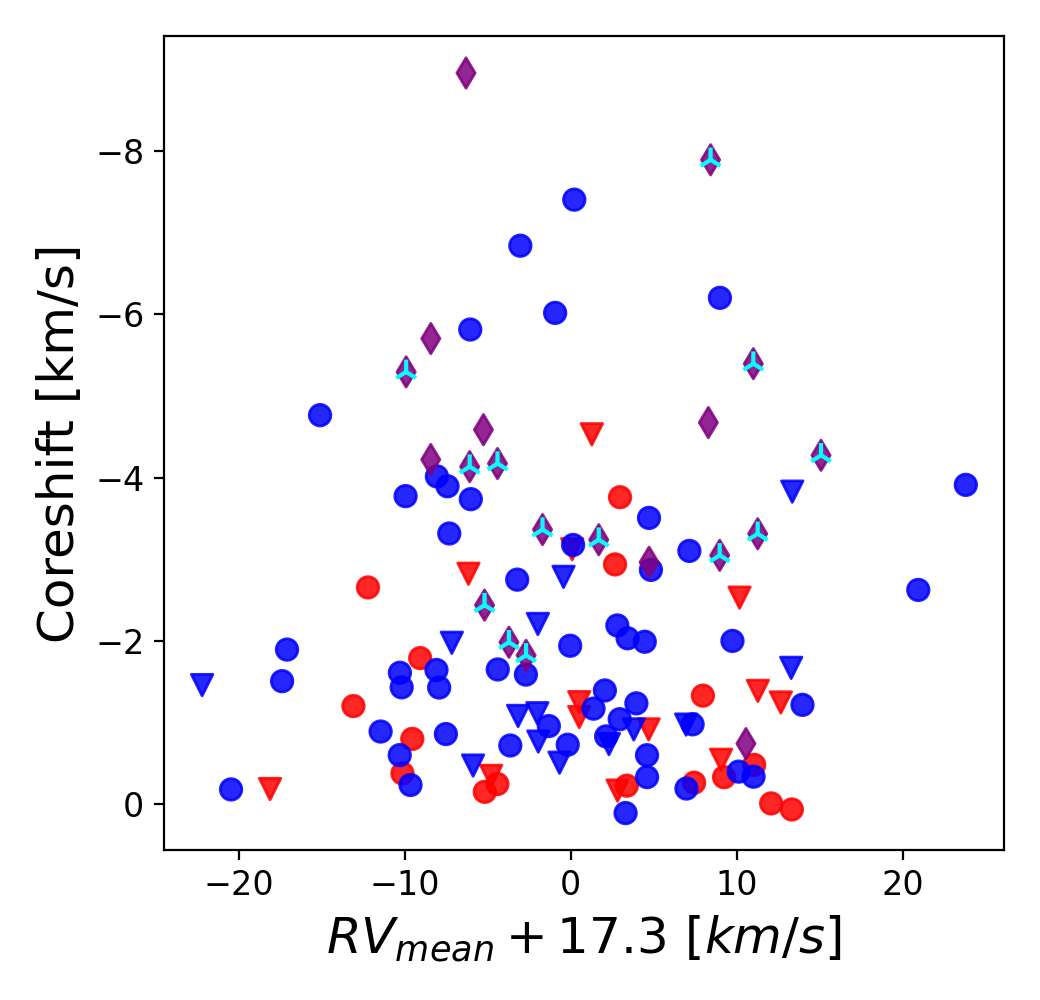}
     \includegraphics[width=0.5\linewidth,trim=0 0 0 0, clip]{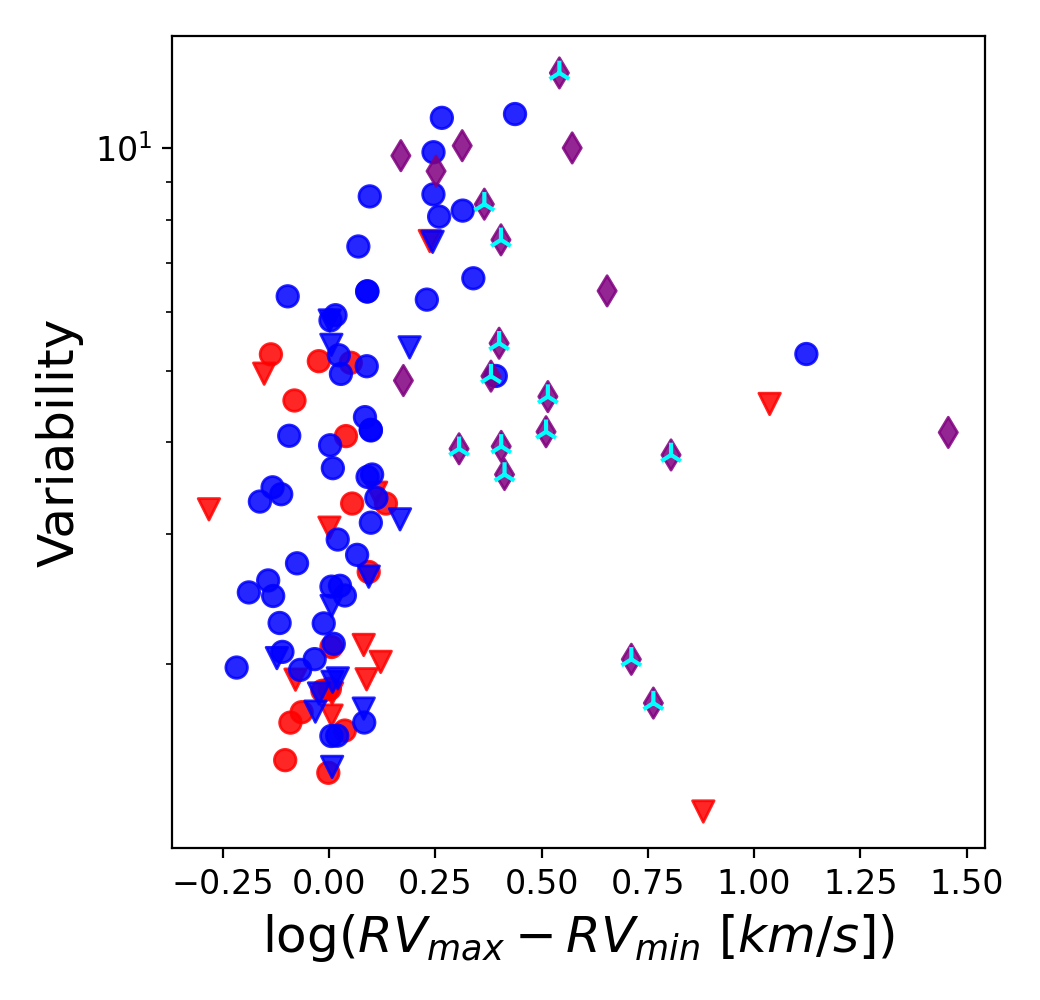}
     \includegraphics[width=0.428\linewidth,trim=54 0 0 0, clip]{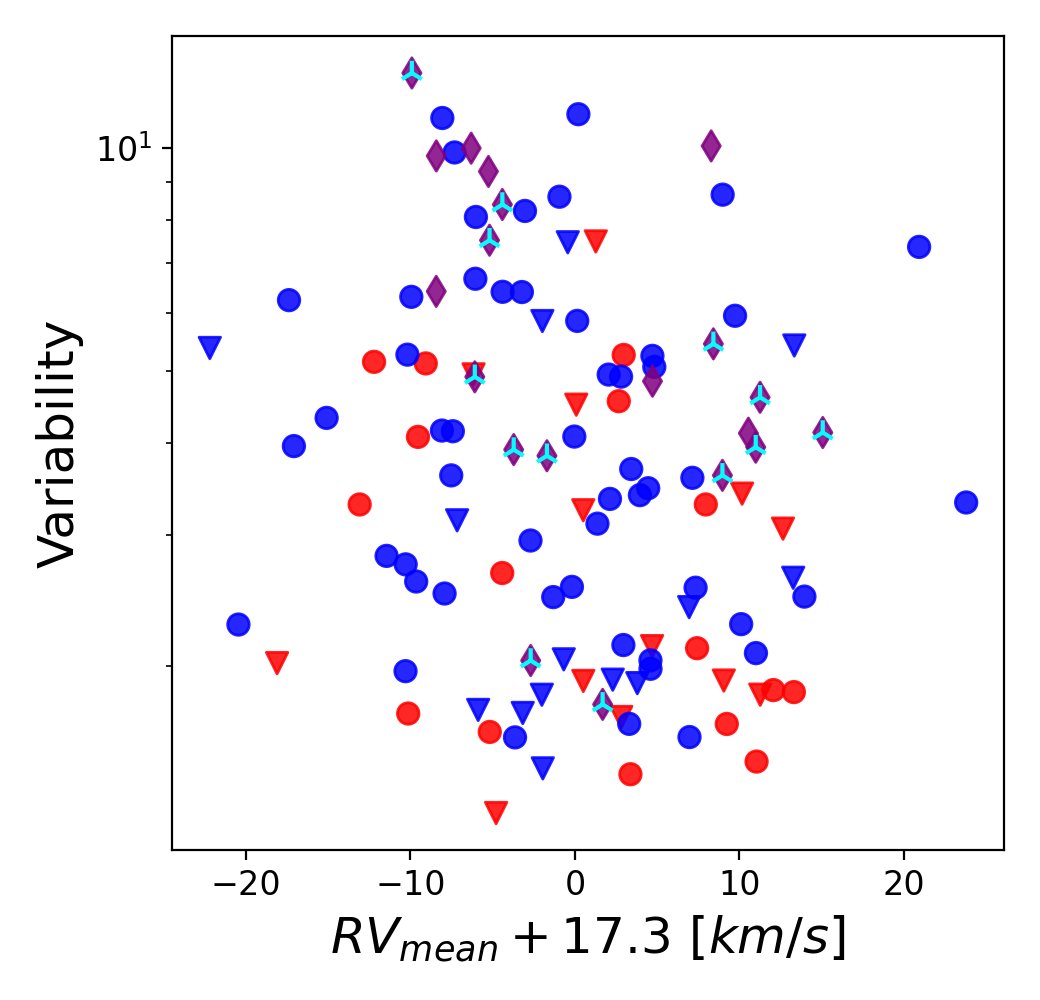}
      \caption{
    Mean {${H\alpha}\ $}coreshift and variability for each star, measured across all epochs, as a function of the photospheric RV amplitude and mean. Colors and symbols are the same as in Fig.~\ref{fig:CMD-kiel}.}
    \label{fig:rvs}
\end{figure}

All our stars show some variation in their photospheric RVs, these variations seems connected with our chromospheric measures (see left panels of Fig.~\ref{fig:rvs}) and appear to display at least two regimes. The first regime is characterized by low-amplitude RV variations, as these slightly increase, so does the variability and the coreshift outflow. This is observed for most stars with $T_\text{eff}\gtrsim3800$~K --stars for which we are able to identify the populations. On the other hand, the high-amplitude RVs variations shows that both, coreshift, and variability tends to decrease. 

Interestingly, while displaying a similar overall behavior, the variability seems to show a more structured trend than the coreshift, even though this latter is the quantity that varies most among the stars (see Fig.~\ref{fig:halpha-profiles}). The coreshift show a slight clustering in low velocity regime $(\lesssim-2\ [km/s])$, beyond this point the correlation with RV amplitude becomes more structured as it does with variability, this point also match with the transition point at $\log(L_\star/L_\odot)\sim2.4$ (see Fig. \ref{fig:halpha-profiles}).

Right panels in Fig.~\ref{fig:rvs} displays the mean radial velocity of our stars minus the mean velocity of our sample (i.e., the cluster velocity), showing no correlation between the mean values and the quantities calculated here, lending credibility to the RV amplitude trends, as no systematic effect is found in our measurements.

\subsection{MPs, RGB, and AGB}

No clear differential behavior is found between populations or evolutionary stages in chromospheric variations or outflows. The E-AGB stars show similar trends to RGB stars at the same luminosities, despite their different $T_\text{eff}$. The H$\alpha$ correlations show no tighter relations with any stellar parameter (see Fig.~\ref{fig:tightness}).


\section{Discussion}
\label{sec:discussion}

The results presented here constitute, to the best of our knowledge, the first systematic characterization of chromospheric variability timescales across the red giants of a GC. Below we place these findings in the context of existing observations and theoretical frameworks, discuss physical scenarios that may explain our observables, and highlighting the limitations.

\subsection{The one-day floor variations}

A ubiquitous result in our sample from our temporal analysis is that all giants, regardless of luminosity, stellar population, or evolutionary status, display statistically significant variations in the {${H\alpha}\ $}profile after approximately one day (see second column in Fig. \ref{fig:variability-on-time}). Chromospheric variability in cool giants on timescales of days has been reported for decades \citep{1983-Cacciari-Freeman-variability-few-days, 1988-Smith-Dupree-wing-emision-h-alpha-variable, 1998-Mullan-variations-on-winds-masive-and-low-mass-giant, 2009-Meszaros-III}, but without a systematic characterization of the threshold timescale. Our one-day floor provides a concrete observational anchor for our luminosity/temperature range.

This floor of variability is broadly flat across the luminosity range of our sample, with only a hint -- below our detection threshold -- that the faintest stars may already reach it on sub-day timescales. Even in integrated disk flux, where contributions from the full stellar surface are averaged together, these variations are detectable. This implies that the responsible structures either cover a substantial fraction of the stellar disk or carry a significant fraction of their total chromospheric energy budget.

In the absence of 3D MHD simulations of low-luminosity red giants, the most natural comparative framework comes from the Sun. Among the most prominent small-scale features of the solar chromosphere are fibrils and spicules -- hair-like structures that outline the chromospheric magnetic canopy \citep{2019-Carlsson-ARA&A-new-view-solar-chromosphere}. These structures exhibit so-called rapid excursions in the spatially resolved disk seen in {${H\alpha}\ $}profiles \citep{2024-Chaurasiya-halpha-sun-spicules}, and may be a good analogy to explain our variations. Although, any analogous structure in red giants likely would be expected to operate on timescales stretched by the ratio of local dynamical timescales. Moreover, the extended low-gravity atmosphere likely buries the magnetic field geometry that organizes the fibrils on the Sun \citep{2003-Ayres-buried-corona-in-chromosphere}, so any solar analog likely will operate under fundamentally different conditions.

Stellar flares offer another possible contribution. While flare activity in cool giants is less well characterized than in active dwarfs, the release of energy in flares is known to produce seismic disturbances on the Sun \citep{2024-Kowalski-stellar-flares}, propagating waves that momentarily alter the chromospheric structure. In a magnetically buried atmosphere, the coupling between a flare event and the chromosphere may be more efficient than in the solar case, potentially amplifying the {${H\alpha}\ $}response. Of course, these are only two of the many solar-surface phenomena \citep[see, e.g.,][]{1964-Beckers-PhDT-structures-chromosphere-sun} that could contribute in different ways to the short-term chromospheric variability we observe. The key point is that the one-day floor must arise from processes energetic enough to imprint themselves on the integrated stellar spectrum, with the solar chromosphere providing a rich phenomenological toolkit for understanding them.

\subsection{The luminosity threshold at \texorpdfstring{$\log(L_\star/L_\odot) \sim$}{log(L/Lsun)\~}2.4}

After $2-4$ days, a different behavior emerges that remains stable over longer timescales. Stars with $\log(L_\star/L_\odot) \lesssim 2.4$ continue to display the flat variability floor seen at shorter timescales, while brighter stars show systematically increasing variability amplitude with luminosity (Fig. \ref{fig:Trends}). Coinciding with this break in variability behavior, the {${H\alpha}\ $}core shifts to systematically more negative values, indicating net outflows in the outer atmosphere (Fig. \ref{fig:halpha-profiles} \& \ref{fig:tightness}). As the variability and the outflows are not perfectly coupled but follow the same trends (left panels Fig~\ref{fig:rvs}) we can picture that the same physical origin produces the changes in the line and just a fraction of the energy is transformed to net outflowing material. Crucially, the inflection is observed at the same point for RGB and E-AGB stars (although our classification may not be certain), and also regardless their population, suggesting an atmospheric structural trend.

\subsubsection{Characteristic timescales}

Two first-order timescales are physically relevant for comparison with the observed $2-4$ day thresholds. The acoustic cut-off period for p modes (the shortest waves that can propagate and persist in the star), which can be approximated as $\nu_\mathrm{cut} = \nu_{\odot}(g/g_\odot)(T_\mathrm{eff}/T_{\mathrm{eff},\odot})^{-0.5}$, \citep{1991-Brown-pmode-cutoff} yields values ranging from sub-day to a few days across the luminosity range of our sample, consistent with the range over which variability becomes detectable, and with the idea that waves propagating through the star could be dissipating on the atmosphere \citep[as done in the Sun through the ``grains''][]{2026-Udnaes-grains-heating-chromosphere-line-behaviour}. The other quantity relevant would be the convective turnover time estimated from mixing-length theory \citep{2012-Kippenhahn-mixing-length-theory-convection-transport} which spans a similar range. Both timescales agree with our observations and can be linked as sources of energy for chromospheric heating and the powering of winds in red giants as previously suggested \citep{2011-Cranmer-mass-loss-theory-cool_stars,2019-Mullan-new-wind-clumpy, 2025-Macdonald-mullan-plasmoid-mass-loss}. Worthy to mention are magnetic spots as they can alter oscillation frequencies in the star \citep{2015-Khomenko-waves-oscillations-in-sunspots} and vary the shape of the lines as they appear and move through the surface \citep{2016-Kochukhov-dopler-imaging-spot} introducing additional variability, but this is too poorly constrained for red giants to attempt a first order approximation.

\subsubsection{Two scenarios for the knee appearance}

\label{sec:two-scenarios}

Considering that the outflows --- although largely variable --- display a systematic tendency toward blueshifted signatures, we can discuss the physical mechanism responsible for the onset of outflow/wind signatures above $\log(L_\star/L_\odot)\sim2.4$, which we interpret as broadly consistent with two limiting scenarios.

\paragraph{Scenario A (observational limit):} The underlying mechanism is present across the full luminosity range of our sample, but below $\log(L_\star/L_\odot) \sim 2.4$ the base of the wind lies above the height at which the {${H\alpha}\ $}core forms, making it invisible to our diagnostic. The outflow velocities at the low-luminosity end of our sample, while small, are not consistent with zero, which may support a scenario where the base of the wind progressively becomes more opaque to {${H\alpha}\ $}core. In this view, the observations do not reveal an onset of mass loss, but rather a change in the sensitivity of {${H\alpha}\ $}as a tracer. This interpretation faces a significant difficulty, {${H\alpha}\ $}is widely regarded as a tracer of the chromosphere, which constitutes the outermost layer in these stars. Contribution function analyses in solar simulations and observations \citep{2007-Rutten-obervation-chromosphere-sun,2019-Borgen-simulation-sun-RMHD} confirm that part of the {${H\alpha}\ $}line core forms at large atmospheric heights \citep[the first 3D chromospheric model of a red giant does too although they were not able to reproduce chromospheric lines,][]{2017-Wedemeyer-CO5BOLD-first-chromospheric-simulation-red-giant}. It therefore seems difficult to maintain a wind that systematically bypasses the chromosphere without leaving any {${H\alpha}\ $}signature. Otherwise, this would raise uncomfortable questions about the mechanism driving winds that can bypass the chromosphere.

\paragraph{Scenario B (physical onset):} The efficiency of magneto-acoustic wave heating in the chromosphere increases sufficiently above $\log(L_\star/L_\odot) \sim 2.4$ that the chromosphere becomes thick and energized enough to sustain a higher ``net outflow''. We note that this picture is also consistent with the deepening of the {${H\alpha}\ $}core as the stars ascend the giant branches, i.e., consistent with increased chromospheric optical depth. This interpretation finds a qualitative match with the 1D-MHD wind models of \citet{2019-Yasuda-types-of-wind---dusty-slow-more-ML-rate-than-windy-fast}, where were identified four distinct wind regimes along low-mass RGB and AGB evolutionary tracks --although not in our mass/metallicity regime. In the most relevant regime for this scenario (their ``$Type\ I\to Type\ II$''), stars with $T_\mathrm{eff} \gtrsim 4530$ K develop a stable chromospheric structure and a steady, Alfv\'en-wave-driven wind with rapid terminal velocities but a small mass-flux. At cooler temperatures, the chromospheric structure begins to break down, the outer atmosphere is levitated by the wave energy injection producing a terminal velocity slower but with higher amount of material, at the same time the propagation of mechanical energy as coherent waves becomes less effective \citep[see also][]{2025-Suzuki-alfven-waves-driving-ML}. These theoretical results support a picture in which the luminosity range $2.4 \lesssim \log(L_\star/L_\odot) \lesssim 3.0$ corresponds to a gradual transition over which magneto-acoustic heating becomes progressively more efficient, building up the net outflow we detect as a systematic {${H\alpha}\ $}blueshift. Because {${H\alpha}\ $}traces density%
\footnote{%
The {${H\alpha}\ $}core forms as a strong "scattering" line, with a source function set by the ambient radiation field rather than by the local gas temperature \citep{2012-Rutten-halpha-collisionally-controled}. Its opacity is likewise insensitive to temperature: set by the n = 2 population of hydrogen, it tracks the local mass density instead. \citet{2012-Leenaarts-simulation-halpha-chromosphere-sun} demonstrated this quantitatively in 3D solar chromospheric simulations, finding the upper-chromospheric {${H\alpha}\ $}opacity to correlate strongly with mass density and only weakly with temperature.
}%
, this growing blueshift likely reflects a rising mass flux at the base of the wind rather than a faster outflow, consistent with the slower but more mass-loaded cooler regime.

\subsection{The RV amplitude connection and the transition to kinematic energy transfer}

A key additional constraint comes from the comparison between {${H\alpha}\ $}properties and photospheric radial velocity variations (see left panels in Fig. \ref{fig:rvs}). Two distinct regimes are apparent.

In the first regime, occupied by the bulk of our sample (stars with $T_\mathrm{eff} \gtrsim 3800$ K), modest, non-stable RV variations correlate positively with both the {${H\alpha}\ $}variability and the magnitude of the core blueshift, as pulsation amplitude grows, so does the outflow and variability. In this regime, the RV variations could be interpreted to act as a mechanical energy source that drives wave propagation into the chromosphere, and the chromosphere in turn dissipates this energy into heat \citep[as seen for the Sun, e.g.,][]{2026-Udnaes-grains-heating-chromosphere-line-behaviour} and kinetic outflow. This is similar with the shock-dominated 3D chromosphere (although purely hydrodynamic) computed by \citet{2017-Wedemeyer-CO5BOLD-first-chromospheric-simulation-red-giant}, who find a dynamical atmosphere characterized by propagating shock waves and a broad temperature distribution. 

In the second regime, occupied by the most luminous and most strongly radially pulsating variables, the positive correlation between RV amplitude and outflow signature reverses. Stars with large RV amplitudes show smaller core blueshifts and often display {${H\alpha}\ $}barely in absorption. When pulsation amplitudes exceed some threshold, we propose that the stellar atmosphere can no longer convert mechanical energy efficiently as the sources become less abundant (e.g., damping of non-radial pulsations \citealp{2009-Dupret-non-radial-damped-in-brighter-giants}), less homogeneous and more localized (e.g., decrease in the number of granules and increase in their size \citealp{2011-Mathur-Granulation-RGB-increases-with-radii}). Instead of converting pulsation energy efficiently into heat through wave dissipation, energy may become predominantly kinematic. Large-amplitude pulsations mechanically throw material to large heights in the atmosphere. Some of this material may escape as episodic clumpy outflows \citep{2023-Freytag-simulation-dust-wind-AGB, 2025-Suzuki-alfven-waves-driving-ML} favoring the formation of dust, but a substantial fraction will remain gravitationally bound and fall back toward the photosphere. In such case, the atmosphere will no longer have a coherently heated chromospheric layer but rather a dynamically complex zone dominated by alternating compressions and rarefactions, populated by material both on its way outward and falling back inward. This is also similar to the ``$Type\ III,\ IV$'' of \citet{2019-Yasuda-types-of-wind---dusty-slow-more-ML-rate-than-windy-fast}, but rather than mechanically dominated they found through the Alfv\'en-wave driving wind produces the same behavior of quenching the chromosphere through saturation. We stress that this trend is not strictly monotonic, and there is considerable scatter among the variable stars also this is a small sample, but the idea that the chromosphere is not able to be sustained is feasible and worth exploring. 

Also, this would not necessarily imply that emission or spectral features commonly associated with chromospheres would no longer be observable \citep[which are seen e.g.,][]{2018-Rau-some-giants-with-uv-spectra}. Rather, these features would no longer be associated with a well-defined layer, but instead with points of compression and rarefaction between shocks in the atmosphere. This interpretation bears a qualitative resemblance to the concept of the ``clapotisphere'' in the solar context \citep{2021-Rutten-summary}, a region where wave propagation and shock interference dominates, rather than a coherent stratified layer-like structure. We borrow this as an analogy -- as this framework hasn't been applied to red giant atmospheres -- but the underlying physics/radiative transfer is well established \citep{2011-Rutten-clapotisphere-comparison-simulation-leenaarts-NaD,2021-Rutten-summary} and provides a basis for the proposed interpretation. Since {${H\alpha}\ $}largely traces the chromosphere, its quenching implies the directed blueshift should weaken and scatter rather than increase, so the smaller blueshifts of the most variable stars (Fig~\ref{fig:rvs}) follow naturally. The variability, by contrast, would increase in amplitude as {${H\alpha}\ $}would be tracing stronger and more localized shocks within the atmosphere.

\subsection{The \texorpdfstring{$\log(L_\star/L_\odot) \sim$}{log(L/Lsun)\~}3.0 boundary and the broader mass-loss context}

An independent observational anchor for the physical transition we describe comes from the mid-infrared study of 47 Tuc by \citet{2012-Momany-ending-discussion-origlia}, who find that detectable infrared excess due to circumstellar dust is confined to stars with $\log(L_\star/L_\odot) \gtrsim 3.0\ ;\ T_\text{eff}\lesssim3800\ [K]$, coinciding with the onset of long-period variability. This boundary likely marks the point where pulsation-driven dust production starts to become the dominant mass-loss channel \citep[see][]{2018-McDonald-AGB-lowL-ML-dominated-pulsation-mass-of-the-star,2025-Schirmer-radiation-pressure-not-enough}. In our framework, $\log(L_\star/L_\odot) \sim 3.0$ corresponds, we believe, approximately to the upper boundary of the luminosity range over which chromospheric wave driving operates --in the sense that begins to break from a coherently wave-heated outer atmospheric layer and ceases to operate as the primary wind driver. Above this boundary, the pulsation amplitude would increasingly become sufficient to mechanically lift material into the outer envelope (likely in the superposition of propagating waves in the atmosphere). This picture is naturally consistent with the 3D radiation-hydrodynamic models of \citet{2023-Freytag-simulation-dust-wind-AGB}, in which convection and pulsation drive irregular, episodic shocks that lift material to high altitudes.

Another interpretation of the observables in this regime is a chromosphere of larger geometric extent. This would require the projected area of the chromospheric layer to become comparable to that of the stellar disk, and/or the layer to be sufficiently heated, for {${H\alpha}\ $}to appear in emission. In that case, rather than tracing localized shocks as in the mechanism proposed above, {${H\alpha}\ $}would still trace a spatially coherent layer of the star. We cannot assess how likely this is, as no realistic model has tested it directly, and it remains unclear whether such an extended structure could be sustained in time. Across so large a geometric extent, the local radiative cooling time may be shorter than the timescale on which energy can be resupplied and transported through the layer, so that it would break down into the intermittent, shock-dominated regime rather than settle into a steady chromosphere.

\subsection{Metallicity insensitivity and implications}

The comparison with the {${H\alpha}\ $}coreshift measurements of \citet{2008-Meszaros-I} for M15, and \citet{2009-Meszaros-II} for M13 and M92, which span a metallicity range of [Fe/H] $\approx -2.3$ to $-1.6$ dex compared to 47 Tuc at [Fe/H] $\approx -0.7$ dex, shows broadly consistent trends within the shared luminosity range (Fig. \ref{fig:Trends-mezsaros}). While the different analysis methods used in those works prevents a precise quantitative comparison, the similarity in the overall shape and amplitude of the coreshift trend across nearly 1.5 dex in metallicity is notable. It is qualitatively consistent with a magneto-acoustic wave-driving mechanism, whose efficiency is set primarily by the local density, velocity, and magnetic field in the chromosphere rather than by the metallicity of the gas. However, this raises questions on what would be needed in order to reproduce the trends found by \citet{2025-Li-trend-with-metallicity}, who found considerable trends with metallicity, when considering integrated mass-loss between RGB and AGB stars. If the dominant chromospheric mechanism is indeed metallicity insensitive within the luminosity range explored here, reproducing the observations would require either that the RGB tip becomes strongly metallicity/composition dependent -- causing stars to spend significantly different times in the RGB and therefore experience {${ML}\ $}for different periods \citep[see][]{2020-Tailo-ML-on-46-GC-first-generation} --- or that the outflows near the RGB tip (dust-driven and/or pulsation-driven) become strongly metallicity dependent \citep{2015-Mcdonald-n-reimers,2018-Hofner-review-AGB}.


\section{Conclusions}
\label{sec:conclusions} %

We have presented the first systematic characterization of chromospheric variability timescales in red giant stars using multi-epoch {${H\alpha}\ $}spectroscopy of 115 members of 47 Tuc. We found that our giants exhibit statistically significant variability after $\sim$1 day, independently of luminosity, evolutionary stage (RGB or E-AGB), and stellar population, suggesting a ubiquitous chromospheric variability floor. After $2$--$4$ days, a dependent behavior emerged: Above $\log(L_\star/L_\odot)\gtrsim2.4$, the variability amplitude increases together with a systematic {${H\alpha}\ $}blueshift indicative of atmospheric outflows, whereas fainter stars remain consistent with the one-day floor. This threshold is independent of evolutionary stage and stellar population, and a comparison with GCs spanning $\sim1.5$ dex in metallicity suggests that the underlying mechanism is largely metallicity insensitive. Although several interpretations remain possible, we favor a genuine structural evolution of the chromosphere in which increasingly efficient magneto-acoustic heating drives progressively stronger outflows above $\log(L_\star/L_\odot)\sim2.4$, while the transition near $\log(L_\star/L_\odot)\sim3.0$ likely marks the onset of a different atmospheric regime associated with strong pulsations and eventually dust-driven mass loss. Future multi-line monitoring and realistic 3D-radiative-MHD models will be essential to test this picture.

\vspace{-1.5mm}
\section*{\small Data availability}
Table \ref{tab:targets_ext} is only available in electronic form (\ref{tab:targets} is also provided) at the CDS via anonymous ftp to cdsarc.u-strasbg.fr (130.79.128.5) or via \url{http://cdsweb.u-strasbg.fr/cgi-bin/qcat?J/A+A/to-be-filled}.


\begin{acknowledgements}
Based on observations made with European Southern Observatory (ESO) Telescopes at Paranal Observatory. Different public AI tools were used to review/correct the grammar and fluency of this manuscript, also part of the codes were edited by AI, nevertheless all the content was created and developed by humans. We thank the anonymous referee for their useful comments. I.B. gratefully acknowledges the insightful discussions with Andrea Dupree and other colleagues during the \href{https://coolstars23.github.io/index.html}{\color{blue!40!black}{Cool Stars 23}}, which contributed to the refinement of the discussion. I.B. acknowledges financial support from Agencia Nacional de Investigación y Desarrollo (ANID) through the National Doctoral Scholarship, Folio No. 21240449. I.B. also acknowledges support from ESO through the Studentship Programme. L.M. gratefully acknowledges support from ANID-FONDECYT Regular Project n.1251809. I.B., L.M. \& H.F. gratefully acknowledges the support provided by ALMA-ANID funds, No. 31230013. K.O. acknowledges the support of the ANID through the FONDECYT Regular grant 1240301.\newline
{\sf Software:} Python \citep{2021-Python}, Numpy \citep{2020-Numpy-package},
Matplotlib \citep{2007-Matplotlib}, SciPy \citep{2020-scipy}.
\end{acknowledgements}

\FloatBarrier
\vspace{-8mm}
\bibliographystyle{aa} 
\bibliography{bibliography}
\FloatBarrier

\begin{appendix}
\FloatBarrier\newpage
\section{Targets}
\label{app:targets} 

\begin{table}[htb]
    \centering
    \caption{Targets main parameters.}
    \label{tab:targets}
    \resizebox{\linewidth}{!}{%
    \renewcommand{\arraystretch}{1.5}
    \begin{tabular}{l|l|c|c|c|c|c|c|c}
N & Star ID & PoP & Class & $I_0$ [mag] & $(V-I)_0$ [mag] & $\log{(L/L_\odot)}$ & $T_\text{eff}$ [K] & $\log(g\ [cm/s^2])$ \\ \hline
1&MED\_RGB\_3872&2&RGB&-1.198&1.199&2.12$^{+0.01}_{-0.01}$&4483$^{+62}_{-53}$&1.81$^{+0.03}_{-0.02}$\\ 
2&MED\_RGB\_557&1&RGB&-1.242&1.186&2.14$^{+0.01}_{-0.01}$&4500$^{+50}_{-58}$&1.80$^{+0.02}_{-0.02}$\\ 
3&MED\_RGB\_459&1&RGB&-1.238&1.206&2.14$^{+0.01}_{-0.01}$&4469$^{+52}_{-55}$&1.79$^{+0.02}_{-0.02}$\\ 
4&MED\_RGB\_402&2&AGB&-1.258&1.089&2.14$^{+0.01}_{-0.01}$&4670$^{+49}_{-55}$&1.83$^{+0.02}_{-0.02}$\\ 
5&MED\_RGB\_719&2&AGB&-1.260&1.054&2.14$^{+0.01}_{-0.01}$&4718$^{+36}_{-49}$&1.85$^{+0.02}_{-0.02}$\\ 
6&MED\_RGB\_454&1&AGB&-1.260&1.103&2.14$^{+0.01}_{-0.01}$&4649$^{+54}_{-54}$&1.82$^{+0.02}_{-0.03}$\\ 
7&MED\_RGB\_520&1&AGB&-1.261&1.093&2.14$^{+0.01}_{-0.01}$&4666$^{+51}_{-53}$&1.83$^{+0.02}_{-0.03}$\\ 
8&MED\_RGB\_510&2&RGB&-1.259&1.185&2.14$^{+0.01}_{-0.01}$&4502$^{+46}_{-57}$&1.79$^{+0.02}_{-0.02}$\\ 
9&MED\_RGB\_1282&2&RGB&-1.268&1.187&2.15$^{+0.01}_{-0.01}$&4500$^{+50}_{-57}$&1.79$^{+0.02}_{-0.02}$\\ 
10&MED\_RGB\_473&2&RGB&-1.281&1.188&2.15$^{+0.01}_{-0.01}$&4497$^{+52}_{-54}$&1.78$^{+0.02}_{-0.02}$\\ 
11&MED\_RGB\_3500&1&RGB&-1.281&1.195&2.15$^{+0.01}_{-0.01}$&4485$^{+49}_{-56}$&1.77$^{+0.02}_{-0.02}$\\ 
12&MED\_RGB\_3708&2&RGB&-1.293&1.184&2.15$^{+0.01}_{-0.01}$&4506$^{+50}_{-55}$&1.77$^{+0.02}_{-0.02}$\\ 
13&MED\_RGB\_95&1&AGB&-1.310&1.147&2.16$^{+0.01}_{-0.01}$&4569$^{+52}_{-50}$&1.78$^{+0.02}_{-0.02}$\\ 
14&MED\_RGB\_1213&1&AGB&-1.313&1.146&2.16$^{+0.01}_{-0.01}$&4571$^{+52}_{-51}$&1.78$^{+0.02}_{-0.02}$\\ 
15&MED\_RGB\_1265&1&RGB&-1.304&1.214&2.16$^{+0.01}_{-0.01}$&4457$^{+51}_{-53}$&1.76$^{+0.02}_{-0.02}$\\ 
16&MED\_RGB\_1112&2&AGB&-1.327&1.091&2.17$^{+0.01}_{-0.01}$&4661$^{+43}_{-47}$&1.80$^{+0.02}_{-0.02}$\\ 
17&MED\_RGB\_3496&1&AGB&-1.330&1.137&2.17$^{+0.01}_{-0.01}$&4586$^{+52}_{-50}$&1.77$^{+0.02}_{-0.02}$\\ 
18&MED\_RGB\_487&2&AGB&-1.362&1.075&2.18$^{+0.01}_{-0.01}$&4671$^{+37}_{-51}$&1.78$^{+0.02}_{-0.02}$\\ 
19&MED\_RGB\_1522&2&AGB&-1.406&1.091&2.20$^{+0.01}_{-0.01}$&4649$^{+36}_{-46}$&1.76$^{+0.02}_{-0.02}$\\ 
20&MED\_RGB\_1090&2&AGB&-1.419&1.094&2.20$^{+0.01}_{-0.01}$&4645$^{+35}_{-48}$&1.75$^{+0.02}_{-0.02}$\\ 
21&MED\_RGB\_1434&2&RGB&-1.409&1.207&2.20$^{+0.01}_{-0.01}$&4466$^{+51}_{-51}$&1.71$^{+0.03}_{-0.02}$\\ 
22&\textcolor{magenta}{\bf*} MED\_RGB\_663&1&AGB&-1.438&1.081&2.21$^{+0.01}_{-0.01}$&4656$^{+29}_{-43}$&1.75$^{+0.02}_{-0.03}$\\ 
23&MED\_RGB\_656&1&RGB&-1.427&1.212&2.21$^{+0.01}_{-0.01}$&4457$^{+41}_{-55}$&1.70$^{+0.03}_{-0.02}$\\ 
24&MED\_RGB\_921&2&AGB&-1.447&1.098&2.21$^{+0.01}_{-0.01}$&4635$^{+35}_{-48}$&1.73$^{+0.02}_{-0.02}$\\ 
25&MED\_RGB\_3689&2&RGB&-1.442&1.203&2.22$^{+0.01}_{-0.01}$&4471$^{+45}_{-48}$&1.69$^{+0.02}_{-0.02}$\\ 
26&MED\_RGB\_575&1&RGB&-1.446&1.231&2.22$^{+0.01}_{-0.01}$&4429$^{+47}_{-52}$&1.68$^{+0.02}_{-0.02}$\\ 
27&\textcolor{magenta}{\bf*} MED\_RGB\_2430&nan&AGB&-1.450&1.220&2.22$^{+0.01}_{-0.01}$&4448$^{+39}_{-55}$&1.69$^{+0.02}_{-0.02}$\\ 
28&MED\_RGB\_2929&2&AGB&-1.481&1.131&2.23$^{+0.01}_{-0.01}$&4590$^{+43}_{-44}$&1.71$^{+0.02}_{-0.02}$\\ 
29&MED\_RGB\_1588&2&RGB&-1.510&1.205&2.24$^{+0.01}_{-0.01}$&4469$^{+46}_{-49}$&1.66$^{+0.02}_{-0.02}$\\ 
30&MED\_RGB\_938&2&RGB&-1.552&1.230&2.26$^{+0.01}_{-0.01}$&4427$^{+42}_{-50}$&1.63$^{+0.02}_{-0.02}$\\ 
31&MED\_RGB\_1289&2&AGB&-1.612&1.111&2.28$^{+0.01}_{-0.01}$&4589$^{+28}_{-40}$&1.65$^{+0.02}_{-0.02}$\\ 
32&MED\_RGB\_3612&1&AGB&-1.608&1.179&2.28$^{+0.01}_{-0.01}$&4508$^{+48}_{-47}$&1.63$^{+0.02}_{-0.03}$\\ 
33&MED\_RGB\_485&1&AGB&-1.621&1.162&2.29$^{+0.01}_{-0.01}$&4534$^{+45}_{-46}$&1.63$^{+0.02}_{-0.02}$\\ 
34&MED\_RGB\_3531&1&RGB&-1.637&1.267&2.30$^{+0.01}_{-0.01}$&4373$^{+39}_{-48}$&1.58$^{+0.02}_{-0.02}$\\ 
35&MED\_RGB\_604&2&RGB&-1.676&1.258&2.31$^{+0.01}_{-0.01}$&4385$^{+41}_{-42}$&1.56$^{+0.02}_{-0.02}$\\ 
36&MED\_RGB\_370&1&RGB&-1.699&1.267&2.32$^{+0.01}_{-0.01}$&4369$^{+43}_{-41}$&1.55$^{+0.02}_{-0.02}$\\ 
37&MED\_RGB\_3776&1&RGB&-1.704&1.309&2.33$^{+0.01}_{-0.01}$&4315$^{+43}_{-48}$&1.53$^{+0.02}_{-0.02}$\\ 
38&MED\_RGB\_1170&2&AGB&-1.761&1.182&2.34$^{+0.01}_{-0.01}$&4491$^{+39}_{-49}$&1.55$^{+0.02}_{-0.02}$\\ 
39&MED\_RGB\_1132&2&RGB&-1.816&1.269&2.37$^{+0.01}_{-0.01}$&4362$^{+43}_{-40}$&1.49$^{+0.02}_{-0.03}$\\ 
40&MED\_RGB\_538&2&RGB&-1.837&1.287&2.38$^{+0.01}_{-0.01}$&4336$^{+41}_{-39}$&1.48$^{+0.03}_{-0.03}$\\ 
41&MED\_RGB\_1599&1&AGB&-1.853&1.238&2.38$^{+0.01}_{-0.01}$&4407$^{+51}_{-46}$&1.48$^{+0.02}_{-0.02}$\\ 
42&MED\_RGB\_706&2&RGB&-1.860&1.277&2.39$^{+0.01}_{-0.01}$&4348$^{+43}_{-36}$&1.47$^{+0.03}_{-0.02}$\\ 
43&MED\_RGB\_3158&2&RGB&-1.894&1.287&2.40$^{+0.01}_{-0.01}$&4336$^{+37}_{-38}$&1.44$^{+0.02}_{-0.02}$\\ 
44&MED\_RGB\_1340&2&RGB&-1.967&1.297&2.43$^{+0.01}_{-0.01}$&4320$^{+43}_{-35}$&1.41$^{+0.02}_{-0.03}$\\ 
45&MED\_RGB\_1293&2&RGB&-1.988&1.307&2.44$^{+0.01}_{-0.01}$&4307$^{+37}_{-38}$&1.39$^{+0.03}_{-0.02}$\\ 
46&MED\_RGB\_926&2&RGB&-2.020&1.306&2.46$^{+0.01}_{-0.01}$&4307$^{+38}_{-37}$&1.38$^{+0.02}_{-0.02}$\\ 
47&MED\_RGB\_752&2&RGB&-2.037&1.310&2.46$^{+0.01}_{-0.01}$&4301$^{+41}_{-37}$&1.37$^{+0.02}_{-0.02}$\\ 
48&MED\_RGB\_1464&2&RGB&-2.037&1.350&2.47$^{+0.01}_{-0.01}$&4251$^{+33}_{-37}$&1.36$^{+0.02}_{-0.02}$\\ 
49&MED\_RGB\_959&2&RGB&-2.061&1.342&2.48$^{+0.01}_{-0.01}$&4259$^{+35}_{-34}$&1.35$^{+0.02}_{-0.02}$\\ 
50&\textcolor{magenta}{\bf*} MED\_RGB\_3396&2&RGB&-2.075&1.339&2.48$^{+0.01}_{-0.01}$&4262$^{+36}_{-34}$&1.34$^{+0.02}_{-0.03}$\\ 
51&MED\_RGB\_3538&1&RGB&-2.090&1.377&2.49$^{+0.01}_{-0.01}$&4218$^{+31}_{-37}$&1.32$^{+0.03}_{-0.02}$\\ 
52&MED\_RGB\_564&1&RGB&-2.107&1.345&2.49$^{+0.01}_{-0.01}$&4251$^{+40}_{-33}$&1.33$^{+0.03}_{-0.03}$\\ 
53&MED\_RGB\_1257&1&RGB&-2.127&1.320&2.50$^{+0.01}_{-0.01}$&4284$^{+41}_{-38}$&1.33$^{+0.03}_{-0.03}$\\ 
54&MED\_RGB\_838&2&RGB&-2.123&1.340&2.50$^{+0.01}_{-0.01}$&4257$^{+42}_{-35}$&1.32$^{+0.02}_{-0.03}$\\ 
55&MED\_RGB\_183&1&RGB&-2.158&1.328&2.51$^{+0.01}_{-0.01}$&4270$^{+43}_{-37}$&1.31$^{+0.03}_{-0.03}$\\ 
56&MED\_RGB\_3665&2&RGB&-2.160&1.336&2.52$^{+0.01}_{-0.01}$&4259$^{+39}_{-36}$&1.30$^{+0.03}_{-0.03}$\\ 
    \end{tabular}}
\end{table}

\begin{table}[htb]
    \centering
    \vspace{1.08cm}
    \caption*{Table \ref{tab:targets}: continued}
    \resizebox{\linewidth}{!}{%
    \renewcommand{\arraystretch}{1.5}
    \begin{tabular}{l|l|c|c|c|c|c|c|c}
N & Star ID & PoP & Class & $I_0$ [mag] & $(V-I)_0$ [mag] & $\log{(L/L_\odot)}$ & $T_\text{eff}$ [K] & $\log(g\ [cm/s^2])$ \\ \hline
57&MED\_RGB\_3427&2&RGB&-2.178&1.328&2.52$^{+0.01}_{-0.01}$&4269$^{+44}_{-36}$&1.29$^{+0.03}_{-0.02}$\\ 
58&MED\_RGB\_3667&2&RGB&-2.180&1.354&2.52$^{+0.01}_{-0.01}$&4239$^{+37}_{-36}$&1.29$^{+0.03}_{-0.02}$\\ 
59&MED\_RGB\_102&1&AGB&-2.230&1.324&2.54$^{+0.01}_{-0.01}$&4274$^{+41}_{-36}$&1.27$^{+0.02}_{-0.02}$\\ 
60&MED\_RGB\_3046&2&RGB&-2.261&1.362&2.56$^{+0.01}_{-0.01}$&4223$^{+35}_{-33}$&1.24$^{+0.02}_{-0.02}$\\ 
61&MED\_RGB\_1211&2&RGB&-2.306&1.378&2.58$^{+0.01}_{-0.01}$&4202$^{+39}_{-32}$&1.22$^{+0.02}_{-0.02}$\\ 
62&MED\_RGB\_3364&2&RGB&-2.322&1.391&2.59$^{+0.01}_{-0.01}$&4186$^{+34}_{-31}$&1.20$^{+0.02}_{-0.03}$\\ 
63&MED\_RGB\_3597&2&RGB&-2.337&1.392&2.59$^{+0.01}_{-0.01}$&4184$^{+32}_{-34}$&1.20$^{+0.02}_{-0.03}$\\ 
64&MED\_RGB\_1349&2&RGB&-2.344&1.377&2.59$^{+0.01}_{-0.01}$&4201$^{+41}_{-32}$&1.20$^{+0.02}_{-0.03}$\\ 
65&MED\_RGB\_1207&1&RGB&-2.338&1.429&2.60$^{+0.01}_{-0.01}$&4146$^{+35}_{-30}$&1.18$^{+0.02}_{-0.03}$\\ 
66&MED\_RGB\_2999&1&AGB&-2.387&1.340&2.61$^{+0.01}_{-0.01}$&4242$^{+35}_{-37}$&1.19$^{+0.03}_{-0.03}$\\ 
67&MED\_RGB\_2704&2&RGB&-2.425&1.407&2.63$^{+0.01}_{-0.01}$&4162$^{+34}_{-29}$&1.15$^{+0.03}_{-0.02}$\\ 
68&MED\_RGB\_1642&2&AGB&-2.444&1.374&2.63$^{+0.01}_{-0.01}$&4201$^{+35}_{-33}$&1.15$^{+0.02}_{-0.02}$\\ 
69&MED\_RGB\_3327&2&AGB&-2.457&1.385&2.64$^{+0.01}_{-0.01}$&4187$^{+38}_{-33}$&1.14$^{+0.02}_{-0.03}$\\ 
70&MED\_RGB\_3399&2&AGB&-2.581&1.371&2.69$^{+0.01}_{-0.01}$&4190$^{+32}_{-34}$&1.09$^{+0.02}_{-0.03}$\\ 
71&MED\_RGB\_1807&2&RGB&-2.579&1.477&2.70$^{+0.01}_{-0.01}$&4082$^{+27}_{-29}$&1.05$^{+0.03}_{-0.02}$\\ 
72&MED\_RGB\_1645&2&AGB&-2.654&1.428&2.73$^{+0.01}_{-0.01}$&4126$^{+32}_{-31}$&1.03$^{+0.02}_{-0.02}$\\ 
73&MED\_RGB\_1401&2&RGB&-2.648&1.501&2.73$^{+0.01}_{-0.01}$&4055$^{+31}_{-27}$&1.01$^{+0.02}_{-0.03}$\\ 
74&MED\_RGB\_82&2&RGB&-2.706&1.509&2.75$^{+0.01}_{-0.01}$&4043$^{+28}_{-27}$&0.98$^{+0.03}_{-0.02}$\\ 
75&MED\_RGB\_944&2&RGB&-2.713&1.494&2.76$^{+0.01}_{-0.01}$&4056$^{+27}_{-29}$&0.98$^{+0.03}_{-0.02}$\\ 
76&MED\_RGB\_2408&2&RGB&-2.725&1.488&2.76$^{+0.01}_{-0.01}$&4060$^{+28}_{-29}$&0.97$^{+0.03}_{-0.02}$\\ 
77&MED\_RGB\_1229&2&RGB&-2.757&1.511&2.78$^{+0.01}_{-0.01}$&4037$^{+31}_{-26}$&0.95$^{+0.03}_{-0.03}$\\ 
78&MED\_RGB\_1413&2&RGB&-2.781&1.499&2.78$^{+0.01}_{-0.01}$&4046$^{+28}_{-26}$&0.94$^{+0.02}_{-0.03}$\\ 
79&MED\_RGB\_874&1&RGB&-2.792&1.554&2.79$^{+0.01}_{-0.01}$&3997$^{+29}_{-25}$&0.92$^{+0.03}_{-0.02}$\\ 
80&MED\_RGB\_576&2&RGB&-2.812&1.527&2.80$^{+0.01}_{-0.01}$&4019$^{+27}_{-28}$&0.91$^{+0.03}_{-0.02}$\\ 
81&MED\_RGB\_1950&2&RGB&-2.829&1.553&2.81$^{+0.01}_{-0.01}$&3994$^{+27}_{-26}$&0.90$^{+0.02}_{-0.03}$\\ 
82&MED\_RGB\_1329&1&AGB&-2.866&1.497&2.82$^{+0.01}_{-0.01}$&4043$^{+32}_{-29}$&0.90$^{+0.03}_{-0.02}$\\ 
83&MED\_RGB\_816&2&RGB&-2.907&1.607&2.85$^{+0.01}_{-0.01}$&3944$^{+23}_{-26}$&0.84$^{+0.03}_{-0.03}$\\ 
84&MED\_RGB\_931&2&RGB&-2.958&1.604&2.87$^{+0.01}_{-0.01}$&3942$^{+27}_{-25}$&0.81$^{+0.03}_{-0.02}$\\ 
85&\textcolor{magenta}{\bf*} MED\_RGB\_1365&2&RGB&-3.040&1.568&2.90$^{+0.01}_{-0.01}$&3966$^{+28}_{-24}$&0.79$^{+0.02}_{-0.03}$\\ 
86&\textcolor{magenta}{\bf*} MED\_RGB\_953&2&RGB&-3.056&1.572&2.91$^{+0.01}_{-0.01}$&3960$^{+26}_{-25}$&0.78$^{+0.02}_{-0.03}$\\ 
87&MED\_RGB\_3222&2&RGB&-3.051&1.594&2.91$^{+0.01}_{-0.01}$&3942$^{+23}_{-28}$&0.77$^{+0.03}_{-0.02}$\\ 
88&MED\_RGB\_3639&2&RGB&-3.069&1.628&2.92$^{+0.01}_{-0.01}$&3915$^{+27}_{-22}$&0.75$^{+0.03}_{-0.02}$\\ 
89&MED\_RGB\_3285&2&RGB&-3.135&1.621&2.94$^{+0.01}_{-0.01}$&3913$^{+24}_{-25}$&0.72$^{+0.03}_{-0.02}$\\ 
90&\textcolor{magenta}{\bf*} MED\_RGB\_1804&2&RGB&-3.171&1.670&2.96$^{+0.01}_{-0.01}$&3873$^{+23}_{-22}$&0.69$^{+0.03}_{-0.03}$\\ 
91&MED\_RGB\_537&1&RGB&-3.177&1.666&2.96$^{+0.01}_{-0.01}$&3875$^{+23}_{-23}$&0.69$^{+0.03}_{-0.03}$\\ 
92&MED\_RGB\_2818&2&RGB&-3.216&1.667&2.98$^{+0.01}_{-0.01}$&3872$^{+23}_{-24}$&0.67$^{+0.03}_{-0.03}$\\ 
93&MED\_RGB\_1079&2&RGB&-3.215&1.711&2.98$^{+0.01}_{-0.01}$&3843$^{+23}_{-21}$&0.65$^{+0.02}_{-0.02}$\\ 
94&MED\_RGB\_453&2&RGB&-3.230&1.658&2.99$^{+0.01}_{-0.01}$&3878$^{+25}_{-24}$&0.66$^{+0.03}_{-0.02}$\\ 
95&\textcolor{magenta}{\bf*} MED\_RGB\_807&2&RGB&-3.245&1.712&3.00$^{+0.01}_{-0.01}$&3838$^{+21}_{-22}$&0.64$^{+0.02}_{-0.03}$\\ 
96&\textcolor{magenta}{\bf*} MED\_VAR\_581 \textcolor{blue}{\bf|} V13&nan&Var&-3.252&2.098&3.04$^{+0.01}_{-0.02}$&3666$^{+20}_{-16}$&0.54$^{+0.04}_{-0.02}$\\ 
97&\textcolor{magenta}{\bf*} MED\_RGB\_85&1&AGB&-3.377&1.661&3.05$^{+0.01}_{-0.01}$&3864$^{+25}_{-21}$&0.59$^{+0.03}_{-0.03}$\\ 
98&\textcolor{magenta}{\bf*} MED\_RGB\_629&nan&nan&-3.462&1.857&3.10$^{+0.01}_{-0.01}$&3735$^{+16}_{-17}$&0.49$^{+0.02}_{-0.03}$\\ 
99&MED\_RGB\_3283&nan&nan&-3.451&1.937&3.10$^{+0.01}_{-0.01}$&3698$^{+15}_{-15}$&0.47$^{+0.03}_{-0.02}$\\ 
100&MED\_RGB\_1098&nan&nan&-3.459&1.922&3.10$^{+0.01}_{-0.01}$&3705$^{+16}_{-15}$&0.47$^{+0.03}_{-0.02}$\\ 
101&\textcolor{magenta}{\bf*} MED\_VAR\_1163 \textcolor{blue}{\bf|} V23&nan&Var&-3.508&1.857&3.12$^{+0.01}_{-0.01}$&3732$^{+17}_{-19}$&0.46$^{+0.03}_{-0.02}$\\ 
102&MED\_RGB\_1743&nan&nan&-3.517&1.852&3.12$^{+0.01}_{-0.01}$&3734$^{+17}_{-18}$&0.46$^{+0.03}_{-0.02}$\\ 
103&\textcolor{magenta}{\bf*} MED\_RGB\_543&nan&nan&-3.535&1.894&3.13$^{+0.01}_{-0.01}$&3711$^{+17}_{-16}$&0.44$^{+0.02}_{-0.03}$\\ 
104&\textcolor{magenta}{\bf*} MED\_VAR\_1142 \textcolor{blue}{\bf|} V5&nan&Var&-3.571&1.895&3.15$^{+0.01}_{-0.01}$&3708$^{+17}_{-17}$&0.42$^{+0.03}_{-0.02}$\\ 
105&\textcolor{magenta}{\bf*} MED\_VAR\_2879 \textcolor{blue}{\bf|} LW5&nan&Var&-3.603&1.946&3.16$^{+0.01}_{-0.01}$&3682$^{+16}_{-14}$&0.40$^{+0.02}_{-0.02}$\\ 
106&\textcolor{magenta}{\bf*} MED\_VAR\_3094 \textcolor{blue}{\bf|} LW4&nan&Var&-3.615&1.891&3.16$^{+0.01}_{-0.01}$&3706$^{+17}_{-15}$&0.40$^{+0.03}_{-0.02}$\\ 
107&\textcolor{magenta}{\bf*} MED\_VAR\_1445 \textcolor{blue}{\bf|} V6&nan&Var&-3.598&2.043&3.17$^{+0.01}_{-0.01}$&3644$^{+13}_{-13}$&0.38$^{+0.02}_{-0.02}$\\ 
108&\textcolor{magenta}{\bf*} MED\_RGB\_64 \textcolor{blue}{\bf|} V17&nan&Var&-3.591&2.095&3.17$^{+0.01}_{-0.01}$&3626$^{+12}_{-12}$&0.37$^{+0.02}_{-0.02}$\\ 
109&\textcolor{magenta}{\bf*} MED\_RGB\_3779 \textcolor{blue}{\bf|} V15&nan&Var&-3.681&2.154&3.21$^{+0.01}_{-0.01}$&3598$^{+12}_{-12}$&0.32$^{+0.02}_{-0.02}$\\ 
110&\textcolor{magenta}{\bf*} MED\_VAR\_1086 \textcolor{blue}{\bf|} V16&nan&Var&-3.706&2.058&3.21$^{+0.02}_{-0.01}$&3629$^{+15}_{-14}$&0.32$^{+0.02}_{-0.03}$\\ 
111&\textcolor{magenta}{\bf*} MED\_VAR\_618 \textcolor{blue}{\bf|} LW2&nan&Var&-3.791&2.286&3.27$^{+0.02}_{-0.01}$&3545$^{+13}_{-13}$&0.24$^{+0.03}_{-0.03}$\\ 
112&\textcolor{magenta}{\bf*} MED\_VAR\_1399 \textcolor{blue}{\bf|} V7&nan&Var&-3.783&2.333&3.27$^{+0.02}_{-0.01}$&3532$^{+12}_{-13}$&0.23$^{+0.03}_{-0.03}$\\ 
113&\textcolor{magenta}{\bf*} MED\_VAR\_3405 \textcolor{blue}{\bf|} V25&nan&Var&-3.851&2.113&3.28$^{+0.02}_{-0.01}$&3596$^{+14}_{-14}$&0.24$^{+0.02}_{-0.03}$\\ 
114&\textcolor{magenta}{\bf*} MED\_RGB\_2690&nan&nan&-3.821&2.463&3.29$^{+0.02}_{-0.01}$&3494$^{+11}_{-12}$&0.19$^{+0.03}_{-0.02}$\\ 
115&\textcolor{magenta}{\bf*} MED\_VAR\_1373 \textcolor{blue}{\bf|} LW3&nan&Var&-3.825&2.467&3.29$^{+0.02}_{-0.01}$&3493$^{+12}_{-12}$&0.19$^{+0.02}_{-0.03}$\\ \\
    \end{tabular}}
\end{table}

\begin{minipage}[t][0.01cm][c]{0.9\textwidth}
\scriptsize\vspace{16mm}
 {\bf Notes.} PoP: 1 = first population, 2 = second population, nan = below
$T_\text{eff}=3800$~K, where the CN bands no longer
separate the populations. Class: RGB / AGB from the position in the CMD, Var for
known variables, nan where the RGB/AGB assignment is uncertain. Full parameters,
identifiers and radial velocities are in Table~\ref{tab:targets_ext}.\\
\newline
 \textcolor{magenta}{\bf*}  Stars marked with this symbol display significant RV
 variations, defined as a peak-to-peak photospheric RV amplitude
 $\ge2.02$~km\,s$^{-1}$; the threshold is the smallest
 amplitude among the known variables, so every known variable is included. The
 limiting case is MED\_VAR\_1373.\\
 \newline
\textcolor{blue}{\bf|}  Some of the targets have double names, corresponding to
cross-identifications with the variables reported in the catalog of
\cite{2024-Prudil-variables-in-GC} and cross-checked with
\cite{2005-Lebzelter-47tuc-variables-ML}.\\
\end{minipage}
\clearpage
\section{Variability formalism}
\label{app:variability-formalism}

In this Appendix we define and describe the variability established and used through the paper.

Given a set of $N$ spectra obtained within a specific time interval, we define the residual as

\begin{equation}
    \label{eq:residual}
    {re}_{i}(\lambda) =  
    \frac{F_{n,i}(\lambda)}{F_T(\lambda)}-1 
    \ , \
    \textup{with}
    \
    \left\{
      \begin{array}{rcl}
        F_T(\lambda) & = & \frac{1}{N}\sum_{i=1}^{N} F_{n,i}(\lambda) \\[2mm]
        F_{n,i}(\lambda) & = & \frac{F_i(\lambda)}{C_i(\lambda)}
    \end{array}
    \right.
\end{equation}

where $F_{n,i}(\lambda)$ is the normalized spectrum of the $i$-th epoch, $F_i(\lambda)$ is the observed spectrum, and $C_i(\lambda)$ is the estimated pseudo-continuum
\footnote{Throughout this section, we use the terms ``pseudocontinuum'', ``continuum'' and ``continua'' interchangeably, but we are always referring to the pseudocontinuum that can be fitted to the observed spectra, unless otherwise indicated.} %
(in the same units as the un-normalized spectrum). The template spectrum, $F_T(\lambda)$, is defined throughout this work as the mean of the $N$ spectra considered in each case.

Therefore, these residuals quantify deviations from the mean spectral profile at each wavelength. However, their dispersion cannot be compared directly across the spectrum because the fractional noise level depends on the local flux. Assuming photon-counting statistics \citep{2022-Taylor-introduction-error-analysis}, the fluctuations are

\begin{equation}
\sigma_{\mathrm{pos},i} \propto \frac{1}{\sqrt{F_i(\lambda)}},
\end{equation}

which implies that deeper regions of an absorption line naturally exhibit larger fractional scatter, even in the absence of intrinsic variability, because we are computing the ratio with respect to the mean.

Furthermore, the continuum normalization itself may introduce spurious trends between epochs. These trends can be induced by spurious fluctuations in the fitting or correspond to real small instrumental response differences across observations. Regardless of their origin, we can account for these variations by defining the relative fractional fluctuations between epochs of the fitted continuum as

\begin{equation}
    \sigma_{\mathrm{cont},i}(\lambda)=\sqrt{
    \left(\frac{C_i(\lambda)-C_T(\lambda)}{C_T(\lambda)\cdot F_{n,i}}\right)^2
    +\left(\frac{0.005}{F_{n,i}}\right)^2}
\end{equation}

where $C_T$ is the mean of the $N$ continua considered in the respective case. The first term represents the fractional variation of the continuum, and the second term is a conservative floor accounting for at least a $0.5\%$ displacement error between continua, in addition to any differential trends.

With these considerations, we define the error of the residual:

\begin{equation}
    \label{eq:total-err}
    \sigma_{{re},i}(\lambda)=\sqrt{1-\frac{1}{N}}\times
    \sqrt{\sigma_{\mathrm{pos},i}^2(\lambda)+\sigma_{\mathrm{cont},i}^2(\lambda)}
\end{equation}

This expression corresponds to the quadratic sum of the individual error terms multiplied by a correction factor. Because our measurement is defined relative to the mean spectrum, the residuals are not fully independent, which in this simplified symmetric form yields the factor $(1 - 1/N)$ on variances. This simple form assumes that the errors between epochs are comparable.

Having defined the expected random errors of the residual, we calculate the reduced residual as

\begin{equation}
\label{eq:reduced-residual}
    \mathcal{R}_i(\lambda)= \frac{{re}_i(\lambda)}{\sigma_{{re},i}(\lambda)}
\end{equation}

This quantity can be interpreted as an indicator of excess variance on a pixel-by-pixel basis \citep[similar to the TVS technique introduced by][whose residuals are defined as a subtraction rather than a ratio, making the present formulation more sensitive to relative variations]{1996-Fullerton-TVS}, where deviations in sigma units trace fluctuations above the expected noise level. At this stage, however, an additional global scaling factor may still be present because our spectra are not single measurements, but combinations of $3$--$5$ individual exposures. 

Then, we define for any spectral window $\Lambda$ the empirical dispersion of the reduced residuals as

\begin{equation}
RMS_i(\Lambda)=\sqrt{\frac{1}{N_\lambda}
\sum_{\lambda\in\Lambda}\mathcal{R}_i(\lambda)^2}
\end{equation}

which provides an estimate of the typical fluctuation amplitude once the expected flux dependence has been removed.

Finally, we define a variability metric that compares the dispersion measured between two bands:

\begin{equation}
\label{eq:variability-def}
\mathcal{V}\left(\frac{\Lambda_{\mathrm{target}}}{\Lambda_{\mathrm{ref}}}\right)=
\frac{1}{N}\sum_{i=1}^N
\frac{RMS_i(\Lambda_{\mathrm{target}})}{RMS_i(\Lambda_{\mathrm{ref}})}
\end{equation}

Under the null hypothesis that the observed differential fluctuations arise solely from the considered sources of error, the reduced residual should remain statistically constant across the spectrum. Consequently, the $RMS$ values, which quantify the variance within different spectral bands, should be comparable. Since we are computing the ratio between regions of the same spectrum, any remaining global scaling factor cancels, making this metric S/N independent.

Within the reference bands we intentionally included spectral lines in order to test whether our noise de-trending procedure behaves consistently. As expected, including or excluding these lines does not modify our results, and we decide to include them in order to mitigate the effect --if present-- of the differential core depth. Furthermore, because the reference and target bands are observed simultaneously and are subject to the same instrumental conditions, this differential metric naturally suppresses global systematic effects and isolates variations intrinsic to the chromospheric line.

If the fluctuations across the spectrum are fully consistent with the expected noise scaling, both regions should exhibit comparable dispersions, yielding $\mathcal{V}\approx1$. Conversely, $\mathcal{V}>1$ indicates excess dispersion within the target region. To derive a confidence interval we make use of the F-distribution \citep{1995-johnson-distribution-book1,1996-Walck-handbook-distributions}. We identify the degrees of freedom (d.o.f.) should correspond approximately to the resolution elements rather than the individual points itself since adjacent pixels are correlated by the spectrograph's line spread function and/or due to oversampling and therefore do not constitute independent measurements. Therefore each resolution element is treated as a single independent carrier of spectral information. Since the wavelength sampling is uniform across the entire spectrum, the oversampling factor should be common to both bands and cancels in the ratio $RMS_{\mathrm{tar}}^2/RMS_{\mathrm{ref}}^2$, consequently, this preserves the ratio of independent resolution elements, and this ratio can be described as an F-distribution, i.e:

\begin{equation*}
    \frac{RMS_{\mathrm{tar}}^2}{RMS_{\mathrm{ref}}^2}=
    \frac{\chi^2_{\mathrm{tar}}/\nu_{\mathrm{tar}}}
         {\chi^2_{\mathrm{ref}}/\nu_{\mathrm{ref}}}
    = F(\nu_{\mathrm{tar}},\nu_{\mathrm{ref}})
    \implies\mathcal{V}\approx\sqrt{F(\nu_{\mathrm{tar}},\nu_{\mathrm{ref}})}
\end{equation*}

The absolute values of $\nu$ are estimated from the number of independent resolution elements within each band. Adopting the instrumental resolution, for the adopted bands\vspace{2mm} 

\hspace{-3mm}\begin{tabular}{lc}
    {BLUE BAND:}  & $[6549.569 \ ;\ 6560.319]$ \AA $\implies d.o.f.\approx55$ \\
    {{${H\alpha}\ $}BAND:} & $[6561.569 \ ;\ 6564.069]$ \AA $\implies d.o.f.\approx14$ \\
    {RED BAND:}   & $[6565.319 \ ;\ 6576.069]$ \AA $\implies d.o.f.\approx58$ \\
\end{tabular}

\vspace{2mm}
We have assumed constant values of d.o.f. for the bands. Although the exact number of independent resolution elements varies slightly from epoch to epoch because of the masking of small sky-line regions, these variations do not significantly affect the confidence interval. The difference in d.o.f. between the BLUE and RED bands is due to more pixels being masked by the presence of small sky lines, conversely this can also be used to estimate the error of the individual measures. The metric described here is summarized in Fig \ref{fig:variability-spectra}, where it is easier to see the defined bands, the regions where the sky affect our observations across all epochs, and the quantities defined in eq \ref{eq:residual} and \ref{eq:total-err}.

\FloatBarrier
\section{Parameters of NGC 104/47 Tuc stars}\label{appendix-ngc104}

Being one of the closest GCs, 47 Tuc is also one of the most studied. Therefore, it is not a surprise that a difference in the intrinsic properties of this cluster have long been debated. For a more extensive review, we highly recommend the reader see the discussions in \cite{2014-Thygesen-47tuc-0.78,2021-Baumgardt-Vasiliev-Distance-Clusters,2020-Thompson-CASES-ages-GC-binaries,2017-Brogaard-age-47-tuc-V69} and \cite{2018-Fu-alpha-enhanced-Parsec-47Tuc}.

\begin{table}[ht]
    \renewcommand{\arraystretch}{1.2} 
    {\centering
    \caption{Properties of 47 Tuc as a GC. }
    \label{table-ng104-properties}
    \resizebox{\linewidth}{!}{%
    \begin{tabular}{r|ccccc}
    \hline\hline
       Source & $[Fe/H]$  & $[M/H]$  & Age     & Distance &E(B-V)\\
              & ${dex}$   & ${dex}$  & ${Gyr}$ & ${kpc}$  & $mag$\\ \hline\hline
    This Work 
        & $-0.68^\star$  & $-0.55\pm0.05$ & $\sim11.5\pm0.3$& $4.72$ & $0.01^{\star\star}$ \\
    \cite{2025-Villanova-47Tuc-CMD-mult-pop}
        & $-0.83,-0.98^\star$  & $-0.7,-0.85$ & $13-14$& $4.29$ & $0.02^{\star\star}$ \\
    \cite{2025-Ying-Absolute-Age-GC-MW}
        & $-0.78$  & $-0.64^\star$ & $\sim11.3-12$& $4.75$ & $0.02$ \\
    \cite{2024-Vandenberg-zero-age-HB-GC}
        & $-0.76$  & $-0.62^\star$ & $\cdots$& $4.57$ & $0.03$ \\
    \cite{2023-Baumgardt-luminosity-and-mass-function-GC}
        & $-0.76$  & $-0.62^\star$ & $12.4$& $4.54$ & $0.05$ \\
    \cite{2022-Larsen-chemical-composition-GC}
        & $-0.735$  & $-0.59^\star$ & $11$& $\cdots$ & $\cdots$ \\
    \cite{2021-Baumgardt-Vasiliev-Distance-Clusters}
        & $-0.7$  & $-0.56^\star$ & $11.75$& $4.52$ & $0.04$ \\
    \cite{2020-Tailo-ML-on-46-GC-first-generation}
        & $-0.72$  & $-0.43^\star$ & $13$& $4.72$ & $0.04$ \\
    \cite{2020-Thompson-CASES-ages-GC-binaries}
        & $<-0.75$  & $<-0.61^\star$ & $12$& $4.50-4.55$ & $0.04$ \\ 
    \cite{2018-Fu-alpha-enhanced-Parsec-47Tuc}
        & $-0.76$  & $-0.43,-0.41$ & $12$& $4.4$ & $\sim0.04^{\star\star}$ \\ 
    \cite{2017-Brogaard-age-47-tuc-V69}
        & $-0.7$  & $-0.41^\star$ & $11.8-13$& $4.36$ & $0.03$ \\ 
    \cite{2014-Thygesen-47tuc-0.78}
        & $-0.78$  & $-0.64^\star$ & $\cdots$& $\cdots$ & $0.04$ \\ 
    \cite{2013-Vandenberg-ages-GC}
        & $-0.76$  & $-0.62^\star$ & $11.75$& $4.65$ & $0.03$ \\ 
    \cite{2013-Hansen-2Gyr-diff-metal-rich-poor-GC}
        & $-0.75$  & $-0.61^\star$ & $9.9$& $4.74$ & $0.04$ \\
    \cite{2012-Milone-47Tuc-MP-HST}
        & $-0.75$  & $-0.45^\star$ & $\cdots$& $4.8$ & $0.04$ \\ 
    \cite{2010-Harris-catalog-GC}
        & $-0.72$  & $-0.58$ & $\cdots$& $4.72$ & $0.04$ \\ 
    \cite{2010-Forbes-acreted-vs-insitu-GC}
        & $-0.78$  & $-0.64^\star$ & $13.06$& $\cdots$ & $\cdots$ \\ 
    \cite{2009-Carreta-metallicity-scale-GC}
        & $-0.76$  & $-0.62^\star$ & $\cdots$& $\cdots$ & $\cdots$ \\ 
    \cite{2009-Marin-ACS-survey-GC-ages}
        & $-0.71,-0.78$  & $-0.64 , -0.57$ & $13.05$& $5.01$ & $0.04$ \\ 
    \cite{2003-Gratton-distance-ages-N104-N6753-N6397}
        & $-0.66$  & $-0.45$ & $11.2$& $5.01$ & $0.04$ \\
    \cite{1999-Ferraro-alpha-enhancement-equation-GC-metallicities}
        & $-0.7$  & $-0.59$ & $20.9$& $4.54-4.61$ & $0.04$ \\ 
    \cite{1997-Gratton-ages-GC}
        & $-0.67$  & $-0.53^\star$ & $9.6-10.8$& $5.29$ & $0.04-0.06$ \\ 
    \cite{1996-Carney-alpha-enhancement-GC}
        & $\sim-0.72$  & $-0.59^\star$ & $16.4$& $\cdots$ & $\cdots$ \\ 
    \cite{1993-Salaris-alpha-enhancement-origin}
        & $-0.71$  & $-0.63$ & $17.6-18$& $4.83$ & $0.04$ \\ \hline\hline \\
    \end{tabular}}}%
    { 
    {\scriptsize {\bf Notes.} Our reported values are based on visual fit adopted and the most likely age and metallicity for the whole sample of stars based in our Bayesian approach (see text below).
    \newline
     $\star$
    We used the conversion $[M/H]=[Fe/H]+\log{(0.638f_\alpha+0.362)}$ with $f_\alpha=10^{[\alpha/Fe]}$ described in \cite{1993-Salaris-alpha-enhancement-origin} and \cite{1999-Ferraro-alpha-enhancement-equation-GC-metallicities}. When the $[\alpha/Fe]$ values were not reported we just assume a value of $[\alpha/Fe]=0.2$ \citep[a representative lower limit taken from][]{1996-Carney-alpha-enhancement-GC,2014-Ventura-formation-of-MP-47Tuc,2018-Fu-alpha-enhanced-Parsec-47Tuc,2025-Ying-Absolute-Age-GC-MW}. 
    }\newline
    { \scriptsize $\star\star$
    These values were converted using $A_v$ (Absorption in the visual) to $A_\lambda$ ratios for values of a G2V star \citep{1989-Cardelli-parsec-values-AyAv-conversion,1994-ODonnell-parsec-values-AyAv-conversion} for $R_v=3.1$ provided by the \href{http://stev.oapd.inaf.it/cgi-bin/cmd_3.8}{PARSEC v2.0 CMD3.8} output page \citep{2022-Nguyen-PARSEC-isochrones}.
    }
    }
\end{table}

We provide in table \ref{table-ng104-properties} a  summary of some relevant
studies that reported values for the cluster's metallicity, age, distance and reddening. The values reported in the table justify the ranges used in the isochrone's fitting procedure to obtain the individual stellar parameters, and might serve as a reference for the interested reader.

\subsection{Stellar parameter estimation via bayesian isochrone fitting}

Because literature estimates for the cluster properties span a broad range, and because a classical single-isochrone fit failed to simultaneously reproduce the turn-off (TO), red giant branch bump (RGBB), horizontal branch (HB), and evolved giant branches, we conclude that a single stellar population model is insufficient for this cluster. In particular, standard PARSEC isochrones cannot fully account for the multiple stellar populations and their different $\alpha$-enhancements. This effect was explored by \cite{2018-Fu-alpha-enhanced-Parsec-47Tuc}, who showed that including $\alpha$-enhancement improves the CMD morphology, especially for bright RGB stars, while preserving agreement in other evolutionary phases.

In our tests (see Fig.~\ref{fig:CMD}), PARSEC isochrones reproduce the main sequence and subgiant branch better at relatively high metallicities, whereas RGB and AGB stars progressively shift toward lower-metallicity trends as they ascend the RGB, before returning to more metal-rich behavior near the tip. Therefore, no single isochrone adequately reproduces the full CMD morphology.

Instead of deriving a unique global cluster solution, we focus on obtaining robust stellar atmospheric and intrinsic parameters for individual stars. We therefore adopted a Bayesian multi-isochrone fitting approach in CMD space, exploiting the fact that degeneracies in CMD coordinates propagate into a much more constrained region in physical parameter space (see Fig.~\ref{fig:kielDG}), particularly for evolved giants.

Formally, for a model point $\theta$ and observed CMD point $\mathcal{D}$, we defined

\begin{equation}
    Prob(\theta|\mathcal{D})\propto\mathcal{L}(\mathcal{D}|\theta)\cdot\mathcal{P}(\theta),
\end{equation}

where $\mathcal{L}$ is the likelihood and $\mathcal{P}$ the prior probability. 

The fitting procedure is semi-automated. The user first selects a representative set of isochrones based on prior knowledge of the system (age, metallicity, distance, and reddening constraints from the literature). The prior is therefore defined as

\begin{equation}
    \mathcal{P}(\theta)=
    \begin{cases}
        1, & \textup{if the model is selected} \\
        0, & \textup{otherwise}.
    \end{cases}
\end{equation}

This effectively restricts the inference to a physically motivated subset of models. Isochrones were generated from the PARSEC library \citep{2022-Nguyen-PARSEC-isochrones}, spanning the range of ages and metallicities reported in the literature, adopting the most metal-rich and most metal-poor measurements as upper and lower limits, respectively.

The initial visual alignment of the isochrone grid with the CMD --defined here as magnitude $m_1$ and color $c=m_2-m_3$-- is constrained using the TO, RGBB, and HB, from which the adopted reddening and distance modulus are determined.

The likelihood is defined through the Mahalanobis distance:

\begin{equation}
    D_M^2 = (\boldsymbol{\Delta})^{\mathsf{T}} \boldsymbol{\Sigma}^{-1} \boldsymbol{\Delta}
\end{equation}

where

\begin{equation}
    \boldsymbol{\Delta}=
    \begin{pmatrix}
    \Delta\mathrm{mag}\\
    \Delta\mathrm{col}
    \end{pmatrix}
\end{equation}

and $\boldsymbol{\Sigma}$ is the covariance matrix:

\begin{equation}
\boldsymbol{\Sigma}=
\begin{pmatrix}
\sigma_m^2 & \mathrm{Cov}(m,c)\\
\mathrm{Cov}(m,c) & \sigma_c^2
\end{pmatrix}
\end{equation}

The variance terms include uncertainties from the photometric measure $(\sigma_{m_x})$, reddening $(\delta_{A_V})$, and distance modulus $(\sigma_\mu)$:

\begin{equation}
\sigma_m^2=
\sigma^2_{m_1,\mathrm{phot}}
+
\Big(\frac{A_{m_1}}{A_V}\delta_{A_V}\Big)^2
+
\sigma_\mu^2
\end{equation}

\begin{equation}
\sigma_c^2=
\sigma^2_{m_2,\mathrm{phot}}
+
\sigma^2_{m_3,\mathrm{phot}}
+
\Big(\frac{A_{m_2}-A_{m_3}}{A_V}\Big)^2\delta_{A_V}^2
\end{equation}

\begin{equation}
\mathrm{Cov}(m,c)=
(\delta_{m_1,m_2}-\delta_{m_1,m_3})
\sigma^2_{m_2,\mathrm{phot}}
+
\Bigg(
\frac{A_{m_1}}{A_V}
\frac{A_{m_2}-A_{m_3}}{A_V}
\Bigg)\delta_{A_V}^2
\end{equation}

where $\delta_{m_i,m_j}$ is the Kronecker delta.

The log-likelihood is then written as

\begin{equation}
    \log\mathcal{L}\propto-\frac{N}{2}D_M^2
\end{equation}

where $N$ is a tunable hyperparameter controlling how rapidly the probability decreases with CMD distance. Larger $N$ values favor the nearest isochrone solutions more strongly. In this work we adopt $N=2$, corresponding to a Gaussian-like decay appropriate for Mahalanobis distances.

\begin{figure}[ht]
\centering
\begin{tikzpicture}[scale=1.8]

\draw[gray!40, dashed] (0,0) circle (1);

\node at (0.9,0.1) {X};
\node at (-0.4 ,0.5) {X};
\node at (-0.42,0.6) {X};

\foreach \x/\y in {0.9/0.1, 0.8/0.2, 0.95/-0.05,-0.4/0.5,-0.42/0.6, -0.35/0.4, -0.6/-0.6, 0.1/0.95, 0.5/0.5, -0.7/0.1} {
    \filldraw[blue!60] (\x,\y) circle (0.02);
}

\foreach \x in {0,20,40,60,80,100,120,140,160} {
    \draw[thick, orange] (-{cos(\x)}, -{sin(\x)}) -- ({cos(\x)}, {sin(\x)});
}

\fill[orange!20, opacity=0.5] (0,0) -- (1,0) arc[start angle=0, end angle=1, radius=1] -- cycle;
\node at (0.5,0.06) {\scriptsize $1^\circ$ cone};

\filldraw[red] (0,0) circle (0.05);

\node[red] at (-1.7,0.3) {Observed Star};
\node[blue] at (-1.7,0.0) {Isochrone points};
\node at (-1.7,-0.3) {Not used point (X)};

\end{tikzpicture}
\caption{Illustration of the angular constraint used to select only one model point per direction in the CMD, within a cone of $1^\circ$.\label{choose-1point-cone}}%
\end{figure}
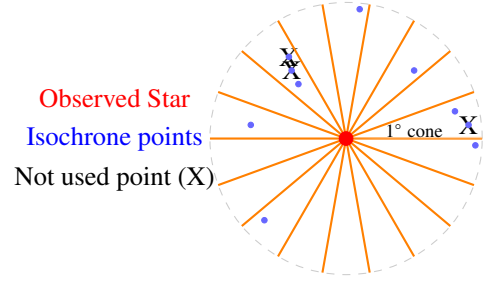

To mitigate biases from non-uniform model sampling along the isochrones, we further restrict the likelihood calculation. Around each observed star, the CMD is divided into angular sectors (360 cones of $1^\circ$ each; see Fig.~\ref{choose-1point-cone}), and only the nearest model point within each cone is retained. This prevents densely sampled evolutionary phases from dominating the posterior distribution.

The final parameter estimates are computed from the posterior-weighted averages:

\begin{equation}
    \bar{X}=
    \frac{\sum_i\omega_i X_i}{\sum_i\omega_i}
\end{equation}

with weights defined as

\begin{equation}
    \omega_i=\exp(\mathcal{L}_i-\max(\mathcal{L}))
\end{equation}

Uncertainties are estimated through asymmetric weighted dispersions:

\begin{equation}
    \sigma_{X_\pm}=
    \sqrt{
    \frac{
    \sum_i\omega_i(X_i-\bar{X})^2
    }{
    \sum_i\omega_i
    }}
\end{equation}

computed independently above and below the mean value.

To evaluate the reliability of the inference, we introduce a heuristic {``skepticism score''}:

\begin{equation}
    SK_{sc}=
    \Bigg|
    \frac{D_{M_{\min}}}{D_{\min\ allowed}}
    \Bigg|
    \Bigg(
    1-\frac{N_{used}}{N_{cones}}
    \Bigg)^{0.5}
\end{equation}

where $D_{M_{\min}}$ is the minimum Mahalanobis distance to the model grid, $D_{\min\ allowed}$ is the maximum accepted CMD distance, $N_{used}$ is the number of contributing model points, and $N_{cones}$ the total number of angular sectors. The first term quantifies proximity to the grid, while the second reflects sampling completeness. Lower values indicate more reliable inferences.

We also define a heuristic scale factor,

\begin{equation}
    \label{eq:scaling-err}
    S_{\text{err}}=1+\sqrt{SK_{sc}}
\end{equation}

used to inflate uncertainties for stars lying in sparsely sampled or extrapolated regions of the CMD, where formal dispersions may underestimate the true uncertainty. This correction was required for only one star in our sample.

    
\begin{figure*}[ht]
    \begin{subfigure}[c]{0.33\textwidth} 
        \caption{}
        \label{cmd+point}
        \hspace{-5mm}\includegraphics[width=1.05\linewidth]{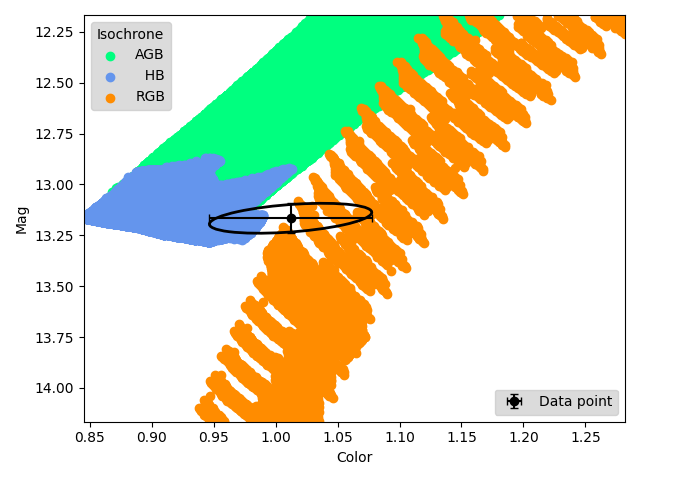}
    \end{subfigure}%
    \begin{subfigure}[c]{0.66\linewidth}
        \begin{subfigure}[t]{0.33\linewidth}  
        \begin{subfigure}[t]{\linewidth}
            \caption{}
            \label{cmd-prob}
            \hspace{-1mm}\includegraphics[width=1.03\linewidth]{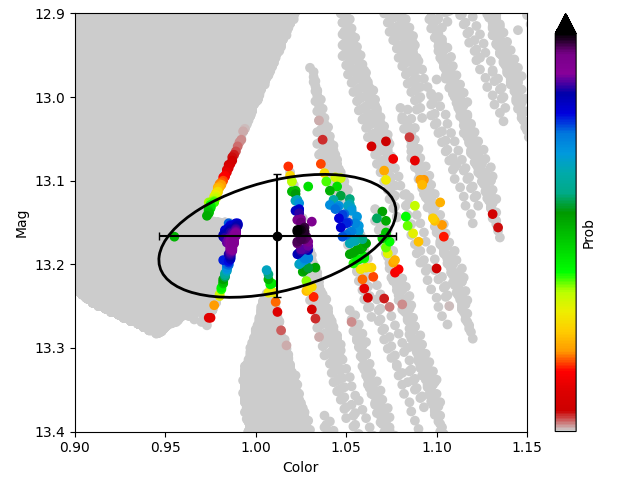}
        \end{subfigure}%

        \begin{subfigure}[t]{\linewidth}
            \caption{}
            \label{lgtf-prob}
            \hspace{3mm}\includegraphics[width=0.94\linewidth]{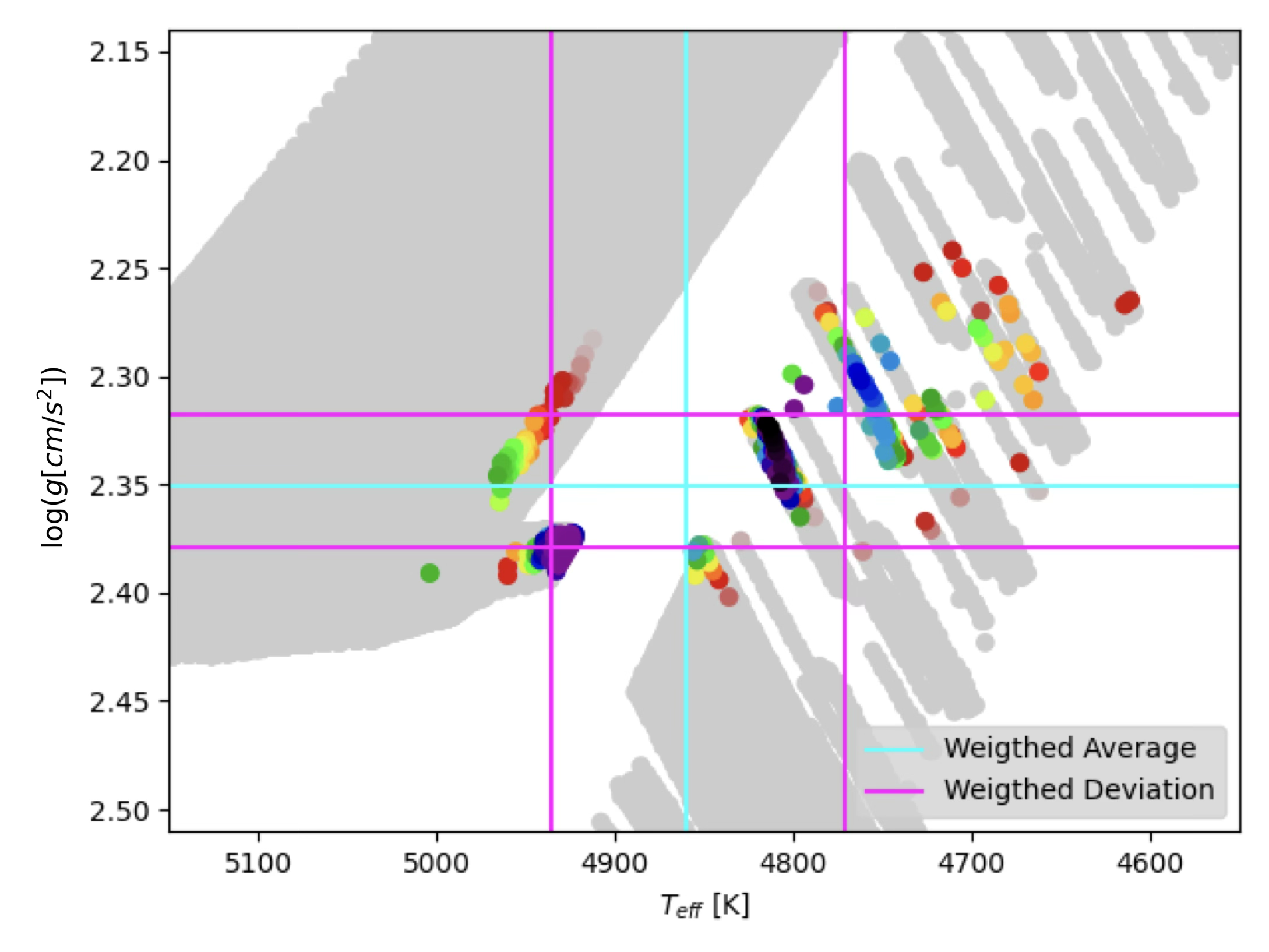}
        \end{subfigure}
        \end{subfigure}%
        \begin{subfigure}[t]{0.33\linewidth}   
        \begin{subfigure}[t]{\linewidth}
            \caption{}
            \label{lg-prob-norm}
            \hspace{-1mm}\includegraphics[width=1.015\linewidth]{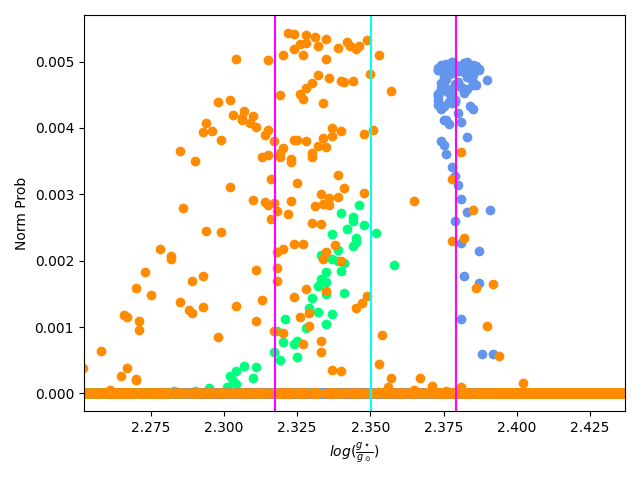}
        \end{subfigure}%
        
        \begin{subfigure}[t]{\linewidth}
            \caption{}
            \label{lg-prob-cum}
            \hspace{3mm}\includegraphics[width=0.98\linewidth]{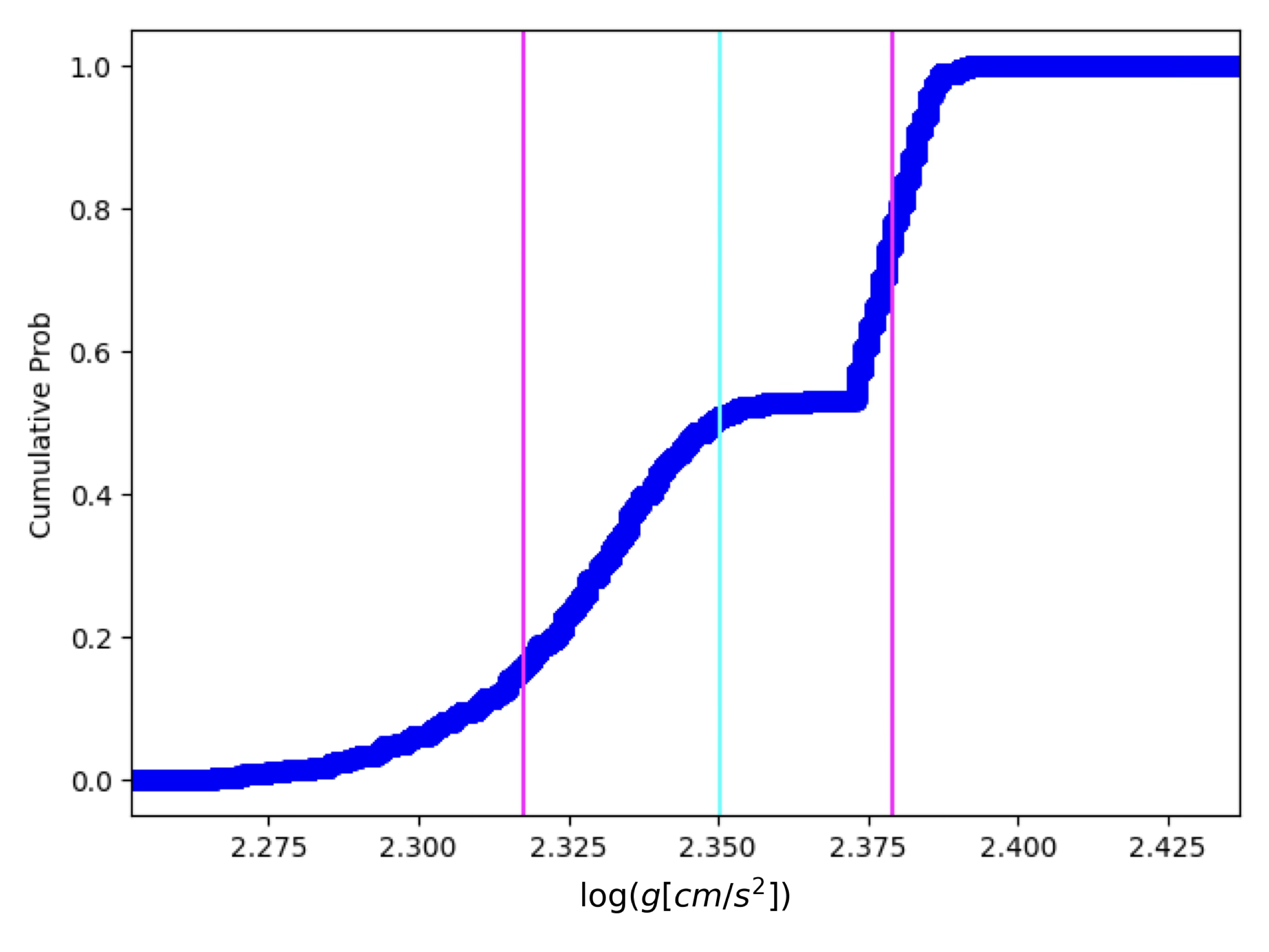}
        \end{subfigure}
        \end{subfigure}%
        \begin{subfigure}[t]{0.33\linewidth}
         \begin{subfigure}[t]{\linewidth}
            \caption{}
            \label{tf-prob-norm}
            \hspace{-1mm}\includegraphics[width=1.015\linewidth]{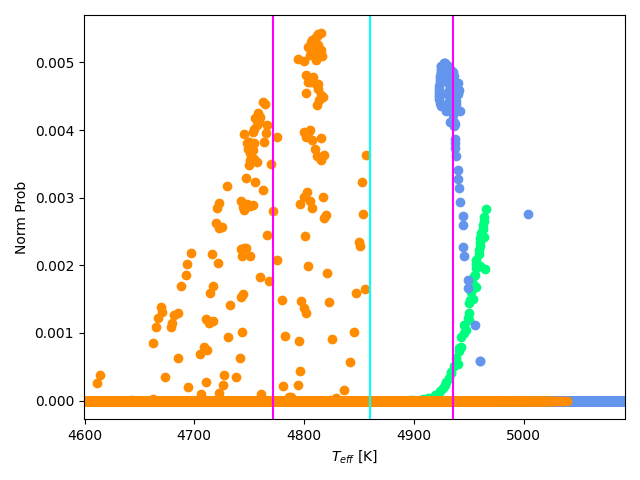}
        \end{subfigure}%
        
        \begin{subfigure}[t]{\linewidth}
            \caption{}
            \label{tf-prob-cum}
            \hspace{3mm}\includegraphics[width=0.98\linewidth]{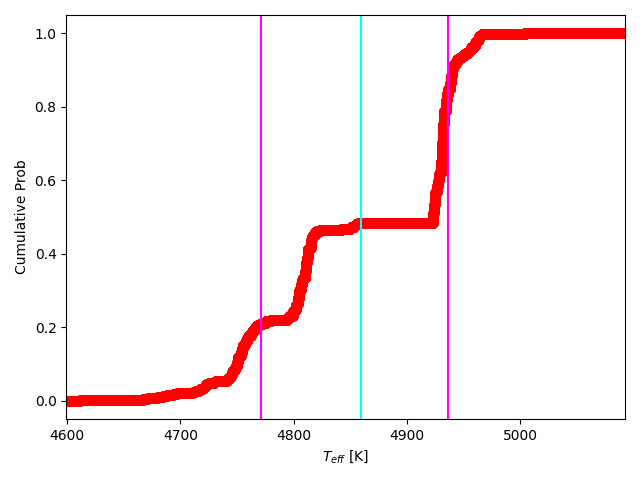}
        \end{subfigure}
        \end{subfigure}%
    \end{subfigure}
    \caption{
    \label{iso-fitting}
    {\bf (a):} I vs V-I Color magnitude diagram (CMD) for a single data point plus a set of multiple isochrones shifted by distance modulus and reddening, according to our best fit of turn-off, RGBB and HB stars (these are considered the priors in our approach). 
    {\bf (b):} Probability associated for each model point according with the data point an their errors in the CMD (Mahalanobis distance).
    {\bf (c):} Probability associated for each model point but in the $T_\text{eff}$ vs $log(g)$ space (Kiel diagram), plus the weighted average and dispersion for each parameter is displayed, in this plot it can be appreciated the no linear propagation of the probability to other parameters spaces.
    {\bf (d) and (f):} Normalized probability for $log(g)$ and  $T_\text{eff}$ respectively, colored accordingly with the panel (a), furthermore we display the weighted average and dispersion. These images show that, for this illustrative case, for this single point, all parameters have an associated probability that is affected by three different distributions reflecting the propagated probability of belonging to the different evolutionary stages (RGB, HB, and AGB) and, within them, which values are most probable for each parameter.   
    {\bf (e) and (g):} Cumulative probability for $log(g)$ and  $T_\text{eff}$ respectively, furthermore we display the weighted average and dispersion. This images illustrate that the method used to determine the most likely value an the dispersion are well suited for the uneven sampled distributions, where within the dispersion we found the most likely values for each parameter are enclosed.
    }
\end{figure*}

Figure~\ref{iso-fitting} illustrates the inference procedure for a representative star. Panel (a) shows the CMD location relative to the selected isochrone grid and adopted reddening/distance shifts. Panel (b) displays the likelihood distribution in CMD space, while panel (c) shows the propagated probability distribution in the $T_{\rm eff}$--$\log(g)$ plane. Panels (d)--(g) illustrate the normalized and cumulative probability distributions for individual parameters, showing how multiple evolutionary solutions contribute to the final posterior.

For our sample we adopt photometric uncertainties of $5\%$ in flux ($\sigma_{m_x}\approx0.05$ mag), a distance modulus uncertainty of $\sigma_\mu\approx0.03$ mag, and a reddening uncertainty of $\delta E(B-V)\approx0.015$ ($\delta A_V\approx0.05$). Using this framework, we infer luminosities, effective temperatures, surface gravities, current masses, metallicities, and ages for the target stars. Final parameters are listed in Table~\ref{table-ng104-properties}.


\section{Data reduction software: Description and public availability}
\label{app:data_reduction_software} 

\begin{figure}[ht]
    \centering
    \includegraphics[width=1.02\linewidth]{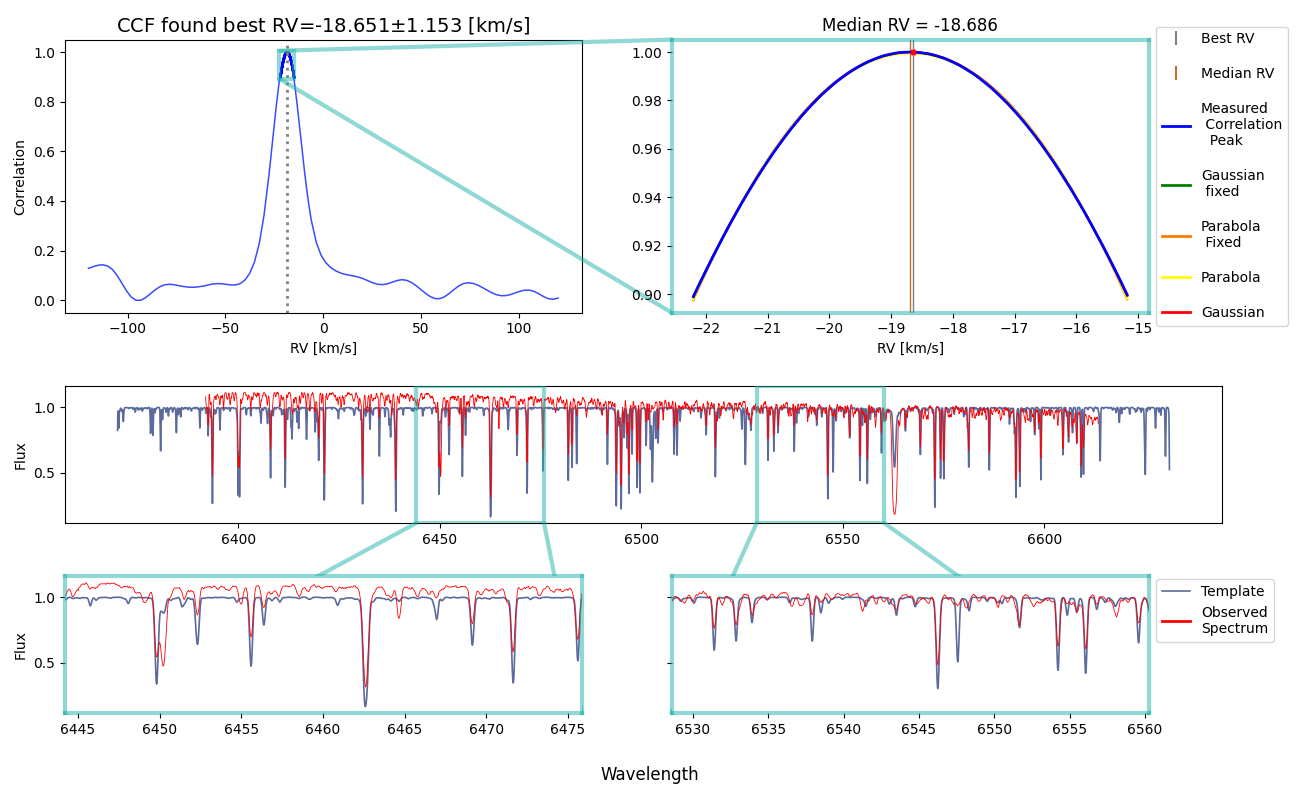}
    \caption{General output of the CCF software for one spectrum from our dataset. Top-left: normalized cross-correlation functions. Top-right: zoom around the correlation peak, where different analytical functions are fitted to estimate the RV and its uncertainty (most fits overlap at this scale). Middle: observed non-normalized spectrum scaled to median$=1$ together with the synthetic template. Bottom panels: zooms of spectral regions used to visually inspect the alignment between the observed spectrum and the template at the derived RV.}
    \label{fig:CCF-software}
\end{figure}

\begin{figure*}[htb]

\begin{subfigure}[c]{1\linewidth} 
    \begin{subfigure}[t]{0.33\linewidth}
        \caption{}
        \label{fig:ideal-spectra}
        \includegraphics[width=0.98\linewidth]{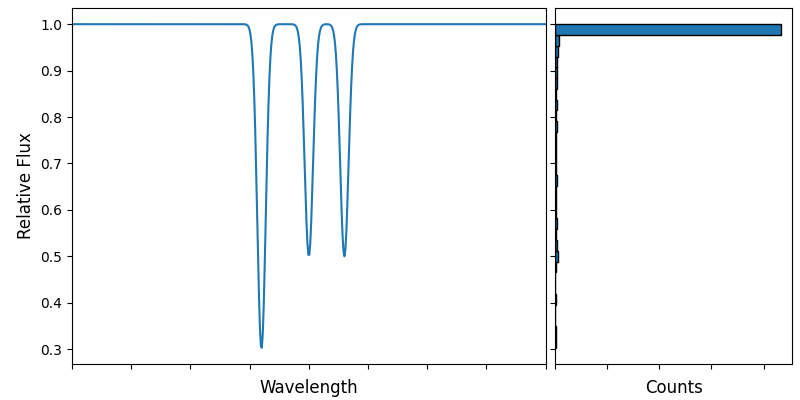}
    \end{subfigure}%
    \begin{subfigure}[t]{0.33\linewidth}
        \caption{}
        \label{fig:noise-spectra}
        \includegraphics[width=0.98\linewidth]{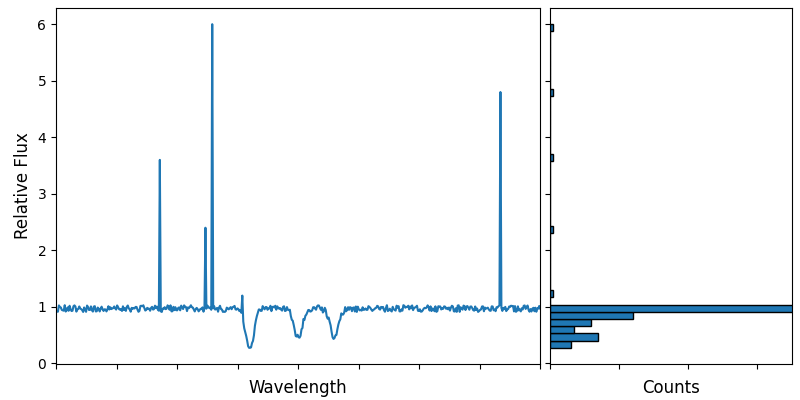}
    \end{subfigure}%
    \begin{subfigure}[t]{0.33\linewidth}
        \caption{}
        \label{fig:instrumental-spectra}
        \includegraphics[width=0.98\linewidth]{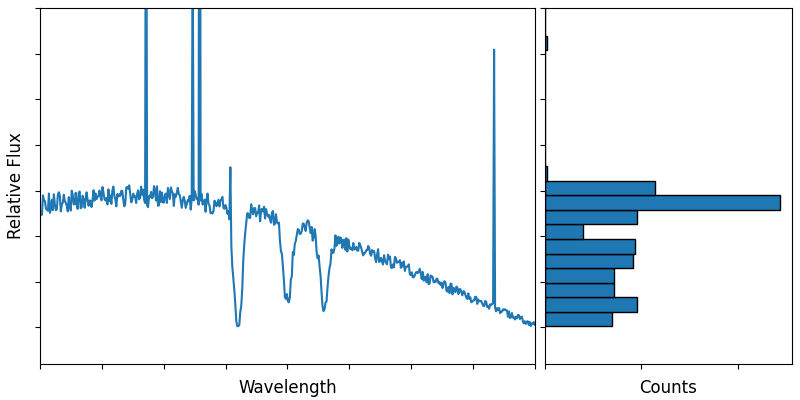}
    \end{subfigure}
\end{subfigure}%

\begin{subfigure}[c]{1.01\linewidth}
    \begin{subfigure}[t]{0.48\linewidth}
        \caption{}
        \label{fig:shifted-spectra}
        \includegraphics[width=1\linewidth]{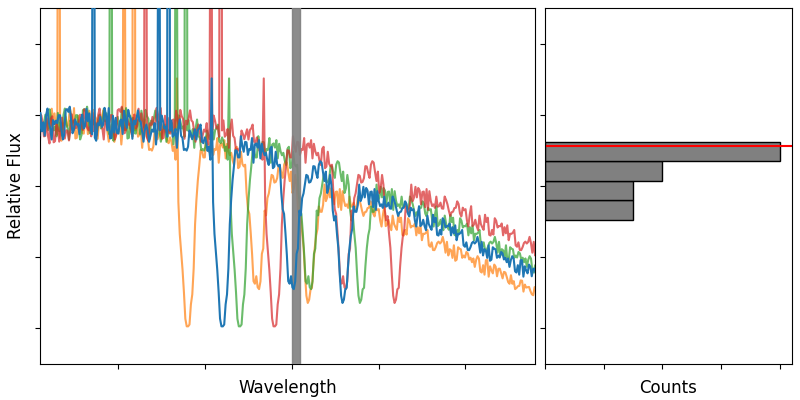}
    \end{subfigure}%
    \begin{subfigure}[t]{0.48\linewidth}
        \caption{}
        \label{fig:1000-shift}
        \includegraphics[width=1.02\linewidth]{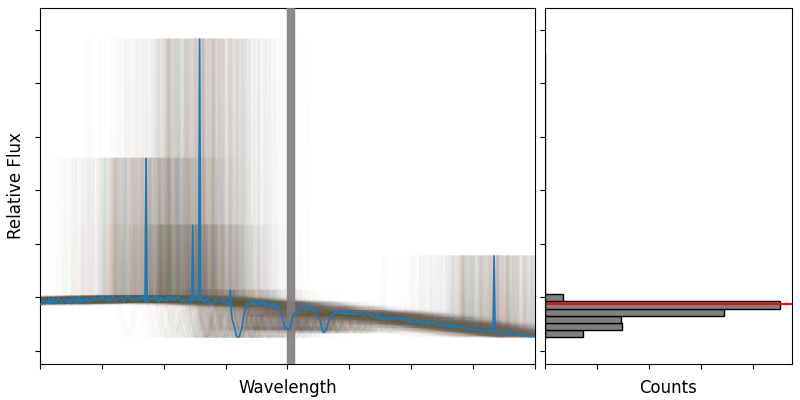}
    \end{subfigure}
\end{subfigure}%

\begin{subfigure}[c]{1.01\linewidth}
     \begin{subfigure}[t]{0.48\linewidth}
        \caption{}
        \label{fig:cont-guess-and-refined}
        \includegraphics[width=1.0\linewidth]{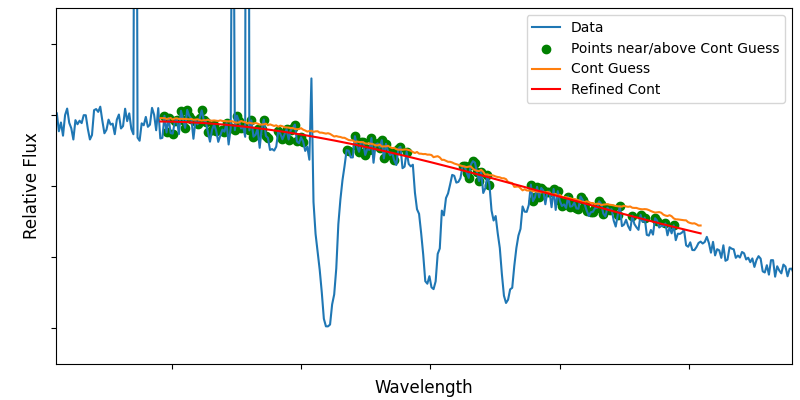}
    \end{subfigure}%
    \begin{subfigure}[t]{0.48\linewidth}
        \caption{}
        \label{fig:cont-sub-residual}
        \includegraphics[width=1.02\linewidth]{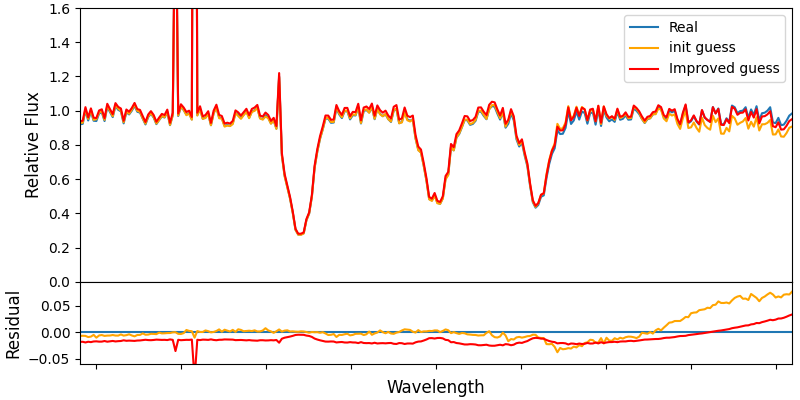}
    \end{subfigure}%
\end{subfigure}

\caption{
\label{fig:cont-fitting}
Illustration of the continuum fitting method.
{\bf (a):} Idealized spectrum with an asymmetric flux distribution ending abruptly at the continuum.
{\bf (b):} Spectrum including noise and cosmic rays.
{\bf (c):} Observed spectrum where instrumental response and intrinsic continuum variations obscure the continuum distribution.
{\bf (d):} The spectrum is repeatedly shifted in wavelength and re-interpolated.
{\bf (e):} Accumulated shifted spectra. The upper envelope traces the continuum level.
{\bf (f):} Initial continuum estimate (orange) obtained from a high percentile of the shifted distributions, and refined spline fit (red).
{\bf (g):} Comparison against the simulated continuum-corrected spectrum and corresponding residuals.
}
\end{figure*}

\begin{figure*}[ht]
    \centering
    \includegraphics[width=0.99\linewidth]{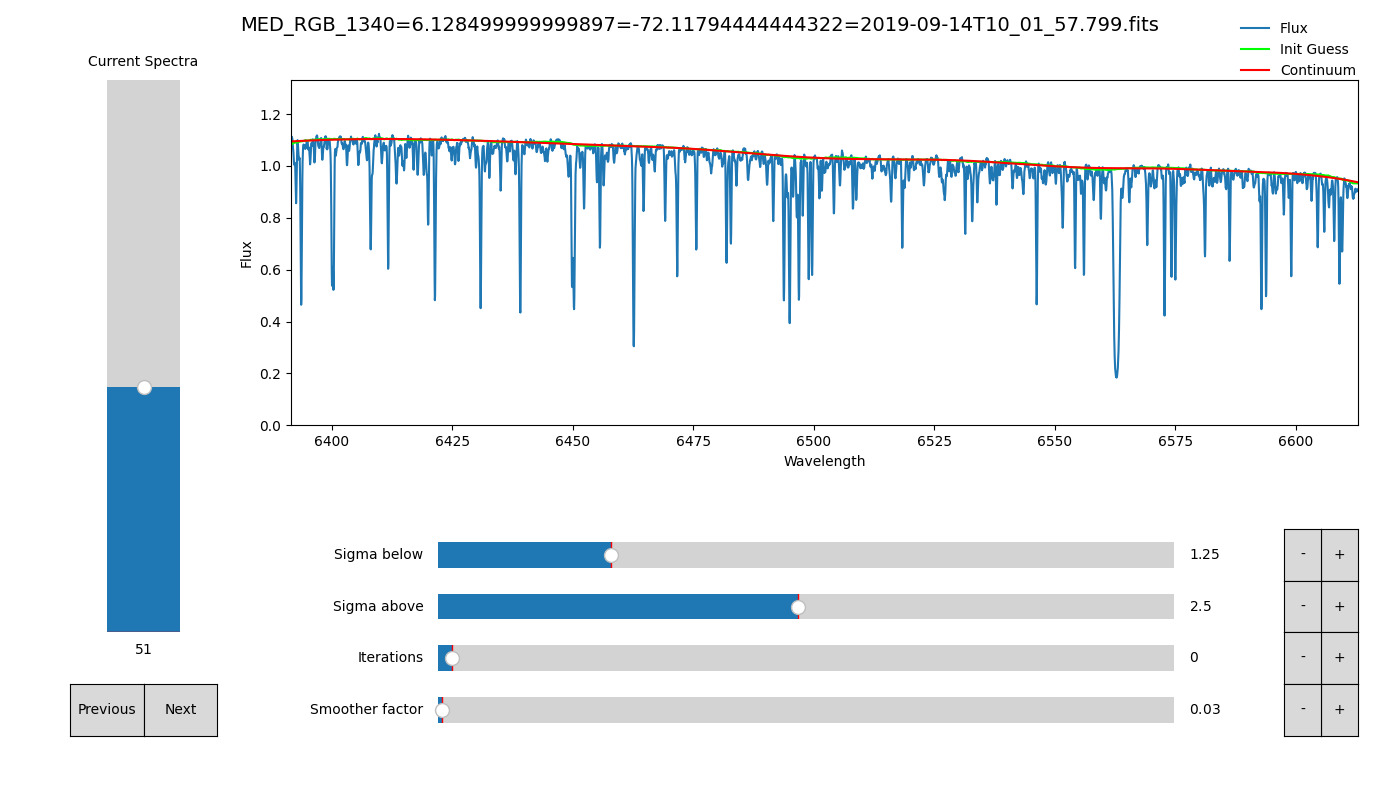}
    \caption{Interactive continuum fitting interface. The main panel displays the current spectrum together with the initial continuum estimate (green) and the refined continuum (red). The left bar indicates the spectrum currently being processed within the dataset, while the lower panels contain the fitting parameters and fine-adjustment controls. Parameters are automatically stored and the fitting session can be resumed later.}
    \label{fig:soft-cont-fit}
\end{figure*}

During this work we developed a suite of {\sf Python} tools for the processing and analysis of multi-epoch astronomical spectra. The software was designed primarily for FLAMES/GIRAFFE and UVES MOS observations, but most modules can be adapted to other spectroscopic datasets with minimal modification. The repository and example datasets will be publicly available\footnote{\url{https://github.com/Inebama/FLAMES-MOS-pipeline-reduction}}.

The pipeline was developed to efficiently handle hundreds of spectra across multiple epochs while preserving enough flexibility for interactive inspection of variable spectral features. The general workflow is:

\begin{enumerate}
    \item Definition of observing blocks (OBs).
    \item Sky subtraction and exposure combination.
    \item Radial velocity estimation through cross-correlation.
    \item Interactive continuum normalization.
\end{enumerate}

Although the data organization tools are instrument-specific, these can be extended by the user. The RV and continuum modules can also operate in a manual mode allowing arbitrary spectra to be processed independently of the original instrument format.

\subsection{Observing block definition}
\label{app:obs_blocks}

Observing blocks group exposures where no significant intrinsic spectral variability is expected. The software automatically associates files according to user-defined criteria (e.g., time separation), with built-in support for ESO GIRAFFE MOSSPECTRA products and ESOREFLEX UVES outputs. 

\subsection{Sky subtraction and exposure merging}
\label{app:sky_subtraction}

Sky subtraction is performed independently for each exposure. Sky fibers are automatically identified and median-combined to create a master sky spectrum, which is then subtracted from each science fiber.

Spectra belonging to the same OB are subsequently combined. Before merging, spectra are shifted to a common reference frame and resampled because barycentric corrections vary between exposures. We adopted a resampling grid with three times the original number of points, allowing slightly different wavelength samplings from individual spectra to contribute additional sub-pixel information. The final OB spectrum is obtained through median combination. Relevant metadata are propagated to the output FITS products.

\subsection{Cross-correlation for radial velocity determination}
\label{app:cross_correlation}

RVs are determined using a cross-correlation function (CCF):

\begin{equation}
    \text{CCF}(v_{\text{shift}}) = \sum_{i} S_{\text{obs}}(\lambda_i)\times S_{\text{template}}(\lambda_i-\Delta\lambda(v_{\text{shift}}))
    \label{eq:CCF}
\end{equation}

where $S_{\text{obs}}$ and $S_{\text{template}}$ are the observed and template spectra, respectively.

Prior to the CCF computation, spectra are scaled close to unity and trimmed if necessary to ensure the template fully covers the wavelength range explored by the velocity shifts. The CCF is computed iteratively:

\begin{enumerate}
    \item A coarse scan identifies the global peak.
    \item A finer scan refines the peak position and searches for nearby asymmetries.
    \item A final high-resolution scan determines the RV.
\end{enumerate}

The RV uncertainty combines:
\begin{itemize}
    \item The scatter of RVs obtained from fitting different analytical functions (two Gaussians and two parabolas) to the normalized CCF peak.
    \item The formal uncertainties from these fits.
    \item A minimum uncertainty estimated from the wavelength sampling around the central wavelength.
\end{itemize}

For high S/N spectra, the sampling term generally dominates the final error budget. Figure \ref{fig:CCF-software} shows a standard output generated by the software.

\subsection{Interactive continuum fitting}
\label{app:continuum_fitting}

We developed an interactive continuum normalization routine based on cubic spline interpolation (see Fig.\,\ref{fig:soft-cont-fit}). The method exploits the asymmetric distribution expected in stellar spectra, where most flux values lie below the continuum.

To estimate an initial continuum, the spectrum is repeatedly shifted in wavelength and re-interpolated at the original pixel positions (Figures \ref{fig:shifted-spectra} and \ref{fig:1000-shift}). Since the continuum varies slowly over small wavelength intervals, the resulting local flux distribution remains dominated by continuum points. A high percentile (typically the 86th) of this distribution is then adopted as the initial continuum estimate (orange line in Figure \ref{fig:cont-guess-and-refined}).

The initial estimate can subsequently be refined interactively using spline fitting and optional sigma-clipping iterations. Residuals increase in regions where the continuum changes rapidly with wavelength, an intrinsic limitation of the shifting approach. Repeating the process iteratively mitigates this effect by improving the overlap between shifted continua.

The iteration parameter has special behaviors:
\begin{itemize}
    \item \texttt{Iteration = -1}: accepts the initial guess directly.
    \item \texttt{Iteration = 0}: smooths the initial guess with a spline.
    \item \texttt{Iteration = 1}: fits a spline to points selected around the initial guess.
    \item \texttt{Iteration > 1}: applies iterative sigma-clipping relative to the previous fit.
\end{itemize}

To reduce memory usage, spectra are loaded individually rather than entirely into RAM. Initial continuum estimates are stored as additional FITS extensions to avoid recomputation.

\subsection{Auxiliary tools}
\label{app:aux_tools}

Additional utilities included in the repository are:

\begin{itemize}
    \item \textbf{Multi-epoch spectral viewer:} interactive inspection and comparison of spectra across epochs.
    \item \textbf{Isochrone interpolator:} graphical interpolation of stellar parameters from PARSEC isochrones, as described in Appendix \ref{appendix-ngc104}.
\end{itemize}

The repository also includes example datasets and test cases for installation verification. Further documentation is currently limited, but questions regarding the software can be addressed upon reasonable request.

\section{Extra resources}
\begin{table}[htb]
    \centering
    \caption{Wavelength intervals of the CN bands} 
    \label{tab:cn_bands}
    \resizebox{\linewidth}{!}{%
    \renewcommand{\arraystretch}{1.15}
    \begin{tabular}{c|c|c|c}
$\lambda$ range [\AA] & $\lambda$ range [\AA] & $\lambda$ range [\AA] & $\lambda$ range [\AA] \\ \hline
6402.60--6403.50 & 6439.60--6440.05 & 6480.00--6481.30 & 6541.00--6541.80 \\
6405.10--6405.75 & 6443.60--6445.40 & 6486.00--6486.90 & 6569.85--6571.60 \\
6410.00--6410.70 & 6451.20--6451.80 & 6505.00--6506.00 & 6575.70--6577.20 \\
6413.00--6414.20 & 6457.00--6457.70 & 6511.85--6512.85 & 6581.70--6583.40 \\
6418.50--6419.74 & 6458.30--6459.30 & 6519.30--6520.25 & 6585.30--6585.80 \\
6425.15--6426.60 & 6465.30--6467.50 & 6522.30--6523.30 & 6588.70--6590.30 \\
6431.30--6432.30 & 6472.80--6473.70 & 6529.00--6530.20 & 6594.70--6596.40 \\
6437.80--6438.40 & 6478.00--6479.20 & 6536.30--6537.60 &  
    \end{tabular}}
    \tablefoot{Bands integrated to separate the two stellar populations (see Sect.~\ref{sec:MPS} and Fig.~\ref{fig:CN-Band}). Wavelengths are given in the rest frame.}
\end{table}

\begin{table*}[t]
    \centering
    \caption{ Description of electronic table
    }
    \label{tab:targets_ext}
    \resizebox{\textwidth}{!}{%
    \renewcommand{\arraystretch}{1.0}
    \begin{tabular}{l|l|l|l}
Quantity & Columns & Units & Notes \\ \hline
Identifiers & \texttt{N}, \texttt{STAR\_ID}, \texttt{GAIA\_DR3}, \texttt{TWOMASS} & -- & row number, internal name, Gaia DR3 \texttt{source\_id}, 2MASS designation \\
Position & \texttt{RA}, \texttt{DEC} & deg & ICRS, epoch J2000 \\
Classification & \texttt{POP}, \texttt{CLASS}, \texttt{MEMB\_PROB} & -- & population (1, 2, NaN), RGB/AGB/Var/nan, membership probability \\
Photometry & \texttt{I0}, \texttt{V\_I0}, \texttt{GMAG}, \texttt{BP\_RP} & mag & dereddened absolute $I$ and $V-I$; Gaia $G$ and $BP-RP$ \\
Stellar parameters & \texttt{LOGL}, \texttt{TEFF}, \texttt{LOGG}, \texttt{RADIUS}, \texttt{MASS\_INI}, \texttt{MASS\_NOW}, \texttt{ERR\_SCALE} & dex, K, dex, $R_\odot$, $M_\odot$, - & from the isochrone fit, each with $\pm$ errors and heuristic scale factor (eq. \ref{eq:scaling-err}) \\
Radial velocity & \texttt{RV\_MEAN}, \texttt{RV\_STD}, \texttt{RV\_AMP}, \texttt{RV\_STRONG} & km\,s$^{-1}$ & mean, epoch scatter, peak-to-peak amplitude, and the variability flag \\
Known variables & \texttt{VAR\_NAME}, \texttt{VAR\_TYPE}, \texttt{VAR\_PERIOD} & --, --, d & designation, type and period \citep{2024-Prudil-variables-in-GC} \\
H$\alpha$ profile &\texttt{N\_EPOCH}, \texttt{HA\_EXP}, \texttt{HA\_WIDTH}, \texttt{HA\_F0}, \texttt{HA\_CORESHIFT}, \texttt{HA\_VARIAB} & --,--, \AA, --, km\,s$^{-1}$, -- & number of epochs and epoch means, each with its epoch scatter (\texttt{\_STD})\\ \hline 
    \end{tabular}
    }
    \tablefoot{Extension \texttt{(TARGETS)} reproduces Table~\ref{tab:targets}, extension \texttt{TARGETS\_EXT} adds columns summarized here, and \texttt{(TARGETS\_EPOCH)} holds the per-epoch H$\alpha$ measurements as vector columns.}
\end{table*}

\begin{figure*}[ht]
    \centering
    \includegraphics[width=\textwidth]{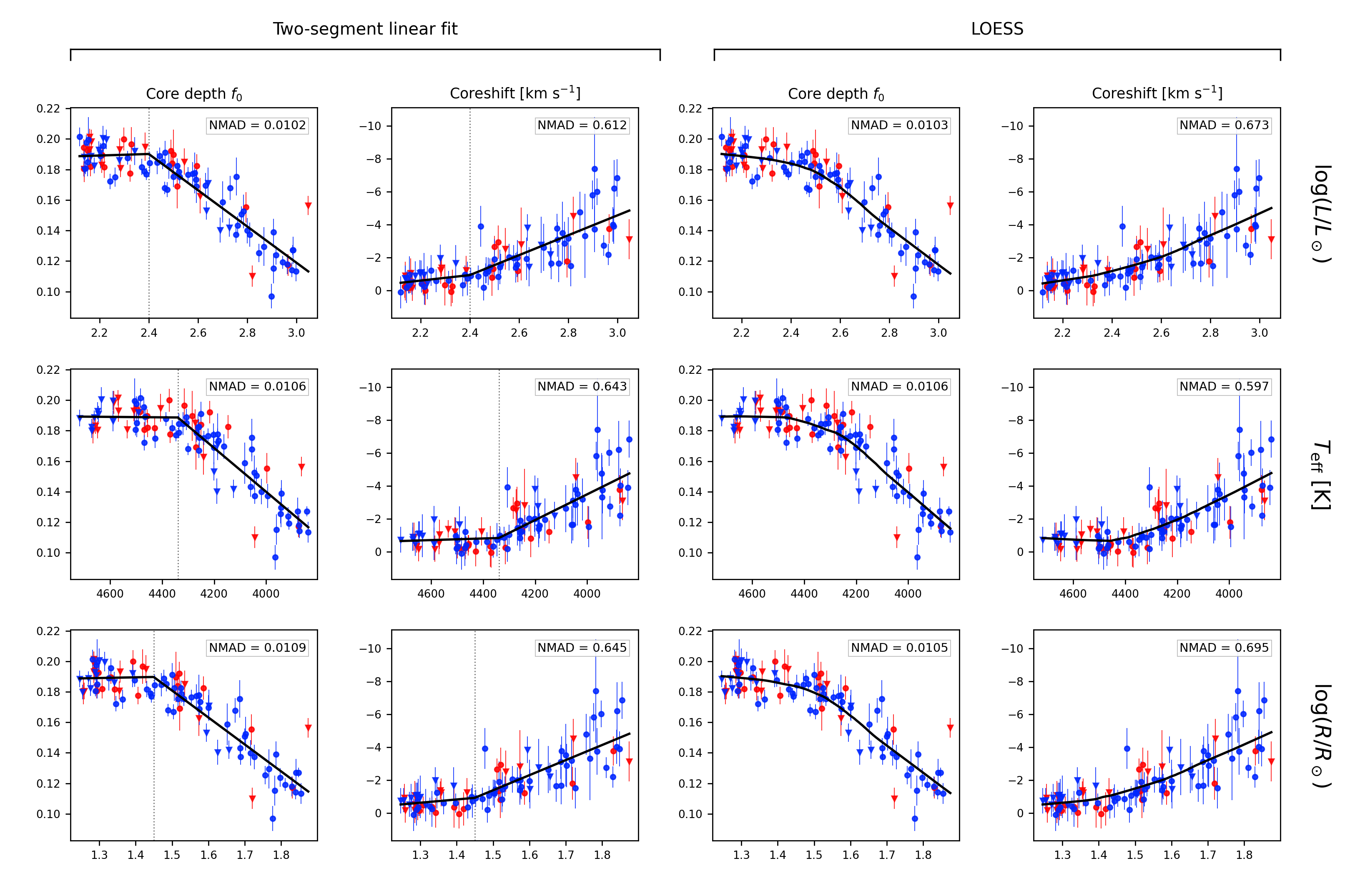}
    \renewcommand{\thefigure}{F.1} 
    \vspace{-6mm}
    \caption{Core depth $f_0$ and H$\alpha$ coreshift fitted against $\log(L/L_\odot)$, $T_\text{eff}$ and $\log(R/R_\odot)$ (rows, labeled on the right), with a continuous two-segment linear model (first pair of columns) and with a LOESS smoother (second pair). Shown are the 95 stars for which the population and the RGB/AGB classification could be determined. The segment break is fixed at $\log(L/L_\odot)=2.4$, converted into each row's own abscissa (dotted line) so that all the left six panels fit exactly the same stars. Error bars are the epoch-to-epoch scatter. Each panel quotes the NMAD of its residuals, in the units of that column. Colors and symbols as in Fig.~\ref{fig:CMD-kiel}.
    }\label{fig:tightness}
\end{figure*}

\section{Correlations with stellar parameters}
\label{app:tightness}

We have tested whether there was a tighter correlation with one stellar parameter over the others with the observables in $H\alpha$. These parameter are correlated between them but in principle the existence of a tighter dependence may hint a dominant process powering chromospheres/winds.

In Fig \ref{app:tightness}\textcolor{magenta}{.1} we show two of the $H\alpha$ parameters fitted displayed in Fig (\ref{fig:halpha-profiles}) against luminosity effective temperature and radius, using two functions: a continuous two-segment linear model, for its simplicity and because it reproduces the change of regime we observe at $\log(L/L_\odot)\sim2.4$, and a LOESS smoother \citep{1988-Cleveland-LOESS}, which does not impose a shape. For this analysis we keep only the stars whose population and RGB/AGB classification could be established. "Tightness" is measured as the normalized median absolute deviation (NMAD) of the residuals.  We have considered the error introduced associated with each point (x and y axis) as $\sigma_\mathrm{intrinsic}^2=\mathrm{NMAD}^2-\sigma_y^2-(b\,\sigma_x)^2$, with $b$ the local slope of the fit. This matters because uncertainties are larger for $T_\text{eff}$ than for $\log L$. Applying this, the intrinsic scatters for $f_0$ are $0.0100/0.0097/0.0106$ (segments) and $0.0101/0.0099/0.0103$ (LOESS) against $\log L$, $T_\text{eff}$ and $\log R$ respectively. And, for the coreshift, $0.57/0.56/0.60$ and $0.63/0.51/0.65$.  With these values --varying few percentage of the actual measure-- we conclude that these trends are not more tied to one parameter than to another or at least that with the present sample a difference cannot be established.

\end{appendix}

\end{document}